\documentclass[sn-nature]{sn-jnl}
\usepackage{iftex}
\usepackage{xparse}
\usepackage{tikz}

\makeatletter
  \@twosidefalse
  \@mparswitchfalse
\makeatother

\newlength{\tablewidth}
\ifpdftex
    \usepackage[T1]{fontenc}
\else
    \usepackage{fontspec}
\fi

\usepackage{amssymb}
\usepackage{amsthm}
\usepackage{amsmath}

\usepackage{booktabs}
\usepackage{multicol}

\usepackage{siunitx}
\usepackage{subcaption}
\usepackage[title]{appendix}

\usepackage[nameinlink, noabbrev]{cleveref}

\RenewDocumentCommand \Re { m } {\operatorname{Re}(#1)}
\RenewDocumentCommand \Im { m } {\operatorname{Im}(#1)}

\NewDocumentCommand \avg { m } {\langle #1 \rangle}

\RenewDocumentCommand \paragraph { m } {\textbf{#1}.}

\newif\ifresponse
\responsefalse

\ifresponse
    \definecolor{ReviewerColor}{HTML}{D95319}
    \NewDocumentCommand \Revision { +m } {\textcolor{ReviewerColor}{#1}}
\else
    \NewDocumentCommand \Revision { +m } {#1}
\fi

\begin{document}


\title{Fractal geometry predicts dynamic differences in structural and functional connectomes}

\author*[1]{\fnm{Anca} \sur{Rădulescu}}
\email{radulesa@newpaltz.edu}
\affil*[1]{\orgdiv{Department of Mathematics},
           \orgname{SUNY New Paltz}, \city{New Paltz}, \state{New York}, 12561, \country{USA}}
\author[2]{\fnm{Eva} \sur{Kaslik}}
\affil[2]{\orgdiv{Department of Computer Science},
          \orgname{West University of Timi\c{s}oara}, \city{Timisoara}, \state{Timi\c{s}}, 300223, \country{Romania}}
\author[3]{\fnm{Alexandru} \sur{Fikl}}
\affil[3]{\orgdiv{Institute for Advanced Environmental Research},
          \orgname{West University of Timi\c{s}oara}, \state{Timi\c{s}}, 300223, \country{Romania}}
\author[4]{\fnm{Johan} \sur{Nakuci}}
\affil[4]{\orgname{U.S. ARMY DEVCOM Army Research Laboratory}, \city{Adelphi}, \state{Maryland}, 20783, \country{USA}}
\author[5]{\fnm{Sarah} \sur{Muldoon}}
\affil[5]{\orgdiv{Department of Mathematics, Institute for Artificial Intelligence and Data Science, and Neuroscience Program},
          \orgname{University at Buffalo}, \city{Buffalo}, \state{New York}, 14260, \country{USA}}
\author[6]{\fnm{Michael} \sur{Anderson}}
\affil[6]{\orgdiv{Department of Data Analytics},
          \orgname{Georgia Tech}, \city{Atlanta}, \state{Georgia}, 30332, \country{USA}}


\abstract{
Understanding the intricate architecture of brain networks and its connection to brain function is essential for deciphering the underlying principles of cognition and disease. While traditional graph-theoretical measures have been widely used to characterize these networks, they often fail to fully capture the emergent properties of large-scale neural dynamics. Here, we introduce an alternative approach to quantify brain networks that is rooted in complex dynamics, fractal geometry, and asymptotic analysis. We apply these concepts to brain connectomes and demonstrate how quadratic iterations and geometric properties of Mandelbrot-like sets can provide novel insights into structural and functional network dynamics. Our findings reveal fundamental distinctions between structural (positive) and functional (signed) connectomes, such as the shift of cusp orientation and the variability in equi-M set geometry. Notably, structural connectomes exhibit more robust, predictable features, while functional connectomes show increased variability for non-trivial tasks. We further demonstrate that traditional graph-theoretical measures, when applied separately to the positive and negative sub-networks of functional connectomes, fail to fully capture their dynamic complexity. Instead, size and shape-based invariants of the equi-M set effectively differentiate between rest and emotional task states, which highlights their potential as superior markers of emergent network dynamics. These results suggest that incorporating fractal-based methods into network neuroscience provides a powerful tool for understanding how information flows in natural systems beyond static connectivity measures, while maintaining their simplicity.
}


\maketitle

\noindent {\bf Lead paragraph.} \emph{Fractal research transformed how we understand complexity, revealing how simple rules can generate intricate, emergent behavior. The Mandelbrot set -- created by iterating a basic quadratic function -- became an iconic example, illustrating self-similarity and nonlinear feedback in complex systems. Building on this legacy, we developed a generalizable framework for analyzing dynamics in complex networks. At its core is the equi-M set, a new object that extends classical complex dynamics into high-dimensional, evolving systems. Unlike traditional network analysis, which emphasizes static features such as node degree or clustering, our approach captures how behavior emerges over time through interactions across the network. This dynamic perspective opens new pathways for understanding complex systems in both natural and engineered contexts.}

\section{Introduction}
\label{sc:intro}

Fractal research transformed how we understand complexity, revealing how simple rules can generate intricate, emergent behavior~\cite{feder2013fractals}. Fractal geometry has long been recognized as a powerful framework for describing self-similar structures in nature, from vascular branching~\cite{gabrys2005fractal,jayalalitha2008fractal} to neuronal morphology~\cite{karperien2016morphology,karperien2024morphology}. In particular, the Mandelbrot set -- created by iterating a basic quadratic function -- became an iconic example, illustrating self-similarity and nonlinear feedback in complex systems~\cite{douady1984exploring,mandelbrot2004fractals}.

Building on this legacy, we previously developed a generalizable framework for analyzing dynamics in complex networks~\cite{Ariel, Simone,AC2,gender,D2}. At its core is the equi-M set, a new object that extends classical complex dynamics into high-dimensional, evolving systems. We explore the role of equi-M sets as a tool for characterizing the global organization of brain networks. By mapping the asymptotic behavior of iterated quadratic dynamics onto empirical connectomes, we generate novel descriptors that integrate connectivity structure with properties of the emerging dynamics. Unlike traditional network analysis, which emphasizes static features such as node degree or clustering, our approach captures how behavior emerges over time through interactions across the network.

Structural and Functional brain connectomes exhibit complex topologies that reflect the brain’s capacity for information processing~\cite{rezaeinia2020identifying}, adaptability~\cite{stampanoni2019synaptic}, and resilience~\cite{puxeddu2024relation}. Applying this methodology to Structural and Functional connectomes derived from the Human Connectome Project, we uncover key differences in the topological and geometric features of equi-M sets. Structural connectomes, which are inherently symmetric and positive, exhibit equi-M sets with robust and consistent cusp-like features that reflect unifying features in network organization. In contrast, Functional connectomes, shaped by the interplay of excitatory and inhibitory interactions, display more diverse and task-dependent fractal structures. Notably, equi-M sets derived from resting-state and emotion-processing Functional networks differ noticeably in their geometry. This suggests that task engagement drives greater dynamical diversity across individuals.

Our findings indicate that fractal dynamics offer a novel lens for studying brain organization that bridges the gap between network topology and emergent neural computation. By leveraging equi-M sets as network dynamics assessment tools, we open new avenues for the dynamical classification of brain states.

\subsection{Complex quadratic dynamics}
\label{sc:intro:complex_maps}

Historically, the study of single map complex iterations emerged at the beginning of the twentieth century, with the work of Pierre Fatou~\cite{fatou1920equations} and Gaston Julia~\cite{julia1918memoire}. It describes (by means of very simple iterative rules) rich, complex systems that produce objects with intricate fractal geometry. Despite the seemingly elementary and natural structure of such systems, investigating them required the development of new areas of mathematics, encompassing real and complex analysis, topology, and computational methods. However, even as important progress has been made, many questions remain open~\cite{devaney2007open}. Although emerging areas of dynamics have been widely applied to understanding nonlinear and chaotic phenomena in various natural systems, the application of complex dynamics has yet to see widespread exploration in many domains, including neuroscience.

Discrete dynamics of single iterated transformations in the complex plane have developed over a few decades into a rich research field~\cite{carleson1996complex}. In particular, quadratic iterations represent a simple, yet incredibly rich, family of examples. They have first been studied on the real line, where they are associated with cascades of period-doubling bifurcations and universality. In turn, complex quadratic iterations in the family $f_c(z) = z^2 + c$, with $c \in \mathbb{C}$, provide textbook recipes for creating asymptotic fractal sets in the complex plane.

For iterations of a single quadratic map, one of the most striking results is that the trajectory (or orbit) of the critical point $z = 0$ under iterations of the map captures global information on the behavior of all other orbits. The property of the critical orbit being bounded is particularly important in the context of long-term iterations. The Mandelbrot set appeared in the 1960s from the need to obtain an overarching description of this property over the whole family of quadratic maps. It is defined as the set of all $c \in \mathbb{C}$ for which the critical orbit of the complex quadratic function $f_c(z) = z^2+c$ is bounded. This is a fractal with fascinating characteristics that have been extensively analyzed since its introduction. The topological organization of the Mandelbrot set into hyperbolic bulbs (leading to its shape and fractal geometry) classifies all possible combinatoric behaviors that orbits can have in the family $f_c$. This makes the Mandelbrot set both a staple in discrete dynamical systems and a canonical representation for more general phenomena.

\subsection{Complex quadratic maps on networks}
\label{sc:intro:cqn}

Our work extends this framework from its original scope of single map iterations to the broader context of complex dynamical networks (CGNs). More specifically, we consider networks in which the coupled nodes are identical complex quadratic maps \Revision{(CQNs)}. \Revision{Quadratic maps were chosen for their historical importance and relative simplicity, both of which are desired properties in this proof-of-principle demonstration of the concept. Alternative families, such as higher-degree polynomials or rational maps, introduce additional critical points and poles, which make the associated theory substantially more complicated. Future studies will extend this approach to a broader class of discrete models for network nodes, with the aim of identifying both differences and universal properties across families.}

The overall evolution of this discrete system takes the form of an iteration in $\mathbb{C}^n$, which acts in each component as
\begin{equation} \label{eq:cqn}
z_k \mapsto \left(\sum_{j = 1}^n A_{kj} g_{kj} z_j\right)^2 + c.
\end{equation}

\Revision{The coupling is specified by the binary adjacency matrix $A \in \mathbb{R}^{n \times n}$, whose binary entries $A_{ij}$ mark an edge between nodes $i$ and $j$ in the graph, and signed weights $g \in \mathbb{R}^{n \times n}$, whose entries $g_{ij}$ provide the weight of the corresponding edge.} Since nodes have the same intrinsic dynamics dictated by the parameter $c$, the emerging network dynamics are a direct reflection of the value of $c$ and of the specific  $(A, g)$ network architecture at play. By analogy to the single map case, we define the \emph{equi-M} set of the network as the set of all $c \in \mathbb{C}$ for which the critical point $0 \in \mathbb{C}^n$ remains bounded under iteration. In this generalized context, many of the results on the properties and structure of the traditional Mandelbrot set are lost. Our theoretical work shows that the equi-M set no longer provides a full combinatorial description of the asymptotic behavior of the map~\cite{D2}. However, our results strongly suggest that equi-M sets have quantifiable topological and fractal properties that reflect the network architecture, making them good candidates for testing and classifying how dynamic patterns evolve for different architectural profiles. In other words, the equi-M set still captures enough of the long-term dynamic behavior to be used as an assessment of the network's dynamic range.

\subsection{Using CQNs to identify patterns in brain networks}
\label{sc:intro:patterns}

Building on previous work, we demonstrate that the size and shape of the equi-M set can capture significant differences across subjects and connectome types, such as gender. This paper specifically focuses on comparing Structural (positive) connectomes with Functional (signed) connectomes, analyzing data from Rest and Emotional task states. We will introduce a new set of geometric landmarks and measures designed to better characterize the shape and geometry of equi-M sets, enabling a more precise analysis of these connectomes.

In previous work~\cite{D2}, we showed that the size and shape of the equi-M set can capture differences between subjects and between predefined connectome types (e.g., they showed significant differences between genders). In this paper, we will focus on addressing differences between Structural (positive) connectomes and Functional (signed) connectomes, corresponding to Rest vs. Emotional task. First, we will describe a collection of new landmarks and measures tailored to capture the shape and geometry of the equi-M set. Then, we use these measures for the Structural connectome data and compute correlations with the traditional graph-theoretical measures. This allows us to investigate to what extent they reflect patterns in the underlying network architecture. For Functional connectomes, this approach presents difficulties, because there is no equivalent set of graph-theoretical measures designed for signed networks. In this case, most network studies use graph-theoretical measures for the positive and negative parts of the network separately. However, we propose that these partial measures are unlikely to serve as effective predictors of the dynamic behavior of the network. To support this, we investigate whether these patterns independently correlate with the shape and geometry of the equi-M sets.

We will next explore the extent to which equi-M sets can distinguish between Structural and Functional data, as well as between Rest and Emotional task states. This analysis will allow us to assess the potential of equi-M sets as a descriptor of network dynamics and to provide insights into the dynamics of Structural and Functional networks.

\section{Methods}
\label{sc:methods}

\subsection{Dataset description}
\label{sc:methods:dataset}

\paragraph{Data collection and preprocessing} Our analysis is based on tractography-derived and functional neural connectivity data obtained from the S1200 Q4 release of the Human Connectome Project (HCP)~\cite{van2013wu}. \Revision{Among others, the data contains Magnetic Resonance Imaging (MRI) scans from approximately 1200 healthy young adults between the ages of 20-40 balanced for sex, who underwent extensive resting-state, task, structural, and diffusion MRI. Conducting the current preliminary analysis on the whole dataset would have posed significant computational challenges. Therefore, this study analyzed a subgroup of 48 subjects, previously used by one of the authors in a statistical network study~\cite{nakuci2025neuroreceptors,nakuci2025quantifying}. The subjects were chosen to belong to the same batch for both functional and diffusion MRI, ensuring consistency in scanner protocols or software version that might affect the imaging. Hence, the subjects are a representative random subsample of the HCP dataset, comparable in size to our previous statistical studies on Structural connectomes. This choice is optimal for a proof-of-principle study, as it ensures computational feasibility without compromising  statistical power. Furthermore, it can be expanded to the larger data set, if needed for validation and testing replicability of the study.} \\

\noindent \paragraph{Structural connectomes} The preprocessed diffusion weighted images from the HCP were used to construct Structural connectomes for each subject. As a first step, the data were reconstructed using Generalized Q-sampling Imaging (GQI)~\cite{yeh2010generalized}. Fiber tracking was performed until 250,000 streamlines were reconstructed with an angular threshold of \ang{50}, step size of \qty{1.25}{\milli\meter}, a minimum length of \qty{10}{\milli\meter} and a maximum length of \qty{400}{\milli\meter}. Fiber tracking was performed using DSI Studio using a modified FACT algorithm~\cite{yeh2013deterministic}. Streamline counts were estimated for parcellation schemes based on the Schaefer atlas containing 200 brain regions~\cite{schaefer2018local}.

For each subject, a weighted nonnegative symmetric structural connectivity matrix was constructed from the connection strengths, based on the number of streamlines connecting two regions. This connectivity matrix was normalized by dividing the number of streamlines between each two coupled regions by the combined volumes of the two regions. Then, the $200 \times 200$ matrix for each subject was globally normalized by dividing by its largest entry, to obtain a positive symmetric matrix with entries in $[0, 1]$. This resulting matrix has been used in the simulation of the CQN~\eqref{eq:cqn} for all Structural connectomes.\\

\noindent \paragraph{HCP Resting-State and Emotion Processing Task} We used preprocessed resting-state functional MRI (fMRI) data from the same 48 participants. For the resting-state analysis, participants underwent four sessions of \qty{15}{\minute} resting-state scanning sessions. The fMRI volumes were recorded using a customized 3T Siemens Connectome Skyra scanner using an echo-planar imaging (EPI) sequence with the acquisition parameters TR = \qty{0.72}{\second}, TE = \qty{33.1}{\milli\second}, 72 slices, \qty{2.0}{\milli\meter} isotropic, and a FOV (field of view) = 208 $\times$ \qty{180}{\milli\meter}.

For the emotion processing task, the participants were presented with either angry or fearful expressions and had to decide which of the two faces presented at the bottom of the screen match the face at the top of the screen~\cite{barch2013function}. Trials were presented in blocks of 6 trials of the same task (face or shape). Whole-brain EPI acquisitions were collected with a 32 channel head coil on a modified 3T Siemens Skyra with parameters TR = \qty{720}{\milli\second}, TE = \qty{33.1}{\milli\second}, flip angle = \ang{52}, BW = \qty{2290}{\hertz} per pixel, in-plane FOV = 208 $\times$ \qty{180}{\milli\meter}, 72 slices, \qty{2.0}{\milli\meter} isotropic voxels and a multiband acceleration factor of 8.

In addition to the preprocessing steps part of the HCP pipeline, we removed the global signal, white matter, and cerebral spinal fluid through standard regression. The data was mapped on the same Schaefer atlas to derive brain activity for 200 regions of interest because voxel-wise estimates can be unstable and noisy.  Next, group-level functional connectivity was estimated using the Pearson correlation on the pooled data. The resulting Functional connectome is a symmetric and signed matrix, in which a positive entry between node $i$ and node $j$ can be interpreted as mutually reinforcing activity in the two nodes, and a negative entry represents opposite trends in temporal activity of the two nodes during the functional scan.

\subsection{Network measures}
\label{sc:methods:graph}

\paragraph{Graph-theoretical measures for Structural connectomes} For Structural networks, we use a collection of graph-theoretical measures that can jointly provide crucial information on the architecture of the connectomes. These have historically been used to differentiate between connectomes at both the individual and group level. The measures are listed below, each with corresponding property of the graph that it aims to capture~\cite{rubinov2010complex}.

\begin{description}
\item[Betweenness Centrality] ($BC$) is a global network measure of how often a node lies on the shortest path between all pairs of nodes in the network.
\\

\item[Assortativity] ($AS$) measures how likely nodes in a network are to connect with other nodes that have the same degree.
\\
\item[Eigencentrality] ($EC_i$) measures how influential a brain region $i$ is in a network. A high value indicates that a brain region is connected to other highly influential regions~\cite{bonacich2007some,lohmann2010eigenvector}.
\\
\item[Weighted Clustering Coefficient] ($CC_i$) for a brain region $i$ is the intensity of triangles in a network and assess the extent of segregation in a network~\cite{rubinov2010complex}. It is calculated as
\[
CC_i = \frac{1}{b_i(b_i-1)} \sum_{j, k = 1}^n \sqrt[3]{w_{ij} w_{ik} w_{jk}},
\]
where $w = A g$ is the weighted adjacency matrix and $b$ is the number of connections for brain region $i$.
\\
\item[Weighted Degree] ($D_i$) of a node $i$ is defined as the sum of all connection weights of the edges adjacent to that node. This measure is an indicator of how important a node is for communication between brain regions.
\\
\item[Global Efficiency] ($GE$) is the average (over all node pairs in the network) of the shortest path length between those two nodes.
\\
\item[Local Efficiency] ($LE_i$) of a network around a vertex $i$ is the average of the inverses of the distances (shortest path lengths) between the node's neighbors through the rest of the network, when the vertex is removed~\cite{sporns2004motifs}.
\\
\item[Transitivity] ($TR$) is the ratio of ``triangles to triplets'' in the network~\cite{sporns2004motifs}.
\\
\item[Strength of 3D Motifs] ($M$) are patterns of network connections among a specified number of nodes forming a sub-graph $g$, which act as building blocks for complex networks~\cite{sporns2004motifs}. Given that our Structural connectivity matrices are undirected, our analysis focused on two types of three-node motifs: \begin{tikzpicture}
  \node[circle, draw, fill=white, inner sep=0pt, minimum size=1mm] (1) at (0,0) {$1$};
  \node[circle, draw, fill=white, inner sep=0pt, minimum size=1mm] (3) at (1.2/2,0) {$3$};
  \node[circle, draw, fill=white, inner sep=0pt, minimum size=1mm] (2) at (1.2/4,1/2) {$2$};

  \draw (1) -- (2) -- (3) -- cycle;
\end{tikzpicture} ($M1$, two links) and
\begin{tikzpicture}
  \node[circle, draw, fill=white, inner sep=0pt, minimum size=1mm] (1) at (0,0) {$1$};
  \node[circle, draw, fill=white, inner sep=0pt, minimum size=1mm] (3) at (1.2/2,0) {$3$};
  \node[circle, draw, fill=white, inner sep=0pt, minimum size=1mm] (2) at (1.2/4,1/2) {$2$};

  \draw (1) -- (2) -- (3) -- (1);
\end{tikzpicture} ($M2$, triangle).

To assess how strongly a brain region participates in a specific motif, we calculated the motif coherence $Q_g$ for each brain region~\cite{onnela2005intensity}. We used motif coherence because it accounts for low probability values due to one connection weight being low versus all connection weights being low~\cite{onnela2005intensity}. For each individual motif pattern, $Q_g$ is defined as
\[
Q_g = |l_g| \frac{\left(\prod_{(i, j) \in g} A_{ij}\right)^{\frac{1}{|l_g|}}}
                 {\sum_{(i, j) \in g} A_{ij}},
\]
where $l_g$ is the set of edges in the specified sub-graph $g$, $|l_g|$ is the number of edges in that sub-graph, and $A_{ij}$ is the connection strength between brain region $i$ and $j$. Motif coherence was calculated using the Brain Connectivity Toolbox~\cite{rubinov2010complex}.
\end{description}

For each of the node-specific measures $X_i$, the network-wide average $\avg{X}$ was computed and used as a global measure in our analysis.\\

\noindent {\bf Graph-theoretical measures for Functional connectomes.} A common approach to analyzing signed networks, particularly in the context of brain connectomes, is to separate the positive and negative sub-networks and examine them independently. \Revision{More specifically, the two sub-networks are constructed using the modified weights
\[
g_{kj}^+ =
\begin{cases}
g_{kj}, & \quad g_{kj} \ge 0, \\
0, & \quad \text{otherwise},
\end{cases}
\quad \text{and} \quad
g_{kj}^- =
\begin{cases}
g_{kj}, & \quad g_{kj} < 0, \\
0, & \quad \text{otherwise}.
\end{cases}
\]
}

\noindent This method assumes that nodes exhibiting strong positive correlations (highly co-activated regions) and those with negative correlations (anti-correlated activity) form distinct functional communities. By analyzing these sub-networks separately, investigators aim to capture the intrinsic dynamics of cooperative and opposing activity patterns within the network~\cite{tang2023signed}. However, this approach implicitly assumes that positive and negative sub-networks can independently encode key aspects of the network's emergent behavior, an approach that has faced scrutiny~\cite{zhan2016importance,zhan2017significance}. In this study, we aim to examine the validity of this assumption by exploring the extent to which overall network dynamics are influenced not only by these segregated sub-networks, but also by the intricate interplay between them. Therefore, for Functional networks, we separately compute the measures for the positive / negative sub-networks and also for the absolute value of the full functional connectome. We will call these the networks associated with the connectome.

\subsection{Construction of equi-M sets}
\label{sc:methods:contruction}

The equi-M sets are computed numerically for the connectome of each of the 48 subjects using the \texttt{netbrot} visualizer \cite{netbrot}. Each group of equi-M sets is rendered in an appropriately scaled bounding box with a resolution of $1200 \times 1200$ pixels. The iteration proceeds for a maximum of $512$ steps using an escape radius $R_e = 100$ until it decides if the chosen point is in the set. The chosen parameters give reasonably accurate numerical representations for the theoretical equi-M set. Illustrations of the corresponding equi-M sets for all 48 subjects are included in \Cref{fig:structural_equi_m_sets,fig:rest_equi_m_sets,fig:emotion_equi_m_sets}, for all Structural and Functional connectomes that were used in this analysis. \\

\noindent \paragraph{Statistical and prototypical representations of equi-M sets} For each of the three types of connectomes (Structural, Rest, and Emotion) we computed statistical equi-M sets over all subjects in the data set. For each parameter point $c$ in the corresponding bounding box, we computed the fraction $f$ of the number of connectomes in the group, such that $c$ is in the equi-M set for that connectome. We illustrate these ``stochastic'' connectomes using \texttt{pcolor} plots, in which the \Revision{``hot'' color gradient (from yellow to black)} represents the fraction $f$. We also took a mean-field approach to the data, and we generated equi-M sets for \emph{prototypical} connectomes, computed as the mean of all connectomes in a data set (group average). These were defined in previous work~\cite{D2} and are described in more detail in \Cref{sc:results}.

\subsection{Geometric measures of equi-M sets}
\label{sc:methods:geometric}

\begin{figure}[h!]
\centering
\includegraphics[width=0.85\linewidth]{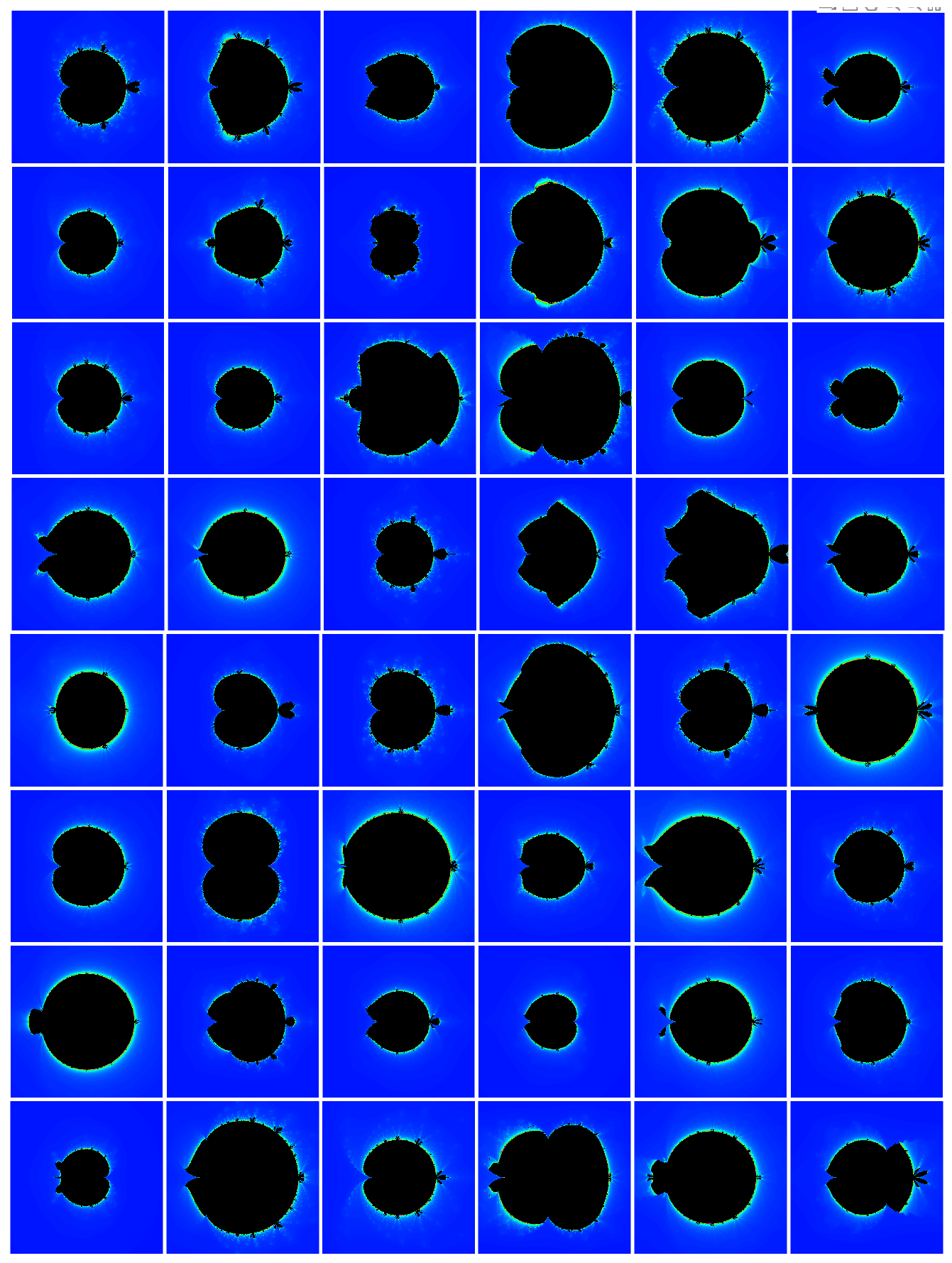}

\caption{\textbf{Equi-M sets for Functional Emotion connectomes} for all subjects in our data set. All sets are represented in the complex region $[-0.03,0.03] \times [-0.03,0.03]$, in resolution $1200 \times 1200$. The equi-M set in each panel appears in black, and the colors represent the number of iterations needed by each point in order to exit the escape disc. The sets show a wide variety of shapes across the data set.}
\label{fig:emotion_equi_m_sets}
\end{figure}

The equi-M sets for all Structural and Functional connectomes in our dataset are shown in \Cref{fig:structural_equi_m_sets,fig:rest_equi_m_sets,fig:emotion_equi_m_sets} in order to allow a general view of their diversity in size and geometry and to facilitate direct visual comparisons. We use a variety of approaches to capture and quantify these visual differences and to understand their relationship to the network architecture.

\begin{figure}[h!]
\centering
\begin{subfigure}{0.5\linewidth}
\centering
\includegraphics[width=0.95\linewidth]{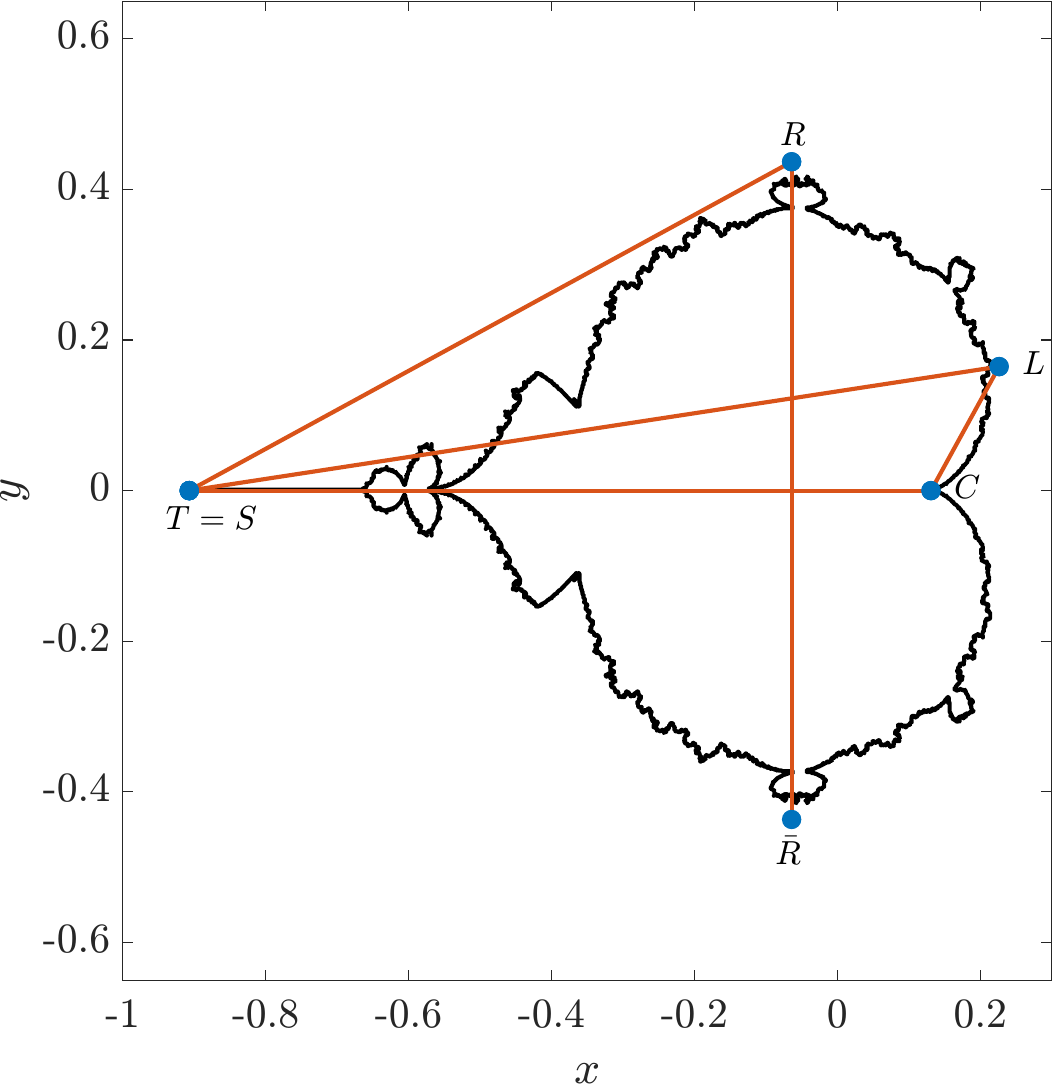}
\caption{}
\end{subfigure}%
\begin{subfigure}{0.5\linewidth}
\centering
\includegraphics[width=0.95\linewidth]{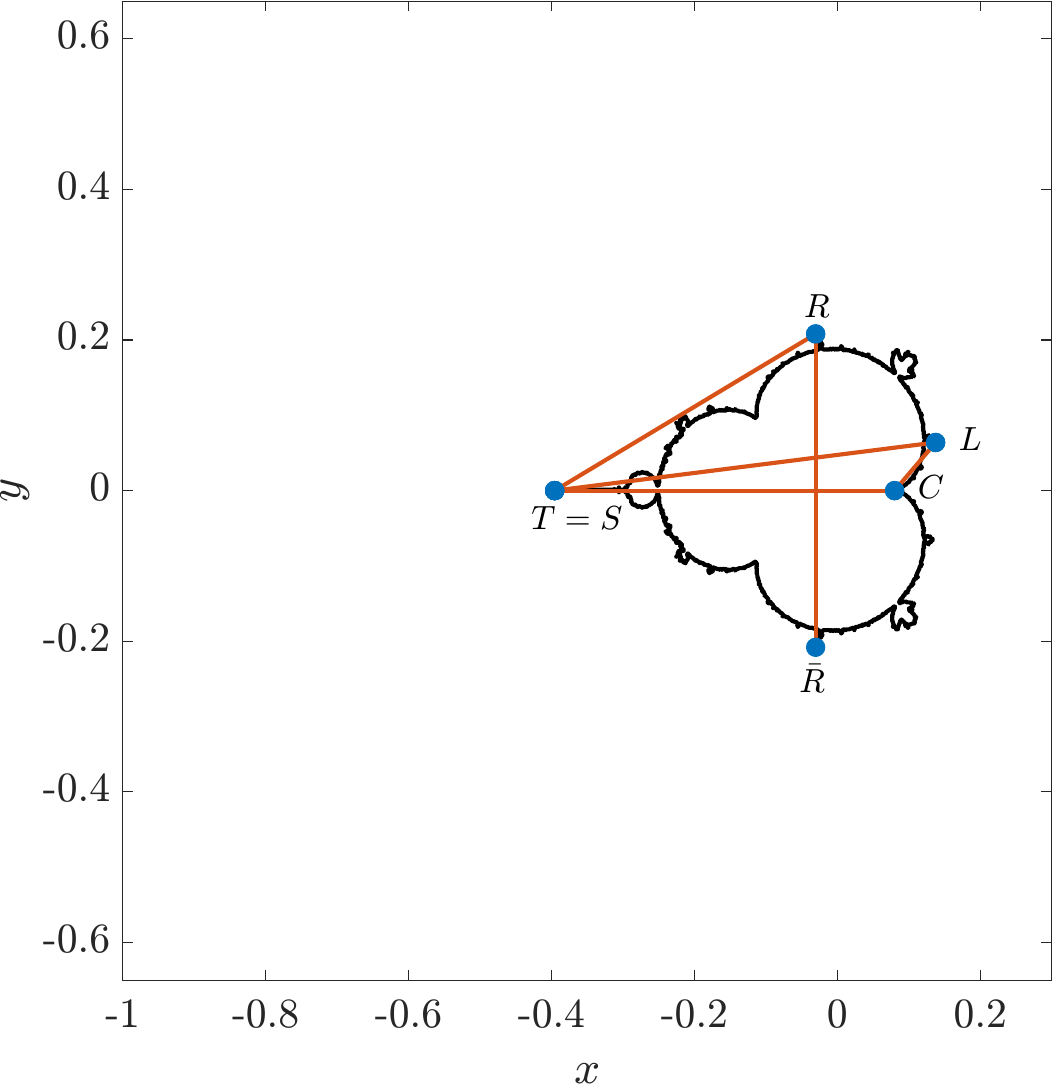}
\caption{}
\end{subfigure}

\caption{\textbf{Measurable differences in equi-M set shape} can be captured by computing a set of geometric measures, illustrated here for two random Structural connectomes in our data set: subject \#22 \Revision{(panel A)} and subject \#5 \Revision{(panel B)}. In the left panel, we have the following landmarks and distances: $C = 0.13$, $T = -0.9$, $L = 0.22 + 0.16 i$, $R = -0.06 + 0.43 i$, $d_{TC} = 1.03$, $d_{TR} = 0.94$, $d_{TL} = 1.14$, $d_{CL} = 0.19$, $d_{R \bar{R}} = 0.41$, $\epsilon = 0.87$. In the right panel: $C = 0.08$, $T = -0.39$, $L = 0.13 + 0.06 i$, $R = -0.03 + 0.2 i$, $d_{TC} = 0.47$, $d_{TR} = 0.42$, $d_{TL} = 0.53$, $d_{CL} = 0.008$, $d_{R \bar{R}} = 0.87$, $\epsilon = 0.84$.}
\label{fig:structural_landmarks}
\end{figure}

\begin{figure}[h!]
\centering
\begin{subfigure}{0.5\linewidth}
\centering
\includegraphics[width=0.95\linewidth]{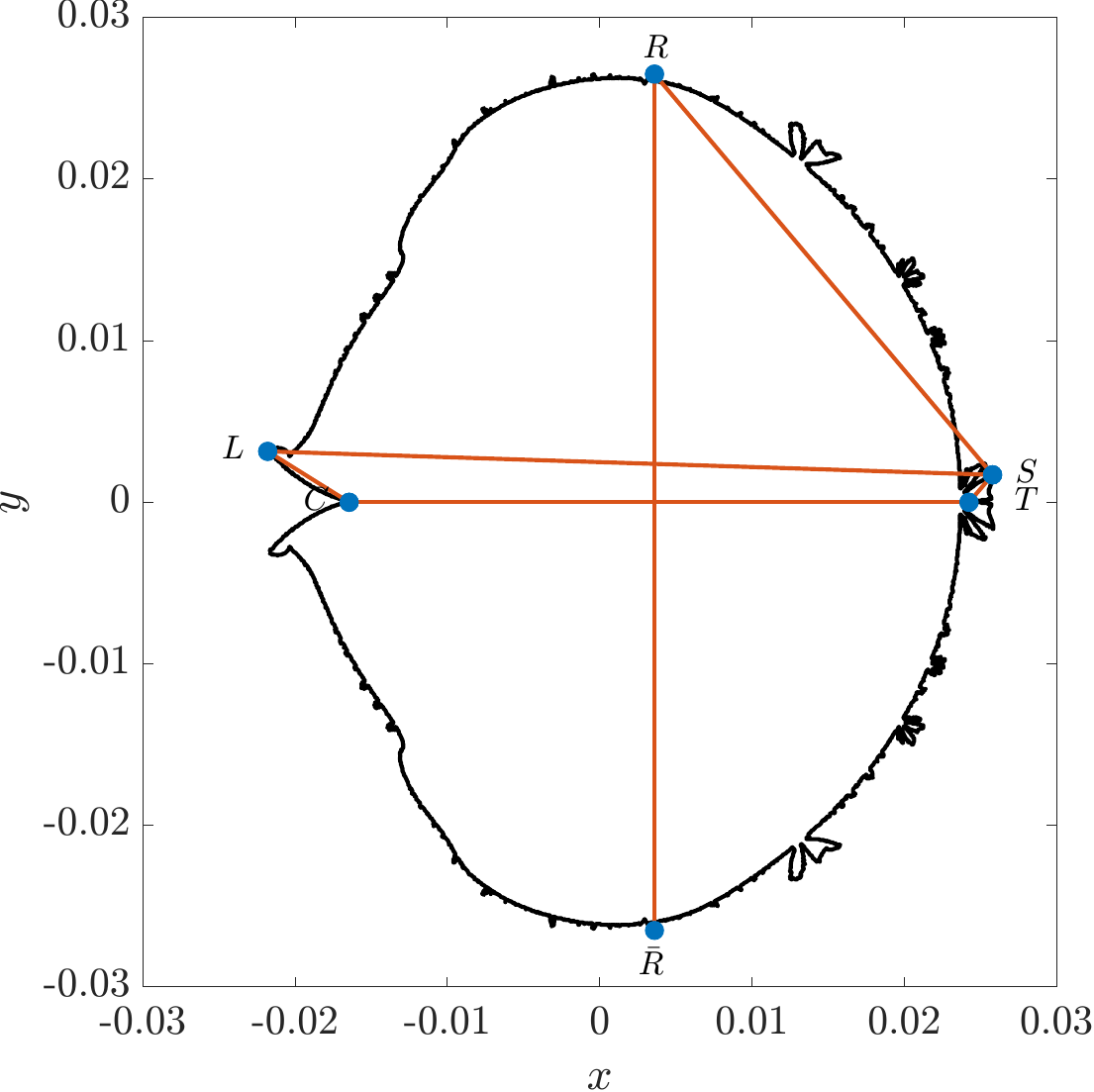}
\caption{}
\end{subfigure}%
\begin{subfigure}{0.5\linewidth}
\centering
\includegraphics[width=0.95\linewidth]{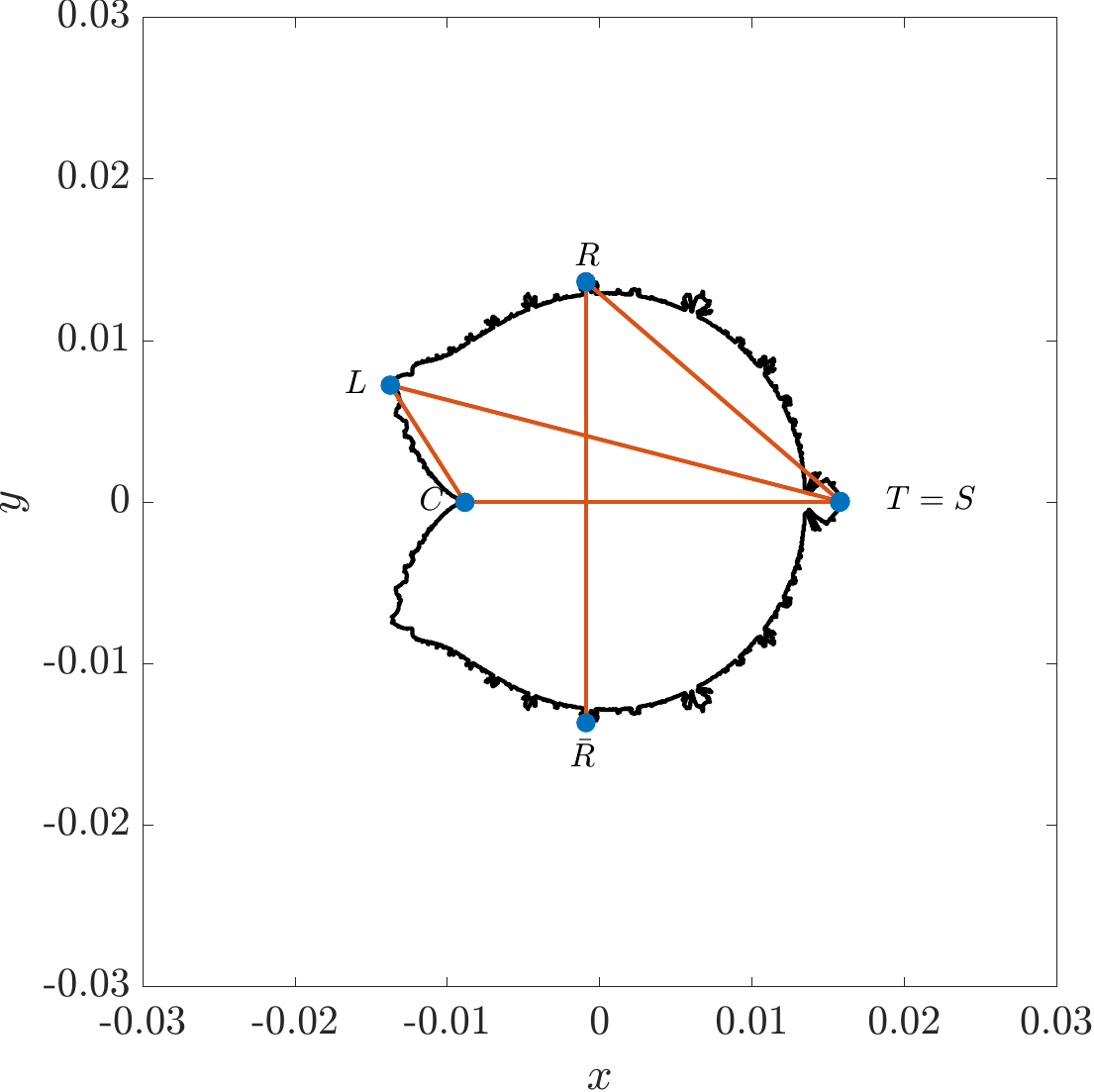}
\caption{}
\end{subfigure}

\caption{\textbf{Measurable differences in equi-M set shape} captured by computing geometric measures for sets derived from Functional connectomes. The two panels illustrate the different geometry landmarks and corresponding measures between two Emotion connectomes: subject \#28 \Revision{(panel A)} and subject \#3 \Revision{(panel B)}. In the left panel, we have the following landmarks and distances: $C = -0.01$, $T = 0.024$, $S = 0.025 + 0.001 i$, $L = -0.021 + 0.003 i$, $R = 0.003 + 0.026 i$, $d_{TC} = 0.04$, $d_{SR} = 0.03$, $d_{SL} = 0.04$, $d_{CL} = 0.006$, $d_{R \bar{R}} = 0.05$, $\epsilon = 1.3$. In the right panel: $C = -0.008$, $T = 0.015$, $S = 0.015$, $L = -0.013 + 0.007 i$, $R = -0.0006 + 0.013 i$, $d_{TC} = 0.024$, $d_{SR} = 0.021$, $d_{SL} = 0.03$, $d_{CL} = 0.008$, $d_{R \bar{R}} = 0.02$, $\epsilon=1.1$.}
\label{fig:emotion_landmarks}
\end{figure}

For each individual equi-M set, we computed a collection of measures that capture details of the 2-dimensional position, size, and shape of the set. To begin with, we noticed that the Structural sets (corresponding to symmetric connectomes with all positive entries) replicate some broad topological features of the traditional equi-M set. This is not surprising, and is explained in more detail in~\cite{D2}. In particular, the sets are symmetric about the real axis and exhibit a main cardioid that delimit the locus of $c$ for which the critical point converges to an attracting fixed point. A feature that appears to be robust among all Structural connectomes is that the main cardioid has a cusp on the right, with varying position and depth. The main cardioid extends to the left with a ``tail'' formation. However, the tail structure is not robust and presents high variability of shapes. The one consistent feature remains that the tip of the tail is the leftmost point of the set. Therefore, we used the position $C$ of the cusp (with $\Im{C} = 0$) and the position $T$ of the tip of the tail (with $\Im{T} = 0$) as landmarks for the stretch of the set along the real axis. In addition, we considered the coordinates of the point $R$ farthest from the real axis (with $\Im{R} > 0$) and the coordinates of the rightmost points (the lip of the cusp) $L$ (with $\Im{L} > 0$). To encode the general shape, we obtained the horizontal diameter $d_{TC} = |T - C|$, the vertical diameter $d_{R \bar{R}} = 2 |\Re{R}|$, the depth of the cusp as $d_{CL} = |C - L|$, the slant diameters from the tail to the highest point $d_{TR} = |T - R|$ and to the lip of the cusp $d_{TL} = |T - L|$, and the ``elongation'' of the set, as the ratio $\epsilon = d_{R \bar{R}} / d_{TC}$. The collection of these measures is expected to encompass key information about variability in shape among Structural equi-M sets (as suggested by~\Cref{fig:structural_landmarks}, in which the measures are illustrated for two example sets).

To identify similar landmarks for the Rest connectomes, we had to take into consideration the impact of the \Revision{negative signs in the} Functional connections on the topological structure. The first striking difference was that the orientation of the sets was ``reversed'' from that of the Structural connectomes. While all Rest sets still had a main cardioid body and a visually identifiable cusp, the cusp was left facing, and the (less prominent) tail formation was extending to the right. \Revision{These differences are encouraging, as they indicate from the outset that the dynamics are qualitatively distinct in positive compared to signed networks.} As before, we identified the cusp $C$, the tip of the tail $T$, the highest point $R$ and the upper lip of the cusp $L$ \Revision{for Rest connectomes}. In addition, we identify a separate marker $S$, which is the rightmost point in the set, similarly to $L$ in the Structural equi-M sets. Together with this new marker, we define a slant diameter $d_{SL} = |S - L|)$ and the depth of the tail $d_{ST} = |S - T|$. We used this expanded set of measures to characterize the shape of the Rest sets.

The Emotion sets presented with even more variability in the shape of the main body (see the complete illustration of all sets in \Cref{fig:emotion_equi_m_sets}). Although most cardioids still had a left cusp, this was replaced in some cases by a more complex formation. Similarly, while most sets had a right tail, this was ``stumpy,'' and in some cases missing or replaced by another cusp. While in our theoretical work we are trying to analytically understand the conditions behind creating main cardioid cusps, this is beyond the scope of this paper. In this project, we simply use the same set of landmarks $(C,T,R,L,S)$ as for the Rest connectomes, and the same combinations between them, to broadly estimate the shape of the sets \Revision{(as suggested by~\Cref{fig:emotion_landmarks}, in which the measures are illustrated for two example sets)}. However, in our Results and Discussion, we further comment on the significance of the differences in cusp and tail orientation and structure between the three types of connectomes.

Note that one of the staple properties of the traditional Mandelbrot set is connectedness. Establishing connectedness of our equi-M sets (for both Structural and Functional connectomes) is a difficult computational problem, due to spatial resolution limitations, and the implications in detecting thin filaments that may connect parts of the sets. Verifying connectedness is also beyond the scope of the current study.\\

\begin{figure}[ht]
\centering
\begin{subfigure}{0.25\linewidth}
\centering
\includegraphics[width=0.95\linewidth]{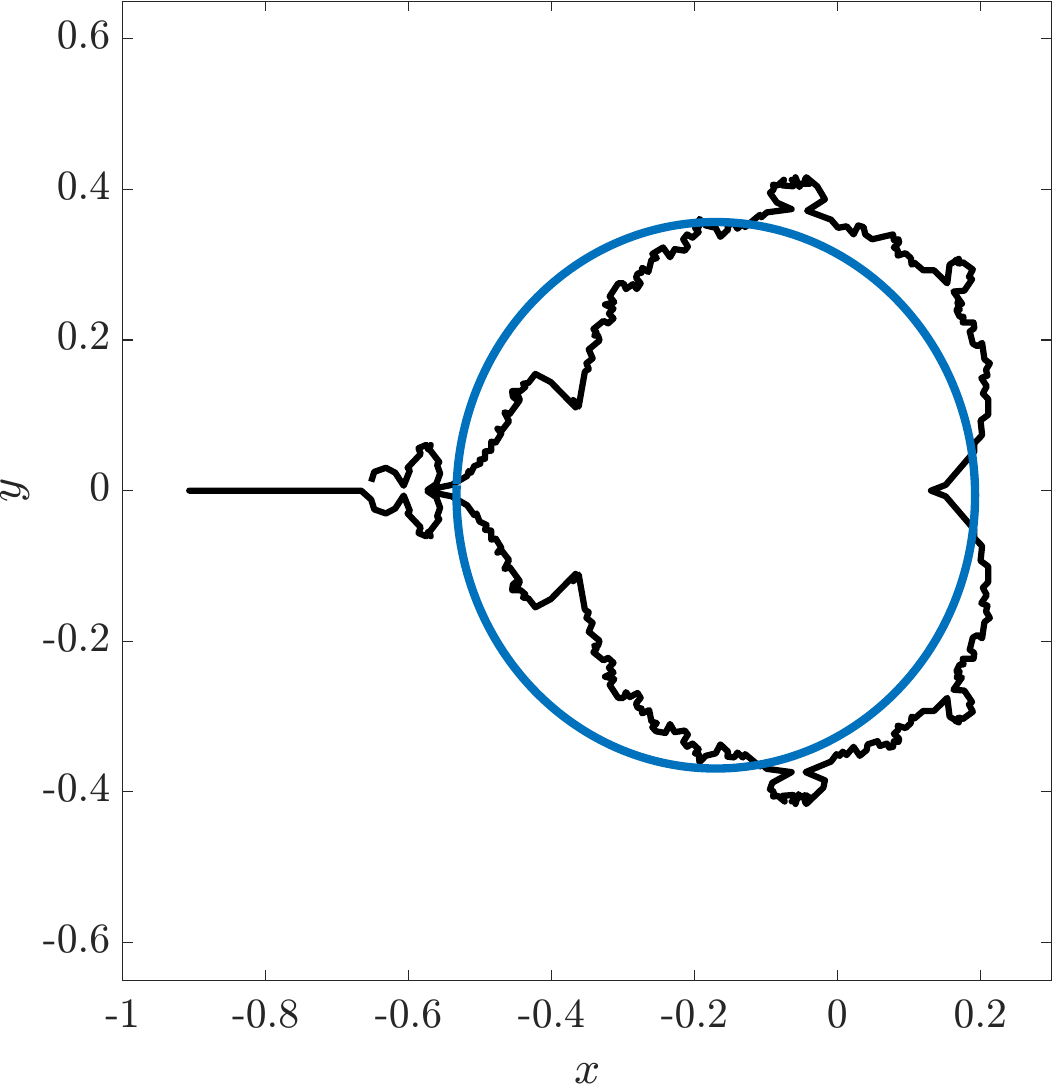}

\includegraphics[width=0.95\linewidth]{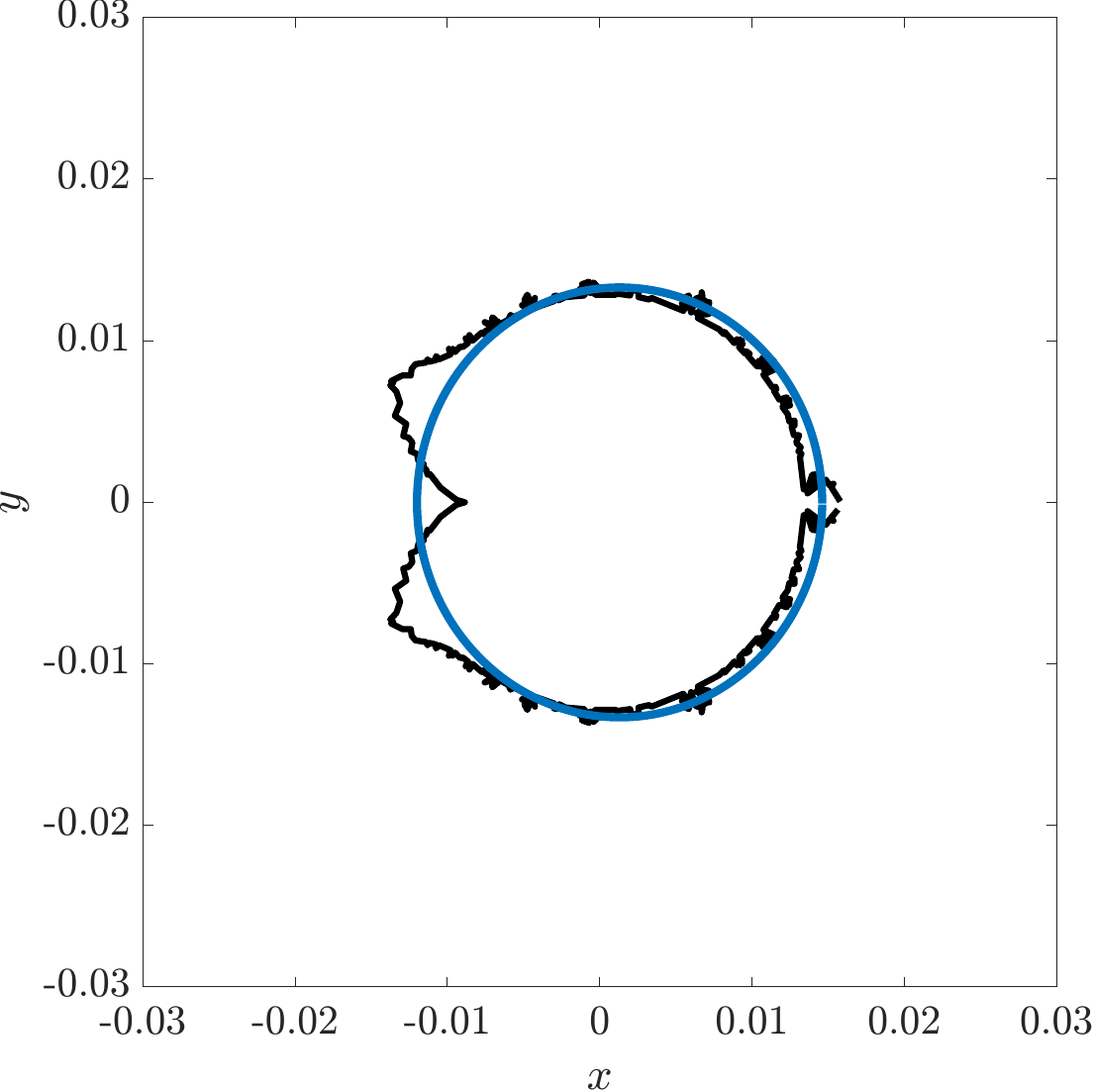}

\caption{N = 2}
\end{subfigure}
\begin{subfigure}{0.25\linewidth}
\centering
\includegraphics[width=0.95\linewidth]{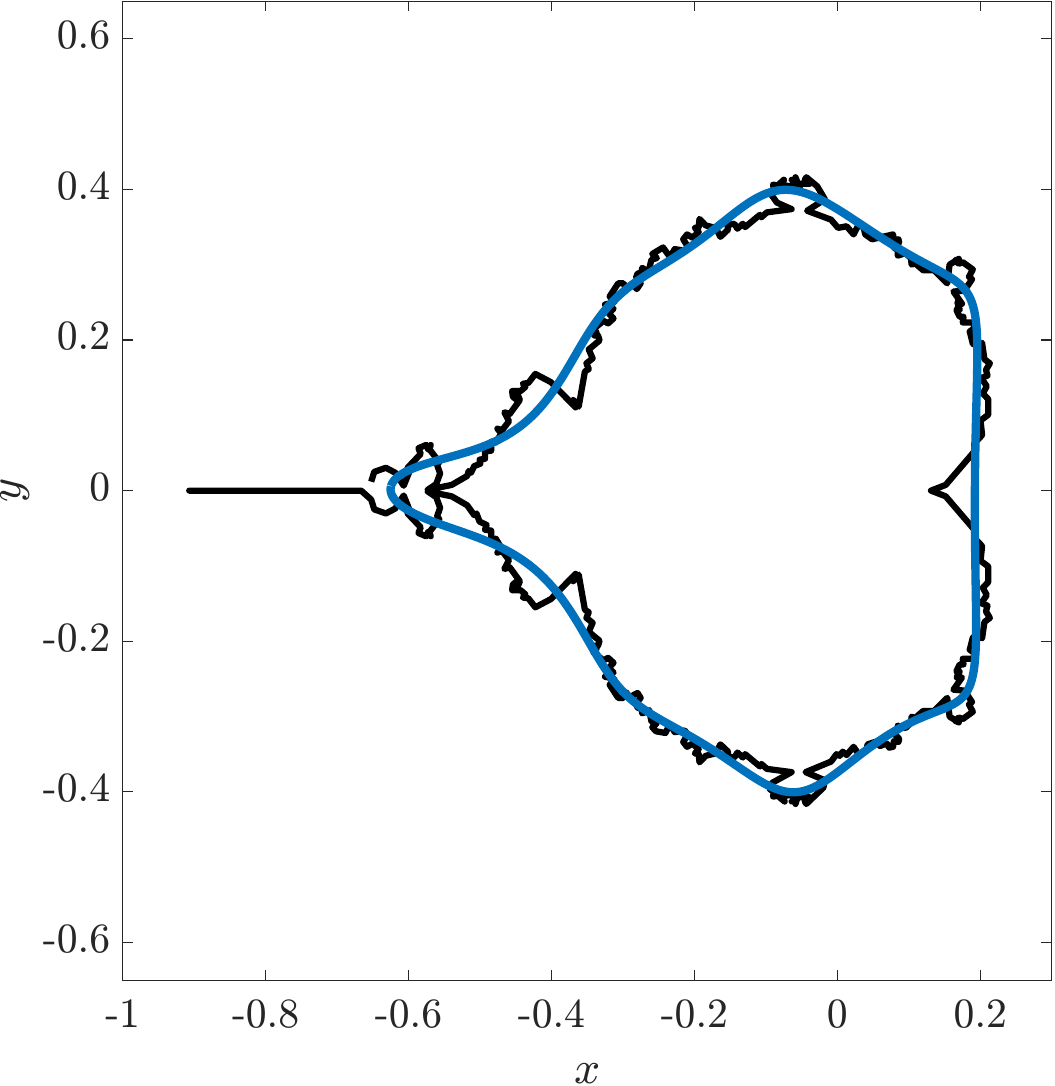}

\includegraphics[width=0.95\linewidth]{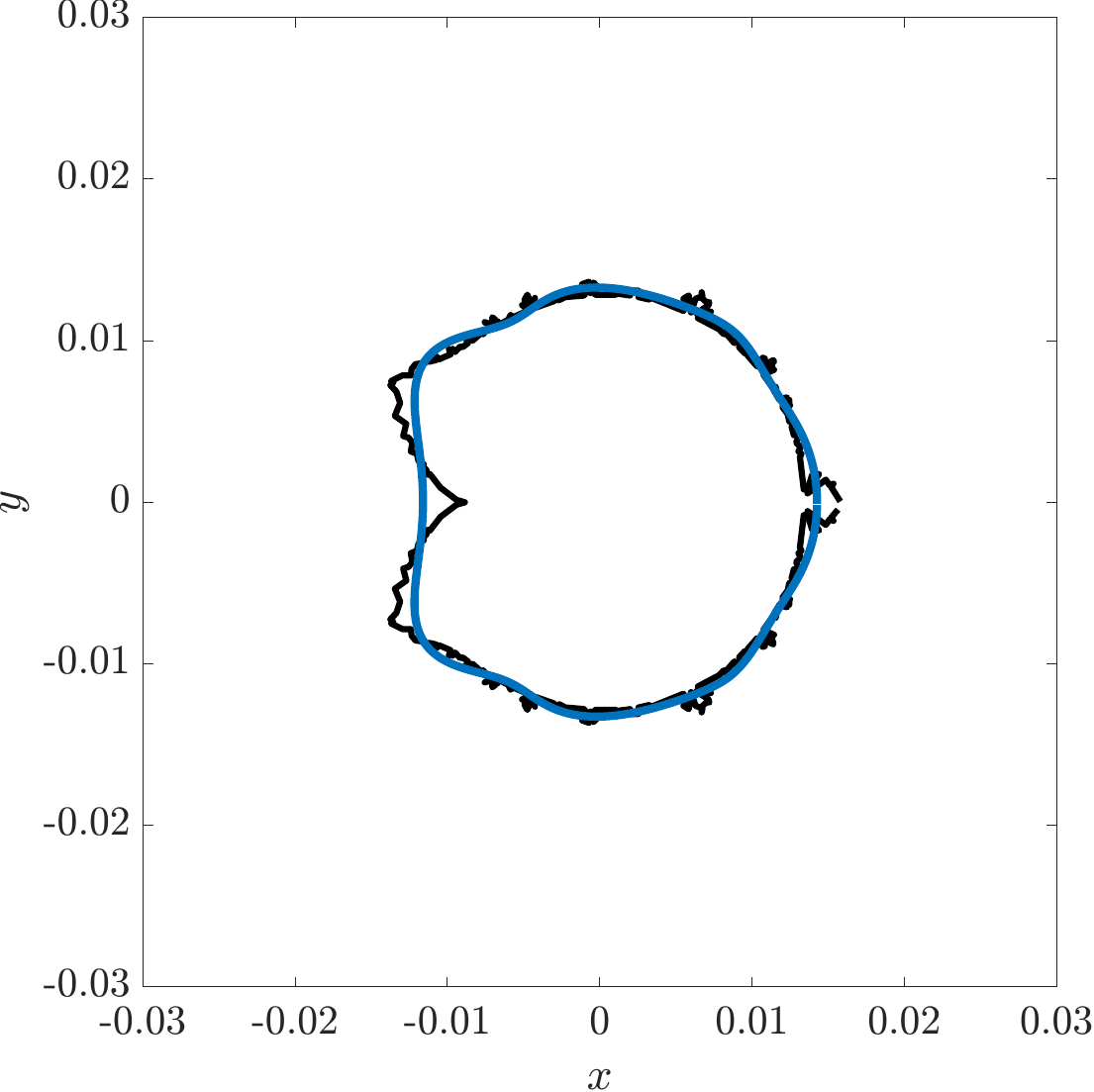}

\caption{N = 16}
\end{subfigure}
\begin{subfigure}{0.25\linewidth}
\centering
\includegraphics[width=0.95\linewidth]{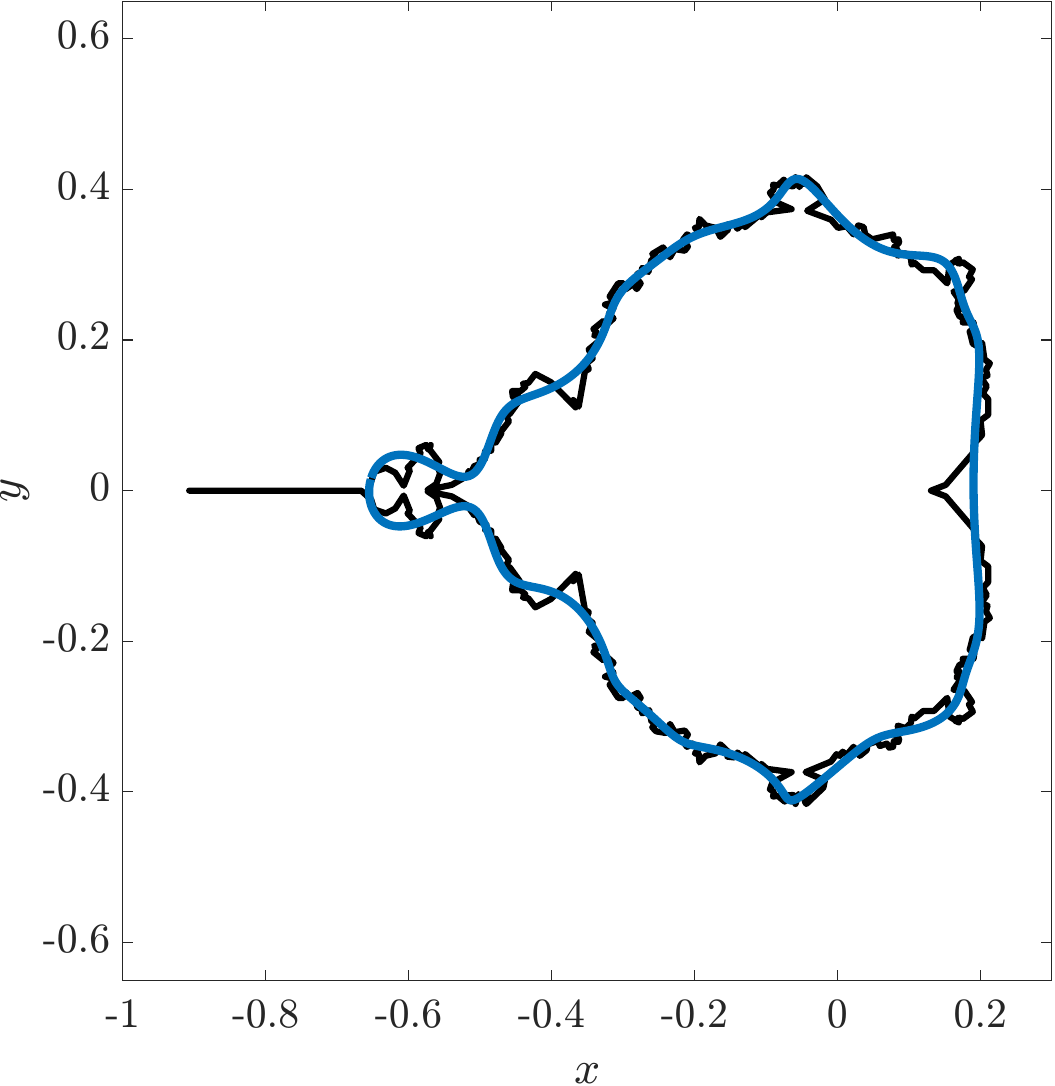}

\includegraphics[width=0.95\linewidth]{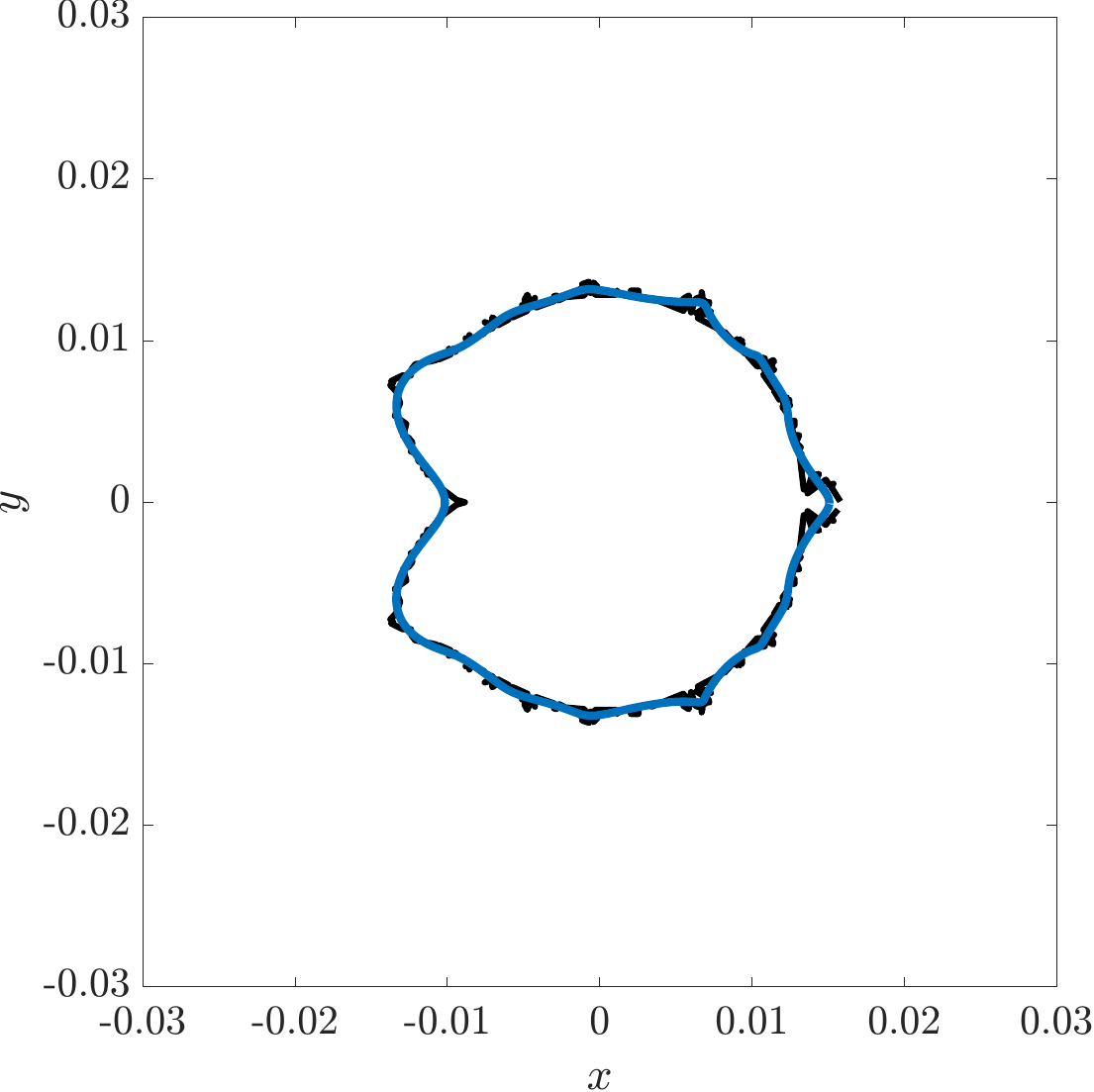}

\caption{N = 32}
\end{subfigure}

\caption{Fourier mode reconstruction of the Structural connectome from
\Cref{fig:structural_landmarks}a (top) and Functional connectome from \Cref{fig:emotion_landmarks}b (bottom) with $N + 1$ modes in $\{-N/2, \dots, N/2\}$.}
\label{fig:fourier_mode_reconstruction}
\end{figure}

\noindent \paragraph{Fourier analysis} Our simulations suggest dramatic between-subject differences in the detail and variability of the equi-M set boundary, beyond the broad geometric measures described in the previous paragraphs. To capture these subtler properties, we constructed a truncated Fourier representation of the equi-M set boundary. For each equi-M set, the boundary was first identified and approximated using the Ramer--Douglas--Peucker algorithm~\cite{douglas1973algorithms} (part of the \texttt{OpenCV} library) to a specified tolerance of $10^{-4} P$, where $P$ represents an approximation of the set perimeter. The resulting polygonal boundary approximation was then used to obtain a truncated Fourier representation (as seen in~\Cref{fig:fourier_mode_reconstruction}). Note that the boundary of an equi-M set is fractal in nature and its Fourier representation does not always converge in the limit. Even so, investigating the resulting Fourier modes (i.e. coefficients in the discrete Fourier expansion) can provide valuable geometric information about the sets. \Revision{Intuitively, lower-order modes capture the broad shape and orientation of the set, while higher-order modes capture the fine-grained complexity and fractality of the boundary.}

We used the Fourier representation (see \cite[Chapter 6]{russ2018image}) of the boundary of the equi-M set to compute several additional measures. In general, the measures are invariant under translations, rotations, and (sometimes) scaling. Among these are the centroid $G$, the area $A$, and the perimeter $P$. Other measures, such as (average) curvature, are not representative due to the presence of cusps. Normalized Fourier modes (also known as Fourier descriptors) can also offer further insight into the large scale structure of the equi-M sets~\cite[Chapter 11]{russ2018image}.

\vspace{3mm}
\noindent \paragraph{Statistical and correlation analysis} For all measures presented below (graph-theoretical and geometric), we computed means and standard deviations to compare the distributions of these measures between connectome types. In addition, we performed correlation analysis between measures in the same category (to investigate to what extent these measures convey similar information about the connectome), as well as correlations between the graph-theoretical measures of the connectomes and the geometric measures of the equi-M sets (to study to what extent our representation of asymptotic dynamics can be predicted directly from the properties of the connectome itself). In the case of the Functional connectomes, we used the graph-theoretical measures for the positive and the negative sub-networks, as well as for the absolute value of the connectomes.

\section{Results}
\label{sc:results}

\subsection{Structural connectomes}
\label{sc:results:structural}

\begin{figure}[h!]
\centering
\begin{subfigure}{0.5\linewidth}
\centering
\includegraphics[width=0.95\linewidth]{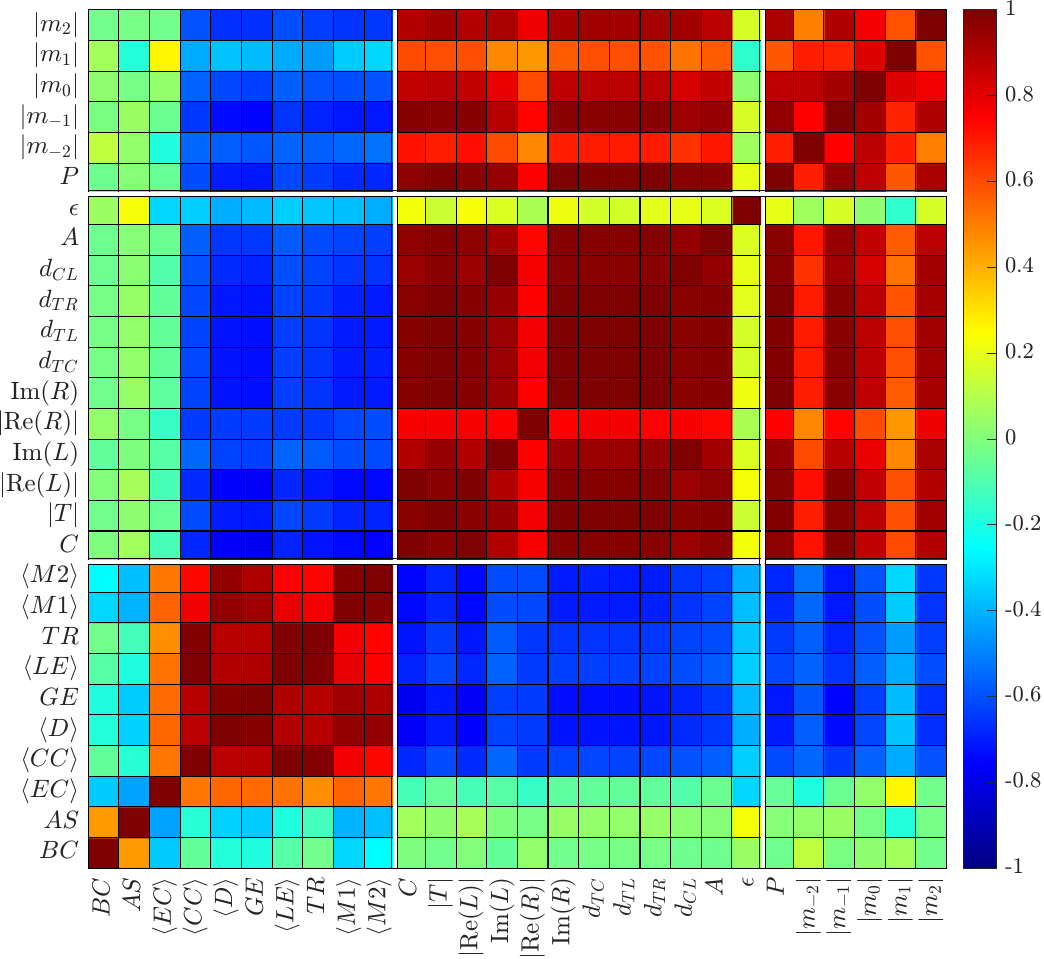}
\caption{}
\end{subfigure}%
\begin{subfigure}{0.5\linewidth}
\centering
\includegraphics[width=0.95\linewidth]{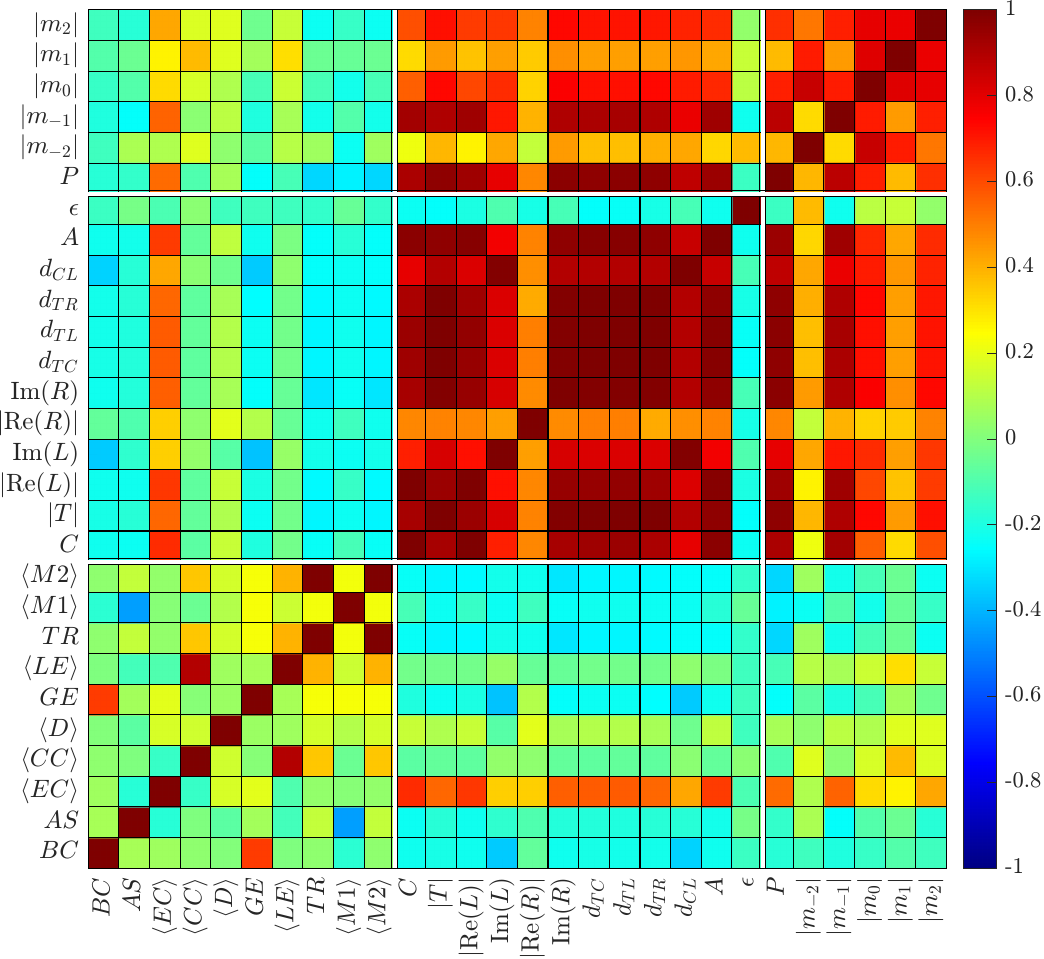}
\caption{}
\end{subfigure}

\caption{\textbf{Correlations} between the graph-theoretical measures and geometric measures for the Structural connectomes \Revision{(panel A)}  and for randomizations of the Structural connectomes that preserve the number of edges into each node \Revision{(panel B)} . The measures are listed along both coordinate axes, in the order found in the tables, and in the order introduced in the text. For $C$, $\Re{R}$ and $\Re{L}$, we used their absolute values, to illustrate the distance to the origin. For the complex modes $m_{-2}$ to $m_2$, we use their moduli (to illustrate their strength). In this scheme, the three diagonal squares show correlations between measures within the same category, and the off-diagonal rectangles show correlations between measures in different categories. For simplicity, we only included the $\rho$ values here.}
\label{fig:structural_correlations}
\end{figure}

\paragraph{Graph-theoretical measures} The graph-theoretical measures described in \Cref{sc:methods:graph} were computed for all Structural connectomes. The mean and standard deviation for each of these measures are provided in~\Cref{tbl:structural_graph_measures}. In \Cref{fig:structural_correlations}, we have the pairwise correlation between the graph-theoretical measures. Notice that, aside from $BC$ and $AS$, the other network measures show significant pairwise positive correlations with significance values $p < 0.0007$.

\begin{table*}[h]
\centering

\caption{\textbf{Network measures statistics for Structural connectomes:} Betweenness Centrality (BC), Assortativity (AS), Clustering Coefficient (CC), Degree (D), Global Efficiency (GE), Local Efficiency (LE), Eigencentrality (EC), Transitivity (TR), Mean 3D Motifs 1 (M1) and 2 (M2). For the node-wise measures, the mean was first computed over all nodes. Then, mean $\mu$ and standard deviation $\sigma$ over the whole subject population are shown in the table for each of the measures used. The strengths of the correlations between these measures ($\rho$ values for Pearson correlations) are illustrated in~\Cref{fig:structural_correlations}.}
\label{tbl:structural_graph_measures}

\begin{tabular}{
    p{0.2\tablewidth}|
    S[table-format=3.3,table-column-width=0.3\tablewidth]
    S[table-format=2.3,table-column-width=0.3\tablewidth]} \toprule
\textbf{Name} \qquad & {$\mu$} & {$\sigma$} \\
\midrule
$\avg{BC}$  & 841.0 & 42.42 \\
$AS$         & 0.022 & 0.036 \\
$\avg{EC}$  & 0.034 & 0.007\\
$\avg{CC}$  & 0.01  & 0.003 \\
$\avg{D}$   & 0.91 & 0.3 \\
$GE$        & 0.03 & 0.01 \\
$\avg{LE}$  & 0.014 & 0.004 \\
$TR$ & 0.008 & 0.002 \\
$\avg{M1}$  & 12.79 & 4.69 \\
$\avg{M2}$  & 5.45 & 2.10 \\
\bottomrule
\end{tabular}
\end{table*}

\vspace{3mm}
\noindent \paragraph{Geometric measures} The measures described in \Cref{sc:methods:geometric} were calculated for each Structural connectome. Their mean and standard deviation are reported in~\Cref{tbl:structural_geometric_measures}. To start with, \Cref{fig:structural_equi_m_sets} readily conveys certain unifying and differentiating features across the Structural equi-M sets. First, we notice relatively broad differences in the size of the sets between subjects. These differences reflect relatively broad distributions in the values of the geometric landmarks and distances (for all diameters $d_{XY}$ that we calculated, the standard deviation is at least half of the mean). The one measure which has a significantly smaller relative variance is the ratio $\epsilon$. This is because the ``eccentricity'' of the sets remains relatively robust across the connectomes, with eccentricities tightly clustered around the average $\mu(\epsilon) = 0.81$, regardless of size.

\begin{table*}[h!]
\centering

\caption{\textbf{Geometric measure statistics for Structural connectomes.} Then mean $\mu$ and standard deviation $\sigma$ over the whole subject population were computed for each geometric measure. For the modes, the standard deviation $\sigma$ was computed by treating the real and imaginary parts as independent and calculating the standard deviation as the square root of the sum of the variances of the real and imaginary parts. The strengths of the correlations between these measures ($\rho$ values for Pearson correlations) are illustrated in~\Cref{fig:structural_correlations}.}
\label{tbl:structural_geometric_measures}

\begin{tabular}[t]{
    p{0.37\tablewidth}|
    S[table-format=2.3,table-column-width=0.459\tablewidth]
    S[table-format=2.3,table-column-width=0.15\tablewidth]} \toprule
\textbf{Name} & {$\mu$} & {$\sigma$} \\
\midrule
$C$ & 0.06 & 0.03 \\
$T$ & -0.38 & 0.21 \\
$\Re{L}$ & 0.10 & 0.05 \\
$\Im{L}$ & 0.06 & 0.03 \\
$\Re{R}$ & -0.02 & 0.01 \\
$\Im{R}$ & 0.18 & 0.10 \\
$d_{TC}$ & 0.44 & 0.24 \\
$d_{TL}$ & 0.49 & 0.26 \\
$d_{TR}$ & 0.40 & 0.21 \\
$d_{CL}$ & 0.07 & 0.04 \\
$A$ & 0.10 & 0.10 \\
$\epsilon$ & 0.81 & 0.02 \\
Centroid ($G$) & -0.06 & 0.03 \\
Perimeter ($P$) & 2.90 & 1.71 \\
\bottomrule
\end{tabular}%
\quad
\begin{tabular}[t]{
    p{0.365\tablewidth}|
    S[table-format=2.3,table-column-width=0.2\tablewidth]
    S[table-format=2.3,table-column-width=0.2\tablewidth]
    S[table-format=2.3,table-column-width=0.15\tablewidth]} \toprule
\textbf{Name} & {$\Re{\mu}$} & {$\Im{\mu}$} & {$\sigma$} \\
\midrule
Mode $m_{-2}$ & 7.79 & -2.59 &  6.33 \\
Mode $m_{-1}$ & -66.97 & 10.56 & 32.28 \\
Mode $m_0$ & -39.05 & -0.09 & 20.15\\
Mode $m_1$ & -10.52 & -1.79 & 6.38\\
Mode $m_2$ & -9.42 & -3.15 & 6.19\\
\bottomrule
\end{tabular}
\end{table*}

Another robust feature for all the equi-M sets is the presence of a cusp at their rightmost point on the real axis. We have shown in our theoretical work~\cite{D2,radulescu2023computing} that this feature, reminiscent of the traditional Mandelbrot set, is not invariably present for any network (with counterexamples from low to high dimensions). The consistent presence and position of the cusp, for all Structural connectomes, becomes a differentiating feature when compared to the equi-M sets computed for Rest and Emotion connectomes.\\

\begin{figure}[h!]
\centering
\begin{subfigure}{0.5\linewidth}
\centering
\includegraphics[width=0.95\linewidth]{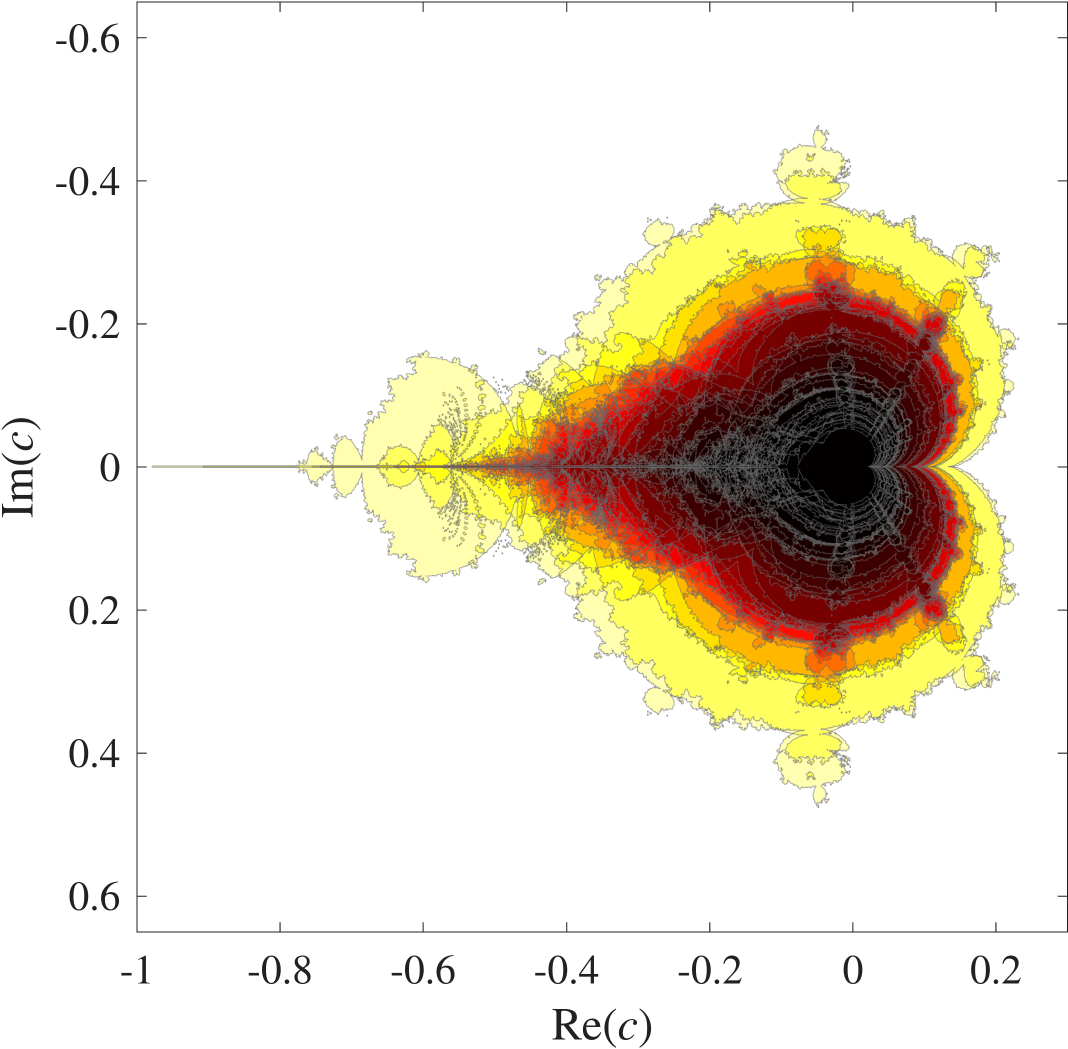}
\caption{}
\end{subfigure}%
\begin{subfigure}{0.5\linewidth}
\centering
\includegraphics[width=0.95\linewidth]{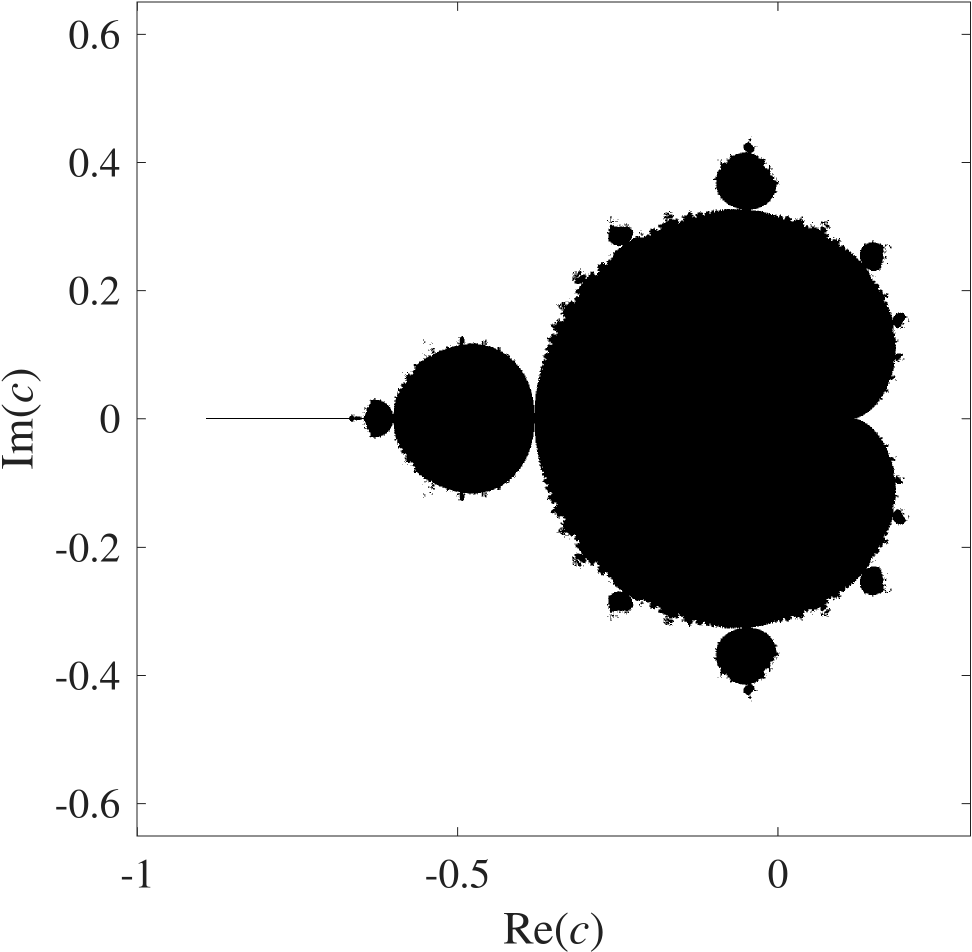}
\caption{}
\end{subfigure}

\caption{\textbf{Different approaches to the Structural connectome data} using the equi-M set. \textbf{(a)} Statistical equi-M set: the light to dark gradient shows the fraction of equi-M sets that the respective point in $\mathbb{C}$ belongs to. \textbf{(b)} The equi-M set computed for the prototypical Structural connectome (i.e., the matrix computed as the mean over all Structural connectomes).}
\label{fig:stats_and_prototypes}
\end{figure}

\noindent \paragraph{Statistical and prototypical equi-M sets} As discussed in~\Cref{sc:methods:contruction}, we approach the collection of sets obtained from connectome data in a statistical way by of considering the distribution of equi-M sets over all connectomes under consideration. We did this by computing the equi-M set for each individual connectome and assigning to each point in the complex plane the fraction of individual equi-M sets that contain the point (effectively measuring how ``likely'' it is for the given point to be in any equi-M set generated by our data). \Cref{fig:stats_and_prototypes}a illustrates this statistical equi-M set with the fraction represented by the color intensity, increasing from light to dark. To emphasize the underpinnings of this representation, we overlaid on it the contours of all equi-M sets for Structural connectomes (in gray), outlining the basis of how these statistics were computed. Consistent with our previous comment, all these contours exhibit a cusp on the right side along the real axis. Note, in particular, that the black center region is the $c$-locus that is common to the equi-M sets for \emph{all} Structural connectomes.

In~\Cref{fig:stats_and_prototypes}b, we show the equi-M set for the \emph{prototypical} Structural connectome, computed as the mean of all Structural connectomes in the data set (group average). This is a different way of checking that, while details in the size and  shape of equi-M sets vary across connectomes, the presence of the cusp on the right appears to be a robust feature for Structural connectomes, for all individual and group representations.\\

\noindent \paragraph{Correlations between graph properties and the geometry of the equi-M set} As described in the previous paragraphs, we calculated mutual Pearson correlations between the measures in each category, but also across categories. This allows us to evaluate to what extent the architecture of the Structural connectome can predict patterns in asymptotic dynamics, as reflected in the geometric properties of the equi-M set. \Cref{fig:structural_correlations} describes all these correlations in a table in which the colors represent positive and negative correlation values.

To reiterate, let us notice that there is a significantly positively-correlated block of graph-theoretical measures (all except $BC$ and $AS$) -- we call this block $\mathcal{G}$. There is also a highly correlated block of geometric measures (all except $\epsilon$) -- we  call this block $\mathcal{T}$. Finally, we noted that there is also some degree of correlation within the collection of seven Fourier-derived measures (less significant as modes increase). In addition to this, \Cref{fig:structural_correlations} shows that there are strong negative correlations between the measures in $\mathcal{G}$ and the measures in $\mathcal{T}$. The only ones that fail the significance level $\alpha = 0.01$ are the correlations with the average Local Efficiency $\avg{LE}$. Since the geometric measures in $\mathcal{G}$ all relate in some way to the size of the set, this result supports an idea that we had previously discussed in our work on both theoretical and Structural connectomes: higher network wiring connectivity and efficiency (higher node degree and centrality, stronger motifs) are associated with smaller equi-M sets. In turn, the graph measures in $\mathcal{G}$ are also negatively correlated with the Fourier measures, with the most significant values for the centroid distance $|G|$ and the perimeter $P$ (which is not unexpected, since these are also measures of ``size'' and show strong positive correlations with the other measures of size in $\mathcal{T}$). What is interesting is that the moduli of the modes also show similar correlation patterns, even though the strength of the correlations weakens with increasing mode index (and significance decreases). This suggests a tie between the wiring of the underlying graph and the geometry and variability of the boundary of the equi-M set, tie which transcends simple size-related properties for Structural connectomes. We will revisit this point for perspective, after we discuss the same relationships in Functional connectomes in \Cref{sc:results:functional}.

\begin{table}
\begin{minipage}{0.36\textwidth}
\begin{tabular}{l|l|l} \toprule
\textbf{Name} & \textbf{Mean} $~\mu$ & \textbf{Std} $~\sigma$ \\
\midrule
$\langle BC \rangle$ & 616.2 & 12.45 \\
$AS$ & -0.03 & 0.01 \\
$\avg{EC}$ & 0.035 & 0.006\\
$\avg{CC}$ & 0.002  & 0.000\\
$\avg{D}$ & 0.70 & 0.000 \\
$GE$ & 0.03 & 0.000 \\
$\avg{LE}$ & 0.006 & 0.000 \\
$T$ & 0.002 & 0.000 \\
$\avg{M1}$ & 16.23 & 0.14 \\
$\avg{M2}$ & 1.14 & 0.02 \\
\bottomrule
\end{tabular}
\end{minipage}
\quad
\begin{minipage}{0.58\textwidth}
\caption{\small \emph{\textbf{Network measure statistics for randomized Structural connectomes:} Betweenness Centrality (BC); Assortativity (AS); Clustering Coefficient (CC); Degree (D); Global efficiency (GE); Local efficiency (LE); Eigencentrality (EC); Transitivity (TR); Mean 3D Motifs 1 (M1) and 2 (M2). For the node-wise measures, the mean was first computed over all nodes. Then, means and standard deviations over the whole subject population are shown in the table for each of the measures used. The strengths of the correlations between these measures ($\rho$ values for Pearson correlations) are illustrated in Figure~\ref{fig:structural_correlations}.}}
\label{structural_random_graph_measures}
\end{minipage}
\end{table}

\begin{table}
\begin{center}
\begin{subtable}{0.35\textwidth}
\centering
\begin{tabular}{l|l|l} \toprule
\textbf{Name} & \textbf{Mean} $~\mu$ & \textbf{Std} $~\sigma$ \\
\midrule
$C$ & 0.11 & 0.02 \\
$T$ & -0.74 & 0.17 \\
$\Re{L}$ & 0.19 & 0.03 \\
$\Im{L}$ & 0.11 & 0.03 \\
$\Re{R}$ & -0.03 & 0.02 \\
$\Im{R}$ & 0.36 & 0.08 \\
$d_{TC}$ & 0.85 & 0.19 \\
$d_{TL}$ & 0.94 & 0.21 \\
$d_{TR}$ & 0.79 & 0.18 \\
$d_{CL}$ & 0.14 & 0.04 \\
$A$ & 0.32 & 0.12 \\
$\epsilon$ & 0.84 & 0.02 \\
\bottomrule
\end{tabular}
\end{subtable}
\hspace{3mm}
\begin{subtable}[h]{0.45\textwidth}
\vspace{-7pt}
\begin{tabular}{l|l|l}
\toprule
\textbf{Name} & \textbf{Mean} $~\mu$ & \textbf{Std} $~\sigma$\\
\midrule
Perimeter ($P$) & 5.76 & 1.26 \\
Mode $m_{-2}$ & $12.20-4.64i$ & 10.27\\
Mode $m_{-1}$ & $-128.93+21,60i$ & 34.43\\
Mode $m_0$ & $-67.18+0.09i$ & 21.42\\
Mode $m_1$ & $-14.92-3.14i$ & 9.58\\
Mode $m_2$ & $-19.39-6.98i$ & 9.27\\
\bottomrule
\end{tabular}
\vspace{1.2cm}
\end{subtable}
\end{center}
\caption{\small \emph{\textbf{Geometric measure statistics for randomized Structural connectomes.} The mean and standard deviation over the whole subject population were computed for each geometric measure. The strengths of the correlations between these measures ($\rho$ values for Pearson correlations) are illustrated in Figure~\ref{fig:structural_correlations}b.}}
\label{structural_random_geometric_measures}
\end{table}

\begin{figure}[h!]
\centering
\begin{subfigure}{0.5\linewidth}
\centering
\includegraphics[width=0.95\linewidth]{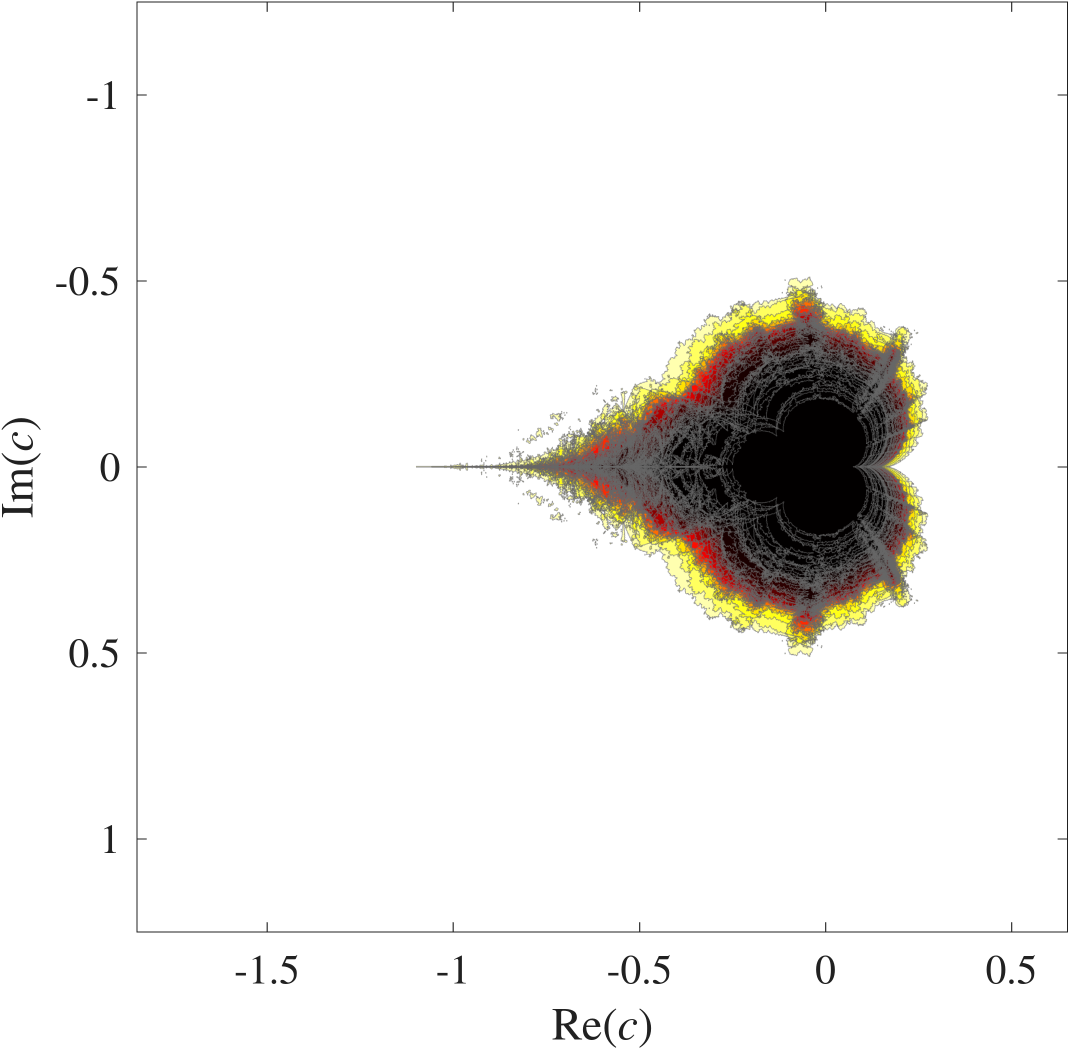}
\caption{}
\end{subfigure}%
\begin{subfigure}{0.5\linewidth}
\centering
\includegraphics[width=0.95\linewidth]{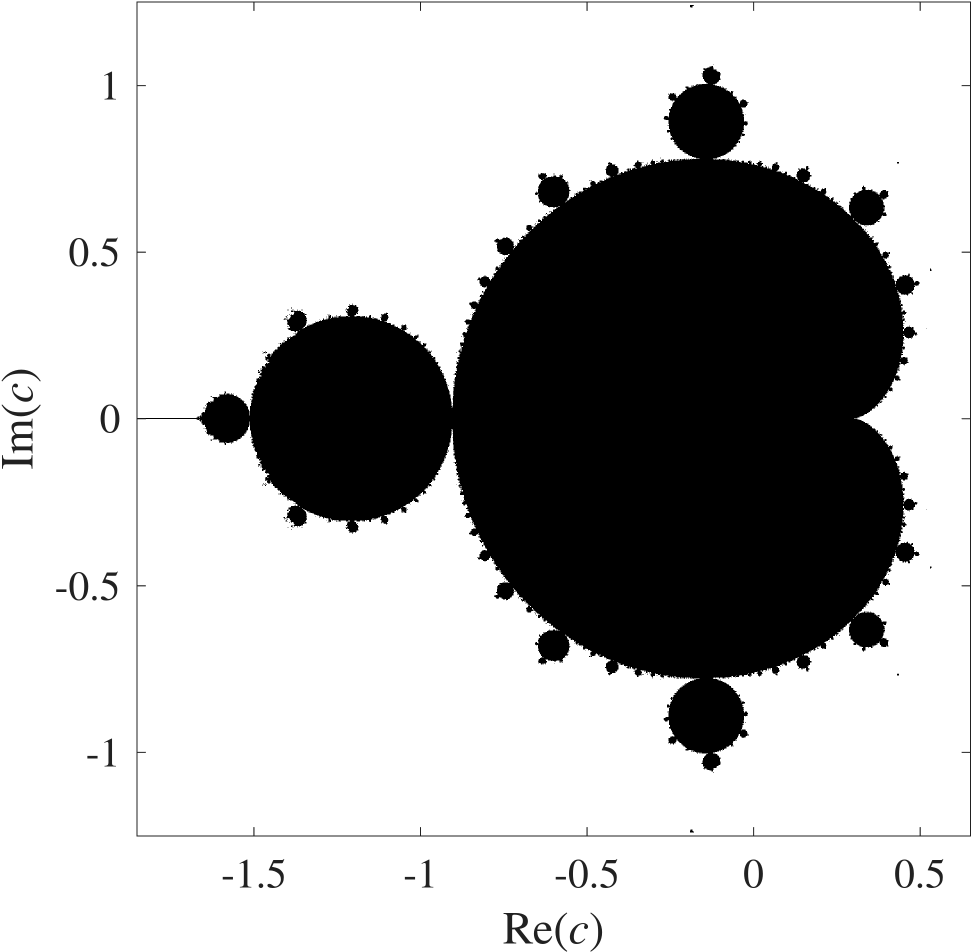}
\caption{}
\end{subfigure}

\caption{\textbf{Visualizations of equi-M set statistics for the randomly generated Structural connectome data}. \textbf{(a)} Statistical equi-M set: the light to dark gradient shows the fraction of equi-M sets that the respective point in $\mathbb{C}$ belongs to. \textbf{(b)} The equi-M set computed for the corresponding prototypical connectome.}
\label{fig:random_stats_and_prototypes}
\end{figure}

Finally, let us point out that Betweenness Centrality and Assortativity don't have any significant correlations with any of the equi-M measures we used. The natural next question is whether any one of the strongly correlated measures in $\mathcal{G}$ holds the key to the shape of the equi-M set. To explore this question further, we simulated an in silico experiment aimed to investigate to what extent the average node Degree dictates in and off itself the measures of the equi-M set. To do so, we chose a random Structural connectome (subject \#2), and generated a set of 47 additional surrogate connectomes with the same node degree distribution as the original. The 48 resulting sets are shown in~\Cref{fig:random_stats_and_prototypes} (as contours), together with the prototypical set for the group of random connectomes. The statistics of geometric and Fourier measures are shown in \Revision{Tables~\ref{structural_random_graph_measures} and~\ref{structural_random_geometric_measures}}.

Notice that the random networks created this way are quite different from the original ones, with different graph-theoretical statistics. They exhibit collectively narrower spread, more generally than just for the Degree, and with distinct global properties (e.g., they are slightly disassortative). In turn, the landmark landscape of the corresponding equi-M sets is different, and the labels generally identify larger sets (as visible from the contours in~\Cref{fig:random_stats_and_prototypes}a), and with significantly more information contained in the first two modes. Interestingly, notice that the eccentricities are comparable to those in the original distribution, suggesting that eccentricity may be robust to Degree-preserving alterations of the network. Another interesting fact to note is that the prototypical equi-M set in this case is a rescaled version of the traditional Mandelbrot set. This may reflect eliminating the randomness of the network by averaging over connectomes, with the resulting system behaving equivalently to a single node via a multiplicative weight.

Finally, notice that, for randomly generated connectomes (unlike for the data-driven Structural connectomes) there are no significant correlations between the graph measures and the geometric/Fourier measures (as seen in~\Cref{fig:structural_correlations}b). This suggests that the tie between the structure and the network's asymptotic dynamics may not simply be intrinsic, or achieved by chance, but rather the reflection of the natural design in functional systems.

\subsection{Functional connectomes}
\label{sc:results:functional}

From~\Cref{sc:methods:dataset}, we consider two types of Functional connectomes: one corresponding to a scan with no task (Rest) and one for an Emotional task. We first describe the graph-theoretical characteristics and the geometry of the equi-M sets for both tasks separately. Then, we will compute correlations between them and draw comparisons between the results, as well as with those obtained for the Structural connectomes. As discussed in \Cref{sc:methods:graph}, it is common practice in network science to describe signed connectomes by separating their positive and negative sub-networks. Then, we can produce graph-theoretical descriptions for each sub-network, which are used as a characterization of the overall, signed network. In order to carry on our study of the relationship between network architecture and dynamics, we compute graph-theoretical measures for the positive sub-network, the negative sub-networks and the absolute value of each connectome. Our analysis of Structural connectomes concluded that, in the case of positive symmetric connectomes, graph-theoretical measures are predictive of geometric measures. A crucial part of our investigation of signed connectomes is to establish whether these partial representations of the network, taken together, can perform the same predictive function or if the information they provide is insufficient. If it is not sufficient, we can conclude that a crucial part of the emerging dynamics relies on the \emph{way in which they interact} that is not captured in these partial measures.

Another important point is that segregating the Functional networks into their positive and negative sub-networks does not allow a comparison of graph architectural measures between the Functional connectomes and the Structural connectomes (which, since being positive, were considered as a whole). However, the equi-M set is computed analogously for all connectomes, thus remaining an efficient across-the-board assessment of asymptotic dynamics and a tool for quantifying differences.

\subsubsection{Measures for Functional connectomes}

\noindent \paragraph{Graph-theoretical measures} For Rest and Emotion connectomes, graph theoretical measures are described in \Cref{tbl:functional_rest_graph_measures,tbl:functional_emotion_graph_measures}. As discussed in the introduction, these measures are not available  for signed networks. We therefore computed them for the associated networks, as done in a typical network study. The table presents the mean and standard deviations for each measure, for each associated network. Further statistical differences between these sets of measures are described and interpreted in~\Cref{sc:results:diff_rest_emotion}.

\begin{table*}[h!]
\centering

\caption{\textbf{Network measure statistics for Rest connectomes.} Graph-theoretical measures were computed separately for the positive and for the negative sub-networks of the connectome, as well as for the absolute value of the connectome: Betweenness Centrality (BC); Assortativity (AS); Clustering Coefficient (CC); Degree (D); Global efficiency (GE); Local efficiency (LE); Eigencentrality (EC); Transitivity (TR); Mean 3D Motifs 1 (M1) and 2 (M2). For the node-wise measures, the mean was first computed over all nodes. Then mean $\mu$ and standard deviation $\sigma$ over the whole subject population of each of the measures used are shown in the table.}
\label{tbl:functional_rest_graph_measures}

\begin{tabular}{
    l|
    S[table-format=3.3,table-column-width=0.12\linewidth]
    S[table-format=3.3,table-column-width=0.12\linewidth]
    S[table-format=3.3,table-column-width=0.12\linewidth]
    S[table-format=3.3,table-column-width=0.12\linewidth]
    S[table-format=3.3,table-column-width=0.12\linewidth]
    S[table-format=3.3,table-column-width=0.12\linewidth]} \toprule
\textbf{Name} & \multicolumn{2}{c}{\textbf{Positive network}}
              & \multicolumn{2}{c}{\textbf{Negative network}}
              & \multicolumn{2}{c}{\textbf{Absolute network}} \\
\cmidrule(lr){2-3} \cmidrule(lr){4-5} \cmidrule(lr){6-7}
& {$\mu$} & {$\sigma$}
& {$\mu$} & {$\sigma$}
& {$\mu$} & {$\sigma$} \\
\midrule
$BC$ & 303.22 & 11.21 & 174.11 & 8.42 & 157.46 & 9.44 \\
$AS$ & 0.06 & 0.03 & -0.06 & 0.02 & -0.05 & <0.001 \\
$\avg{EC}$ & 0.06 & 0.004 & 0.06 & 0.008 & 0.06 & <0.001\\
$\avg{CC}$ & 0.10 & 0.01 & 0.04 & 0.004 & 0.12 & 0.01 \\
$\avg{D}$ & 15.24 & 1.77 & 16.07 & 1.71 & 31.31 & 3.48\\
$GE$ & 0.16 & 0.01 & 0.16 & 0.01 & 0.21 & 0.01 \\
$\avg{LE}$ & 0.12 & 0.01 & 0.08 & 0.008 & 0.12 & 0.01\\
$T$ & 0.09 & 0.01 & 0.04  & 0.004 & 0.12 & 0.01\\
$\avg{M1}$ & 423.65 & 43.57 & 1493.47 & 199.33 & {\textemdash} & {\textemdash} \\
$\avg{M2}$ & 391.96 & 62.80 & 250.32 & 33.16 & 247.68 & 293.77 \\
\bottomrule
\end{tabular}
\end{table*}

\begin{table*}[h!]
\centering

\caption{\textbf{Network measure statistics for Emotion connectomes.} Graph-theoretical measures were computed separately for the positive and for the negative sub-networks of the connectome, as well as for the absolute value of the connectome. Betweenness Centrality (BC); Assortativity (AS); Clustering Coefficient (CC); Degree (D); Global efficiency (GE); Local efficiency (LE); Eigencentrality (EC); Transitivity (TR); Mean 3D Motifs 1 (M1) and 2 (M2). For the node-wise measures, the mean was first computed over all nodes. Then mean $\mu$ and standard deviation $\sigma$ over the whole subject population of each of the measures used are shown in the table.}
\label{tbl:functional_emotion_graph_measures}

\begin{tabular}{
    l|
    S[table-format=3.3,table-column-width=0.12\linewidth]
    S[table-format=3.3,table-column-width=0.12\linewidth]
    S[table-format=3.3,table-column-width=0.12\linewidth]
    S[table-format=3.3,table-column-width=0.12\linewidth]
    S[table-format=3.3,table-column-width=0.12\linewidth]
    S[table-format=3.3,table-column-width=0.12\linewidth]} \toprule
\textbf{Name} & \multicolumn{2}{c}{\textbf{Positive network}}
              & \multicolumn{2}{c}{\textbf{Negative network}}
              & \multicolumn{2}{c}{\textbf{Absolute network}} \\
\cmidrule(lr){2-3} \cmidrule(lr){4-5} \cmidrule(lr){6-7}
& {$\mu$} & {$\sigma$}
& {$\mu$} & {$\sigma$}
& {$\mu$} & {$\sigma$} \\
\midrule
$BC$ & 256.79 & 5.75 & 178.82 & 6.08 & 143.02 & 6.85 \\
$AS$ & 0.05 & 0.04 & -0.07 & 0.03 & -0.005 & <0.001 \\
$CC$ & 0.09 & 0.01 & 0.04 & 0.002 & 0.13 & 0.01\\
$\avg{D}$ & 15.99 & 2.06 & 16.72 & 2.10 & 32.72 & 4.16\\
$GE$ & 0.17 & 0.01 & 0.18 & 0.01 & 0.22 & 0.02\\
$\avg{LE}$ & 0.12 & 0.01 & 0.09 & 0.008 & 0.13 & 0.01\\
$\avg{EC}$ & 0.06 & 0.005 & 0.06 & 0.009 & 0.06 & 0.001\\
$T$ & 0.09 & 0.01 & 0.04 & 0.003 & 0.13 & 0.01\\
$\avg{M1}$ & 529.38 & 34.99 & 1518.38 & 265.36 & {\textemdash} & {\textemdash} \\
$\avg{M2}$ & 425.98 & 80.18 & 237.01 & 20.22 & 2607.47& 356.78\\

\hline
\end{tabular}
\end{table*}

\vspace{3mm}
\noindent {\bf Geometric measures} were calculated for all Rest and Emotion connectomes and are shown in~\Cref{fig:functional_geometric_measures,fig:functional_fourier_measures}. In the table, we show only \Revision{measures for the full, signed network,} since one of our goals is precisely to establish to what extent partial graph theoretical information from the associate networks is relevant to the whole-network dynamics. To support this choice, we also computed the equi-M sets for the associated networks for all Functional connectomes (Rest and Emotion). What we observed is that these sets were virtually all scaled versions of the traditional Mandelbrot set (see for example Figures~\ref{fig:rest_abs_equi_m_sets} and~\ref{fig:emotion_abs_equi_m_sets} in the Appendix for the equi-M sets corresponding to the absolute value connectomes). This suggests that our intuition is correct, and that the associate networks cannot recreate, taken separately,  the complexity of the overall network dynamics. This will be further discussed in the following sections.

\begin{table*}[h!]
\centering

\caption{\textbf{Geometric statistics for Functional connectomes.} The mean $\mu$ and standard deviation $\sigma$ over the whole subject population were computed for each one-dimensional measure, for the Rest connectomes (top) and the Emotion connectomes (bottom). The strengths of the correlations between these measures ($\rho$ values for Pearson correlations) are illustrated in \Cref{fig:functional_correlations}.}
\label{fig:functional_geometric_measures}

\begin{tabular}[t]{
    p{0.2\tablewidth}|
    S[table-format=2.4,table-column-width=0.3\tablewidth]
    S[table-format=2.4,table-column-width=0.3\tablewidth]} \toprule
\textbf{Name} & {$\mu$} & {$\sigma$} \\
\midrule
$C$ & -0.009 & 0.002 \\
$T$ & 0.018 & 0.004 \\
$\Re{L}$ & -0.013 & 0.003 \\
$\Im{L}$ & 0.005 & 0.002 \\
$\Re{R}$ & 0.0003 & 0.003 \\
$\Im{R}$ & 0.017 & 0.003 \\
$\Re{S}$ & 0.019 & 0.004 \\
$\Im{S}$ & 0.0007 & 0.0009 \\
$d_{TC}$ & 0.027 & 0.005 \\
$d_{SL}$ & 0.033 & 0.007 \\
$d_{SR}$ & 0.026 & 0.005 \\
$d_{CL}$ & 0.006 & 0.002 \\
$A$ & 0.0007 & 0.0003 \\
$\epsilon$ & 1.25 & 0.17 \\
\bottomrule
\end{tabular}%
\qquad
\begin{tabular}[t]{
    p{0.2\tablewidth}|
    S[table-format=2.4,table-column-width=0.3\tablewidth]
    S[table-format=2.4,table-column-width=0.3\tablewidth]} \toprule
\textbf{Name} & {$\mu$} & {$\sigma$} \\
\midrule
$C$ & -0.012 & 0.004 \\
$T$ & 0.018 & 0.005 \\
$\Re{L}$ & -0.016 & 0.004 \\
$\Im{L}$ & 0.004 & 0.002 \\
$\Re{R}$ & 0.001 & 0.003 \\
$\Im{R}$ & 0.018 & 0.004 \\
$\Re{S}$ & 0.02 & 0.004 \\
$\Im{S}$ & 0.001 & 0.001 \\
$d_{TC}$ & 0.03 & 0.007 \\
$d_{SL}$ & 0.037 & 0.008 \\
$d_{SR}$ & 0.027 & 0.005 \\
$d_{CL}$ & 0.006 & 0.003 \\
$A$ & 0.0009 & 0.0004 \\
$\epsilon$ & 1.22 & 0.23 \\
\bottomrule
\end{tabular}
\end{table*}

\begin{table*}[h!]
\centering

\caption{\textbf{Fourier-based measure statistics for Functional connectomes.} The mean $\mu$ and standard deviation $\sigma$ over the whole subject population were computed for each one-dimensional measure. For the modes (which are distributions in $\mathbb{C}$), more information on standard deviations and principal components is included later in the section. The strengths of the correlations between these measures ($\rho$ values for Pearson correlations) are illustrated in \Cref{fig:functional_correlations}.}
\label{fig:functional_fourier_measures}

\begin{tabular}{
    p{0.37\tablewidth}|
    S[table-format=2.4,table-column-width=0.2\tablewidth]
    S[table-format=2.4,table-column-width=0.2\tablewidth]
    S[table-format=2.4,table-column-width=0.15\tablewidth]} \toprule
\textbf{Name} & {$\Re{\mu}$} & {$\Im{\mu}$} & {$\sigma$} \\
\midrule
Centroid ($G$) & 0.001 & {---} & 0.0009 \\
Perimeter ($P$) & 0.24 & {---} & 0.06 \\
Mode $m_{-2}$ & -0.107 & 0.0035 & 0.45\\
Mode $m_{-1}$ & 5.27 & 0.0844 & 0.88\\
Mode $m_0$ & 0.6 & -0.0008 & 0.57\\
Mode $m_1$ & -0.19 & -0.003 & 0.24 \\
Mode $m_2$ & 0.22 & 0.005 & 0.14\\
\bottomrule
\end{tabular}%
\qquad
\begin{tabular}{
    p{0.37\tablewidth}|
    S[table-format=2.4,table-column-width=0.2\tablewidth]
    S[table-format=2.4,table-column-width=0.2\tablewidth]
    S[table-format=2.4,table-column-width=0.15\tablewidth]} \toprule
\textbf{Name} & {$\Re{\mu}$} & {$\Im{\mu}$} & {$\sigma$} \\
\midrule
Centroid ($G$) & 0.0004 & {---} & 0.001 \\
Perimeter ($P$) & 0.259 & {---} & 0.059 \\
Mode $m_{-2}$ & 0.03 & 0.002 & 0.24\\
Mode $m_{-1}$ & 0.09 & 0.008 & 0.54 \\
Mode $m_1$ & 0.54 & -0.004 & 0.93 \\
Mode $m_2$ & -0.19 & -0.0004 & 0.76 \\
Mode $m_3$ & 0.14 & 0.007 & 0.34 \\
\bottomrule
\end{tabular}
\end{table*}

\subsubsection{Differences between Structural and Functional connectomes}
\label{sc:results:diff_struct_func}

We first examine differences in the equi-M sets between structural and functional connectomes, as seen in Figures~\ref{fig:structural_equi_m_sets},~\ref{fig:rest_equi_m_sets} and~\ref{fig:emotion_equi_m_sets} in the Appendices.
The first noticeable difference is that, in contrast with equi-M sets for Structural connectomes (which all had a right cusp), all equi-M sets for both Rest and Emotion connectomes have a left cusp. We attribute this to the distribution of negative signs in the Functional connectomes. Further experimental and theoretical work \Revision{not included in this study}~\cite{D2} suggests that replacing positive connections with negative connections eventually leads to a change in the orientation of the cusp (process that may involve intermediate stages of the set with multiple cusps). Of course, at what point and in what form this swap occurs is further determined by the placement of the negative connections. Moreover, since the quadratic function is invariant to the sign of the input, adding more negative connections will eventually swap the cusp back to the right. It is very interesting to observe \Revision{the left-oriented cusps for the Functional connectomes} in our data set (especially because it offers great insight into how different dynamics operate on a positive versus signed connectome). Although we do not have a full analytical explanation for it, we have a candidate mechanism that we are investigating in our theoretical work.

Another, perhaps less striking, but important difference in the shapes of the equi-M sets is the significantly different eccentricities between the sets for Structural connectomes (with a mean of $\mu(\epsilon) = 0.81$) and the Rest and Emotional connectomes (with means of $\mu(\epsilon) = 1.25$ and $\mu(\epsilon) = 1.22$, respectively). More specifically, a \texttt{ranksum} test shows significant differences in eccentricity between Structural and Rest connectomes ($p < 10^{-10})$, between Structural and Emotion connectomes ($p < 10^{-10}$), as well as between Rest ad Emotion connectomes ($p < 0.005$). The difference in eccentricities extends to their variability, with a very tight distribution for the Structural connectomes ($\sigma(\epsilon) = 0.02$) and Rest and Emotion connectomes (with $\sigma(\epsilon) = 0.17$ and $\sigma(\epsilon) = 0.23$, respectively).

Although it is undeniable that there are significant differences between the equi-M sets for Structural and Functional connectomes. This is not unexpected, since the connectomes themselves are very different. What we would like to investigate next is whether the equi-M set can effectively identify differences between different types of Functional connectomes (Rest and Emotion, in our case).

\subsubsection{Differences between Rest and Emotion connectomes}
\label{sc:results:diff_rest_emotion}

First, a basic analysis \Revision{(using the Wilcoxon rank sum test)} found direct statistical differences between the graph-theoretical measures for the Rest and Emotion connectomes. This is far from surprising, but is difficult to interpret, since these measures are computed separately for the associated networks. Moreover, differences tended to appear in scattered patterns across the three types of associated networks considered (positive sub-network, negative sub-network and absolute value network). For example, $BC$ was the only measure that was consistently different for the three associated networks (with $p_{pos}< 10^{-10}$, $p_{neg}=0.004$ and $p_{abs}< 10^{-10}$, respectively). In addition to this, the positive sub-network exhibited significant differences in \Revision{motif} strengths $M1$ and $M2$ (with $p< 10^{-10}$ and $p=0.059$, respectively), the absolute value network shows trend differences in $GE$ and $TR$ ($p=0.059$ and $p=0.045$), and the negative network captured differences in all measures except $A$ and $D$ (with $p<0.05$ across the board). We believe these patterns may be important, but it is very hard to judge their impact, since none of them encompasses the interconnections between the positive and negative sub-networks in assembling the whole connectome.

\begin{figure}[h!]
\centering
\includegraphics[width=0.85\linewidth]{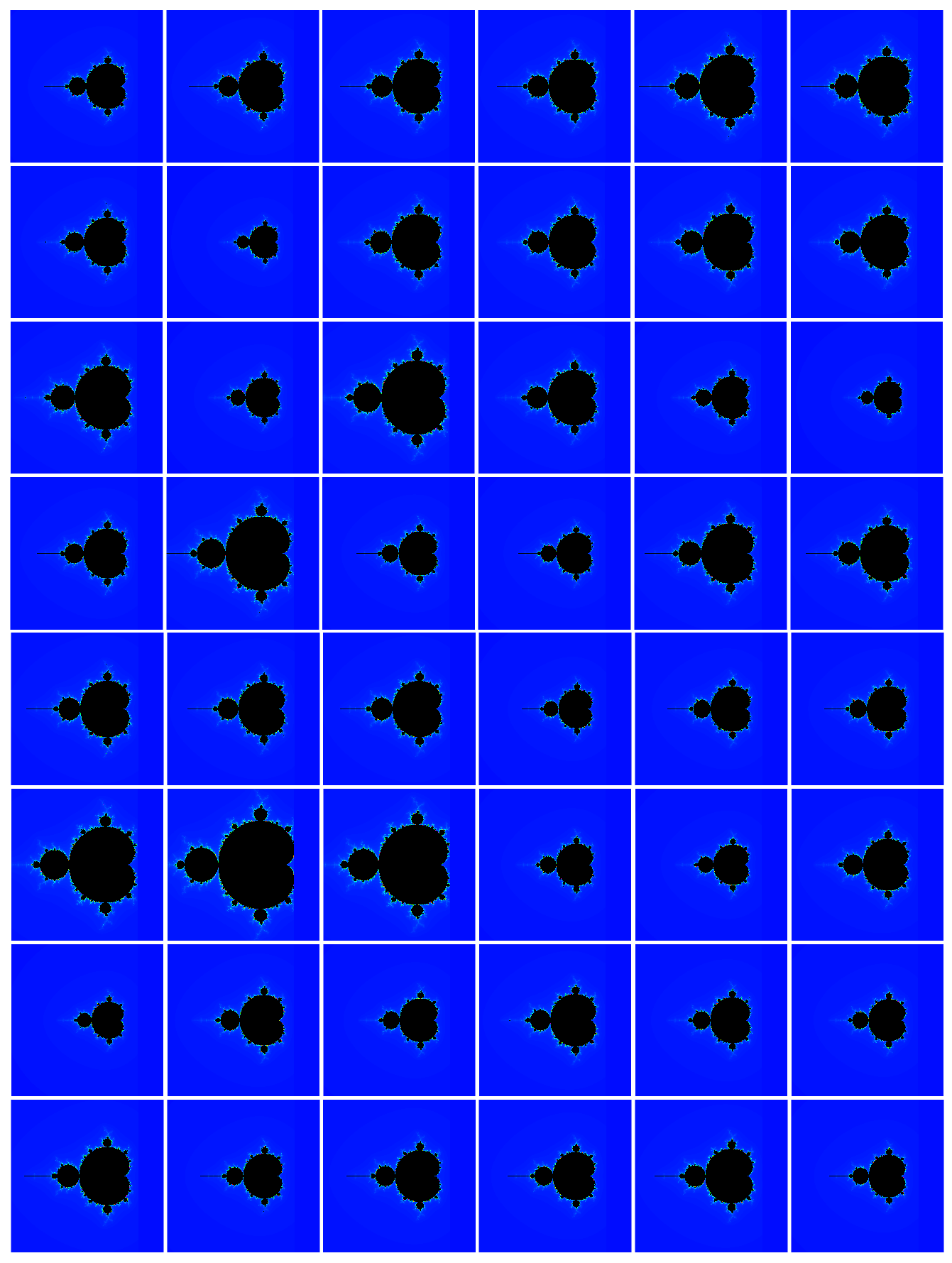}

\caption{\textbf{Equi-M sets for the absolute value associated networks of the Functional Rest connectomes.} All sets are represented in the complex region $[-0.002, 0.0005] \times [-0.0015, 0.0015]$, in resolution $1200 \times 1200$. The equi-M set in each panel appears in black, and the colors represent the number of iterations needed by each point in order to exit the escape disc. When using the absolute value of the functional connectome as a generating network, the resulting sets are scaled copies of the traditional Mandelbrot set.}
\label{fig:rest_abs_equi_m_sets}
\end{figure}

\begin{figure}[h!]
\centering
\includegraphics[width=0.85\linewidth]{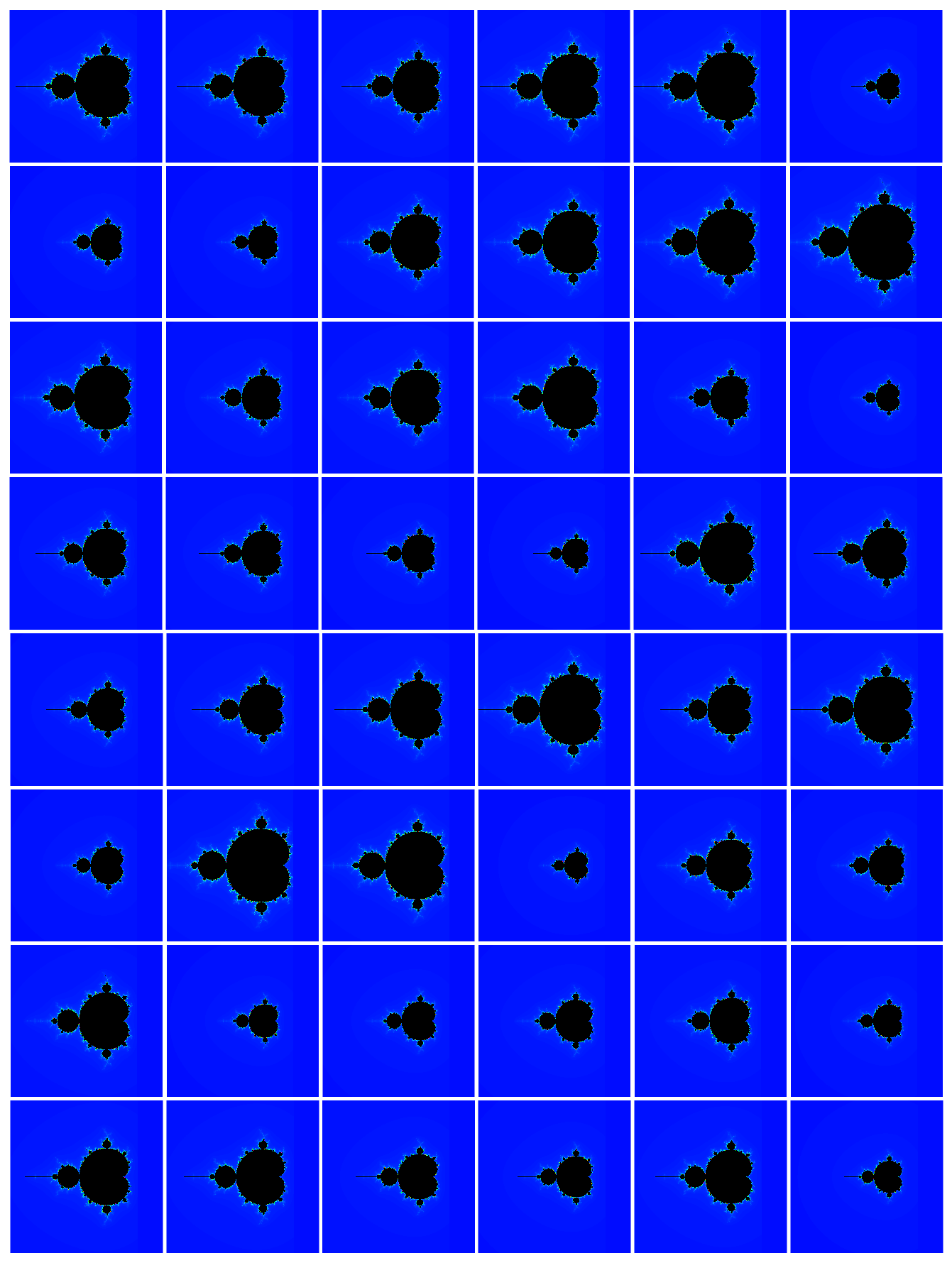}

\caption{\textbf{Equi-M sets for the absolute value associated networks of the Functional Emotion connectomes.} All sets are represented in the complex region $[-0.002, 0.0005] \times [-0.0015, 0.0015]$, in resolution $1200 \times 1200$. The equi-M set in each panel appears in black, and the colors represent the number of iterations needed by each point in order to exit the escape disc. When using the absolute value of the functional connectome as a generating network, the resulting sets are scaled copies of the traditional Mandelbrot set.}
\label{fig:emotion_abs_equi_m_sets}
\end{figure}

One can in fact use the equi-M set to further assess whether/to what extent any of the three associated networks capture the dynamic flow in the original connectome. To do so, we computed the equi-M set for each of the 48 connectomes, for all three associated networks. In all cases, the resulting equi-M sets were all just scaled copies of the traditional equi-M set (the sets for the absolute networks are shown in \Cref{fig:rest_abs_equi_m_sets,fig:emotion_abs_equi_m_sets}, for illustration). This is in itself a powerful result. First, it suggests that the dynamics propagates completely differently in the associated networks when considered separately than in the original network. Second, it shows that significant differences in the structure of these associated networks may lead to indistinguishable dynamics when considered independently. Moreover, the striking differences between the equi-M sets for the original connectomes were completely wiped out and were devoid of network-derived information (all were scaled Mandelbrot sets). Our analysis of the Structural data ensures that this is not the case for all positive connectomes, since the Structural connectomes themselves  produced a wide diversity of equi-M sets. The loss of dynamic diversity must therefore be due to the loss of information on a signed network's architecture when considering only parts of it, or when ignoring the signs. Third, this tells us that the statistical differences found between the associated networks for Rest and Emotion connectomes should not be used to assess the impact of the whole connectome on dynamics.

We next want to show that the equi-M set is a better instrument for such an assessment. We investigate the statistical differences in geometric and Fourier measures between Rest and Emotion connectomes. We use correlation analysis to reexamine the relationship of these measures with the graph-theoretical measures.

\begin{figure}[h!]
\centering
\begin{subfigure}{0.5\linewidth}
\centering
\includegraphics[width=0.95\linewidth]{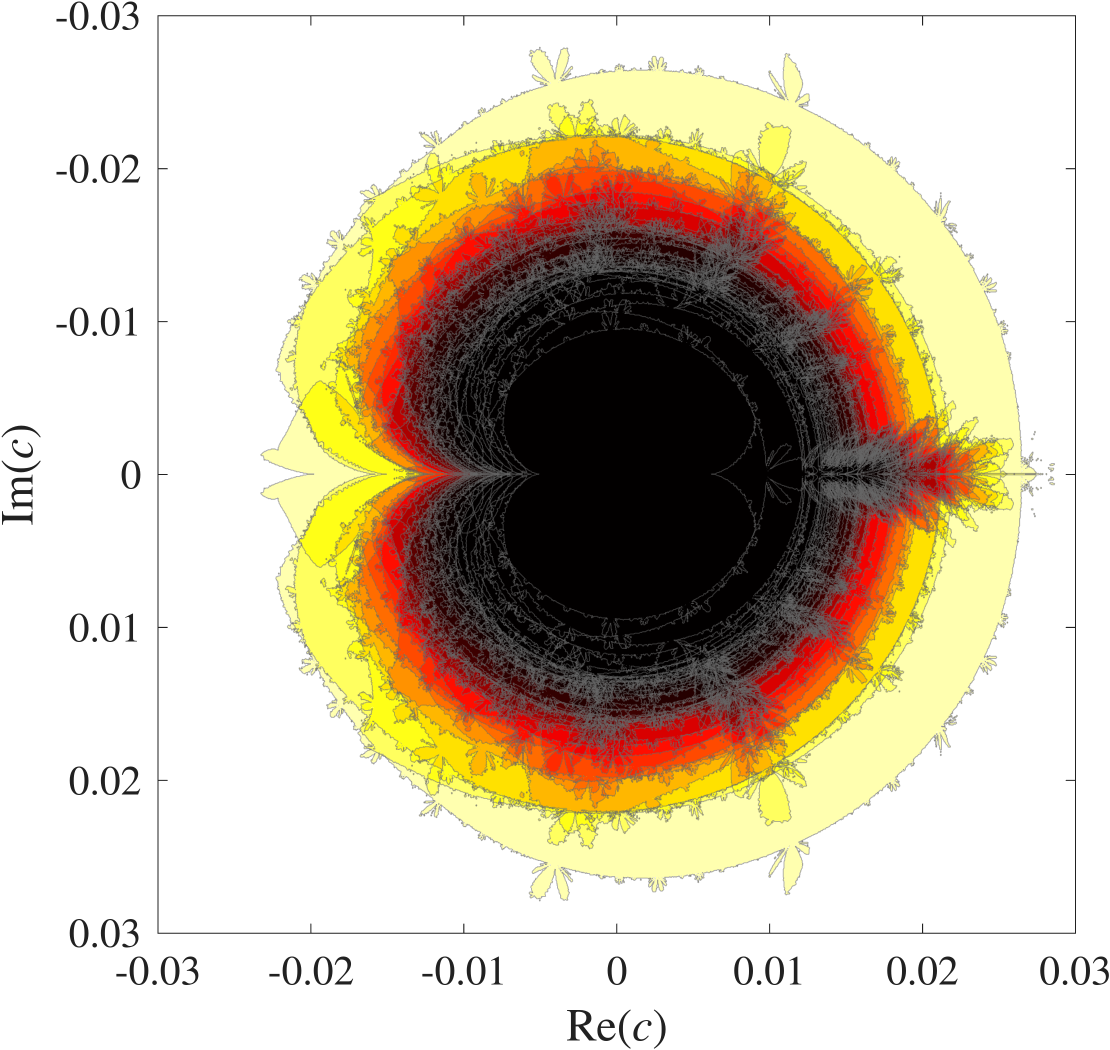}
\caption{}
\end{subfigure}%
\begin{subfigure}{0.5\linewidth}
\centering
\includegraphics[width=0.93\linewidth]{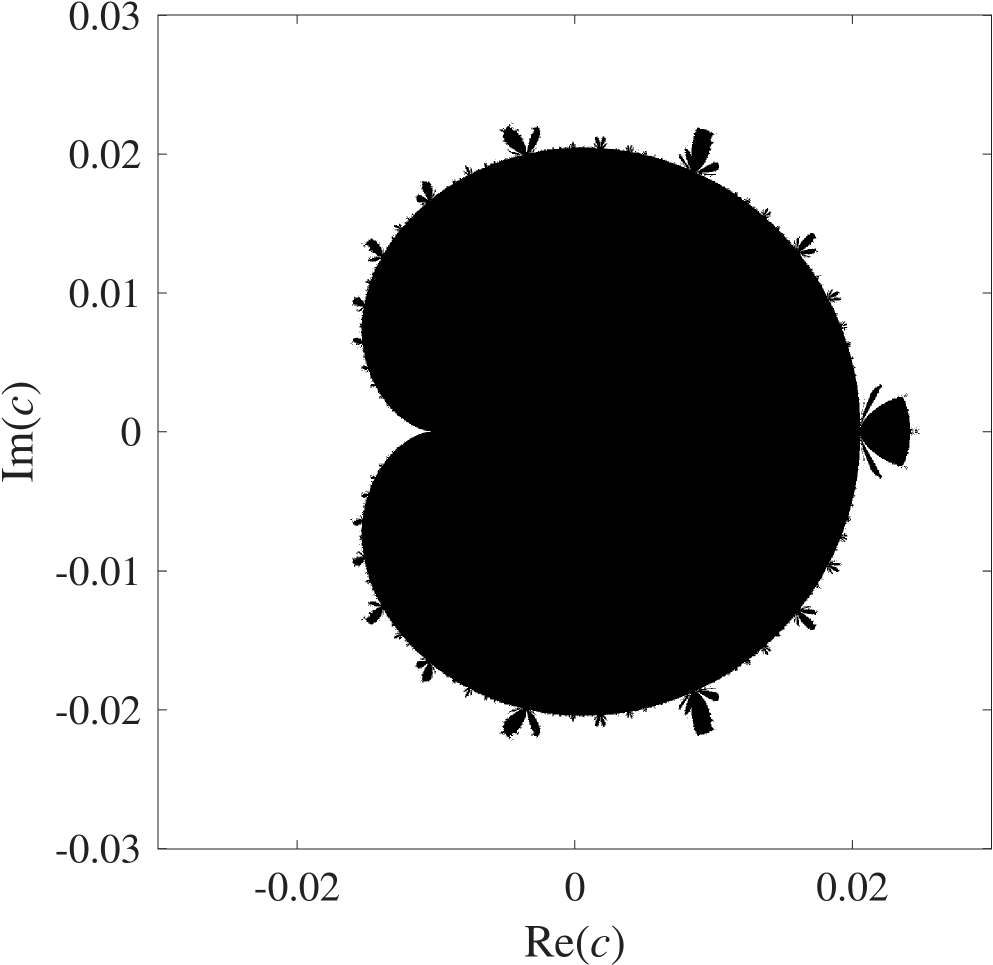}
\caption{}
\end{subfigure}

\caption{\textbf{Different approaches to the Rest connectome data} using the equi-M set. \textbf{(a)} Statistical equi-M set: the light to dark gradient shows the fraction of equi-M sets that the respective point in $\mathbb{C}$ belongs to. \textbf{(b)} The equi-M set computed for the prototypical Rest connectome.}
\label{fig:functional_rest_stats_and_prototypes}
\end{figure}

\begin{figure}[h!]
\centering
\begin{subfigure}{0.5\linewidth}
\centering
\includegraphics[width=0.95\linewidth]{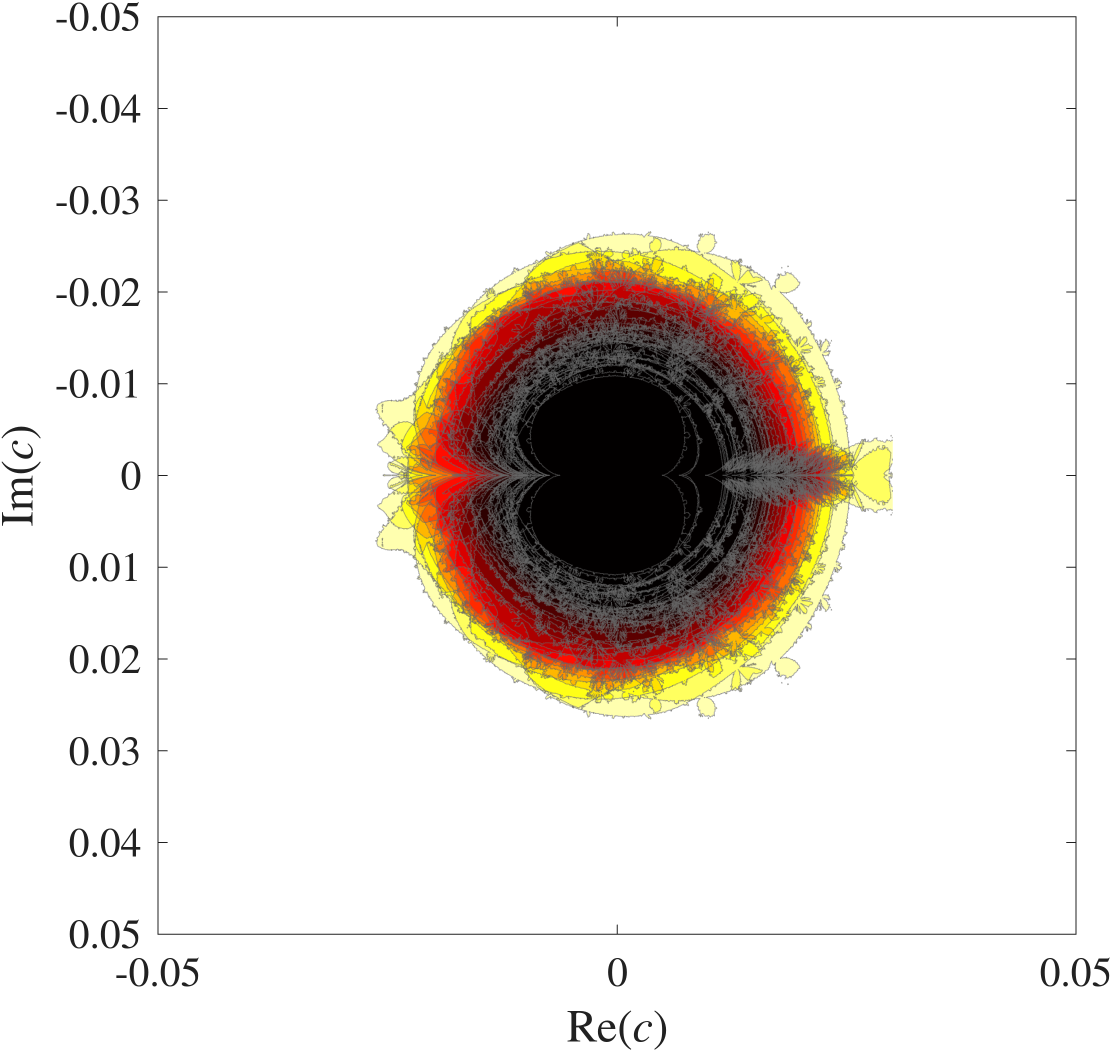}
\caption{}
\end{subfigure}%
\begin{subfigure}{0.5\linewidth}
\centering
\includegraphics[width=0.94\linewidth]{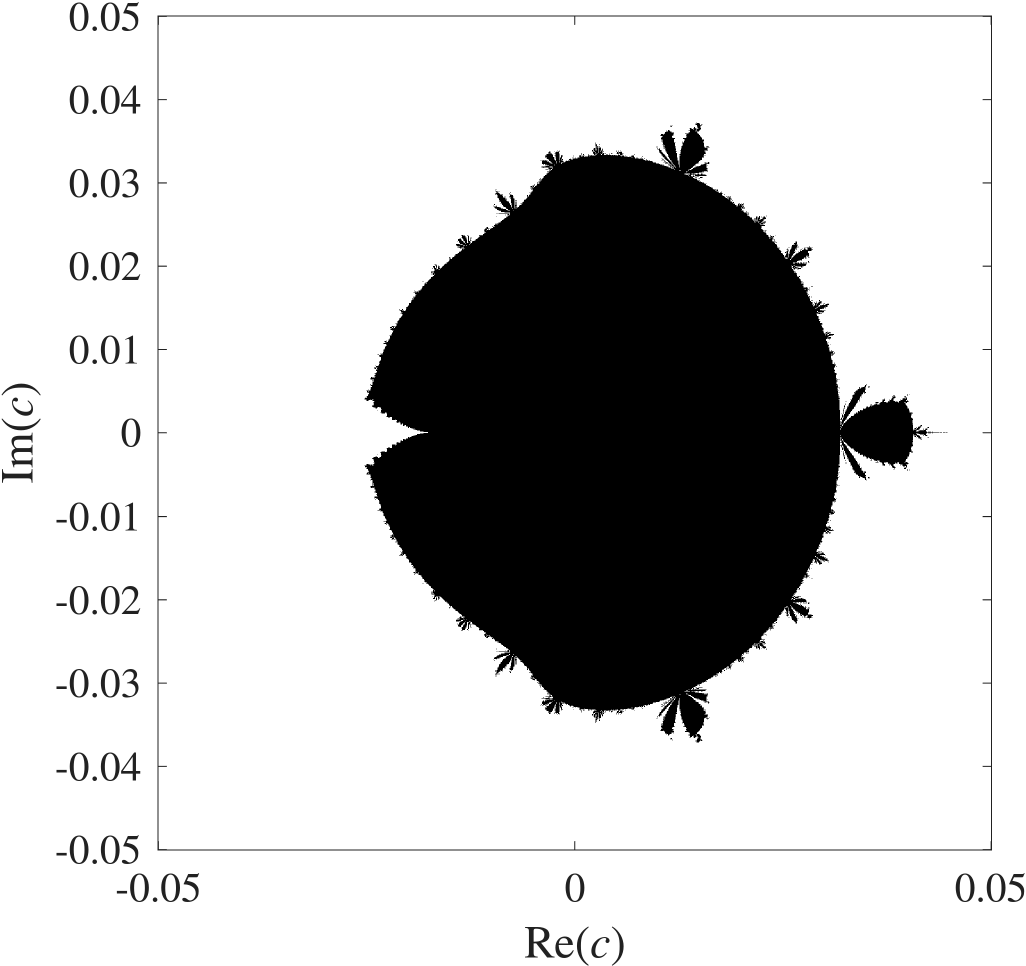}
\caption{}
\end{subfigure}

\caption{\textbf{Different approaches to the Emotion connectome data} using the equi-M set. \textbf{(a)} Statistical equi-M set: the light to dark gradient shows the fraction of equi-M sets that the respective point in $\mathbb{C}$ belongs to. \textbf{(b)} The equi-M set computed for the prototypical Emotion connectome.}
\label{fig:functional_emotion_stats_and_prototypes}
\end{figure}

\begin{figure}[h!]
\centering
\begin{subfigure}{0.5\linewidth}
\centering
\includegraphics[width=0.95\linewidth]{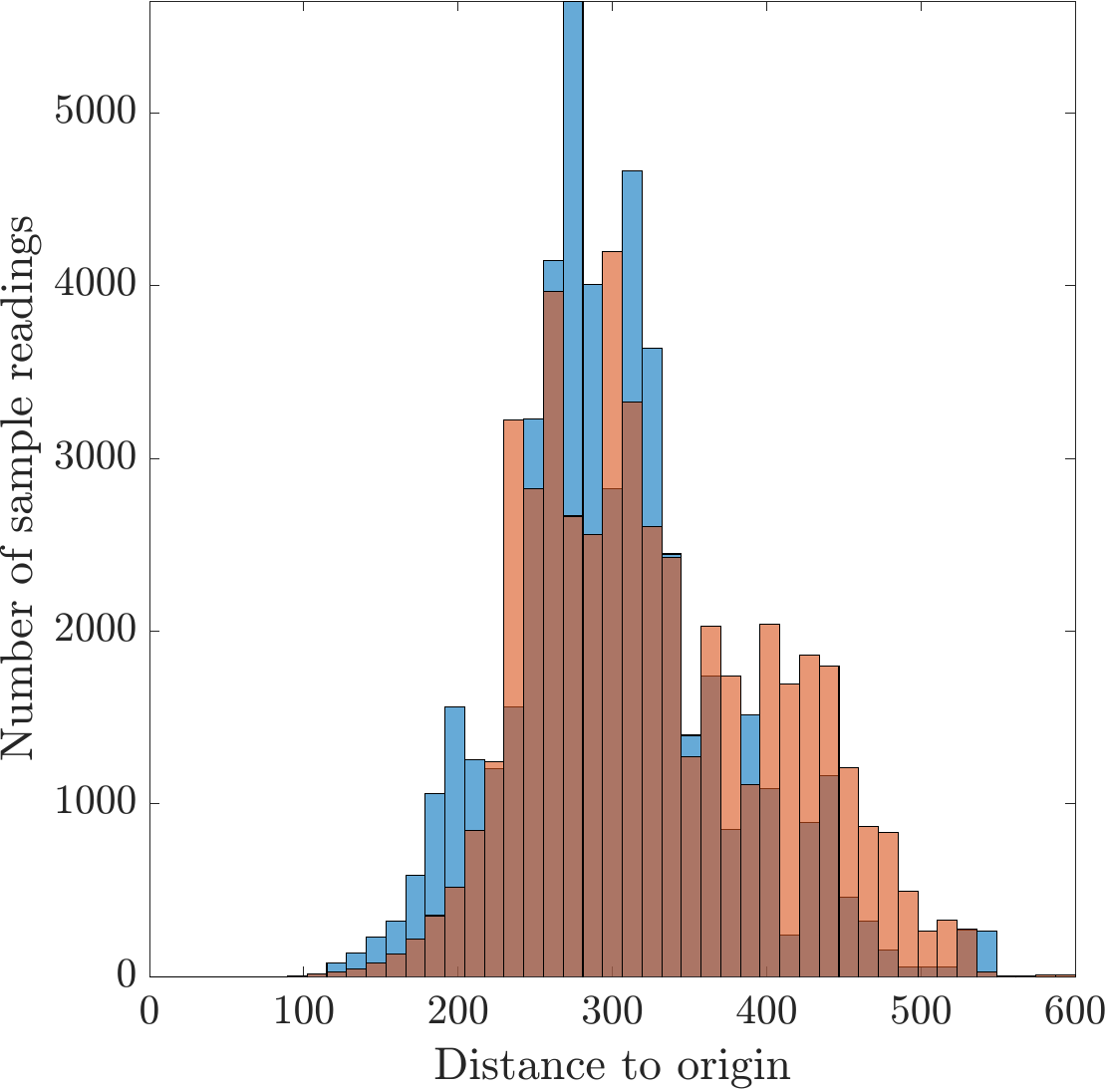}
\caption{}
\end{subfigure}%
\begin{subfigure}{0.5\linewidth}
\centering
\includegraphics[width=0.92\linewidth]{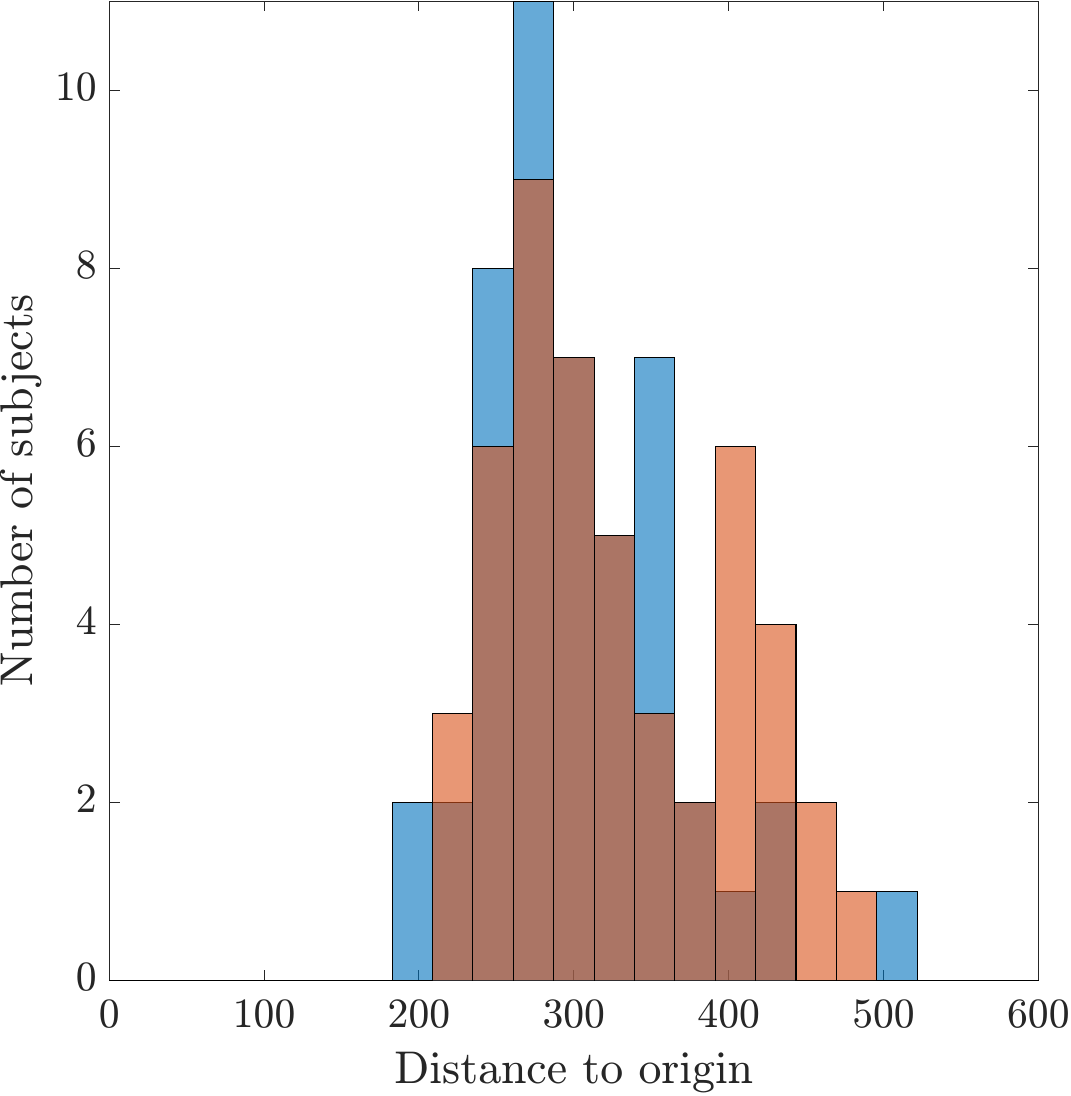}
\caption{}
\end{subfigure}

\caption{\textbf{Difference in the spread of equi-M size between the Rest and Emotion conditions.} (a) Each of the two histograms shows the distribution of $D_{\theta k}$ over all subjects and all sample angles, in blue for Rest and in red for Emotion. (b) Each of the two histograms shows the distribution of $\langle D_{\theta k} \rangle$ over all subjects, in blue for Rest and in red for Emotion.}
\label{fig:functional_spread_distribution}
\end{figure}

For a visual assessment of each individual group, ~\Cref{fig:functional_rest_stats_and_prototypes,fig:functional_emotion_stats_and_prototypes} illustrate the stochastic and the prototypical equi-M set for Rest and Emotion connectomes respectively. One may notice that both statistical and prototypical sets for Emotion are a lot more spread out than those for Rest. This is interesting since it suggests more diverse functional dynamics when processing emotional content and more unifying dynamics when processing neutral input (possibly related to reliance on the default network at rest). To verify this interesting result quantitatively, we sampled the spread of the boundaries of the equi-M set in each case, by taking a rotating radius and keeping track of where it intersects these boundaries. We obtained a distribution $D_{\theta k}$, where $0 \leq \theta \leq 2\pi$ is the angle \Revision{(sampled with a step of 0.001)}, and $1 \leq k \leq 48$ is the subject in the data set. In Figure~\ref{fig:functional_spread_distribution} we compare the Rest and Emotion histograms for $D_{\theta k}$ (considered for all values of $\theta$ and $k$), as well as for the means $\langle D_{\theta k} \rangle$ of these distributions over all sampled $\theta$s. A \Revision{Levene} non-parametric test shows that, while the means of these distributions are not a distinguishing factor between them, they are different in variance. For the overall distributions in the left figure panel, we obtained significantly different variances $\sigma_\text{Rest} = 0.0035$ and $\sigma_{Emotion}=0.0044$, $F=1700$, with $df_{Rest}= 1$, $df_{Emotion}=96094$ and $p<10^{-10}$.
For the mean distribution in the right panel, averaging somewhat obscured the effect, but we still found a trend in differences, with $\sigma_\text{Rest} = 0.0028$ and $\sigma_{Emotion}=0.0038$, $F=3.02$, $df_{Rest}= 1$, $df_{Emotion}=94$ and $p=0.08$). This confirms a generally wider variability on dynamics for Emotional task than for Rest.

\begin{figure*}
\centering
\begin{subfigure}{0.43\linewidth}
\centering
\includegraphics[width=0.9\linewidth]{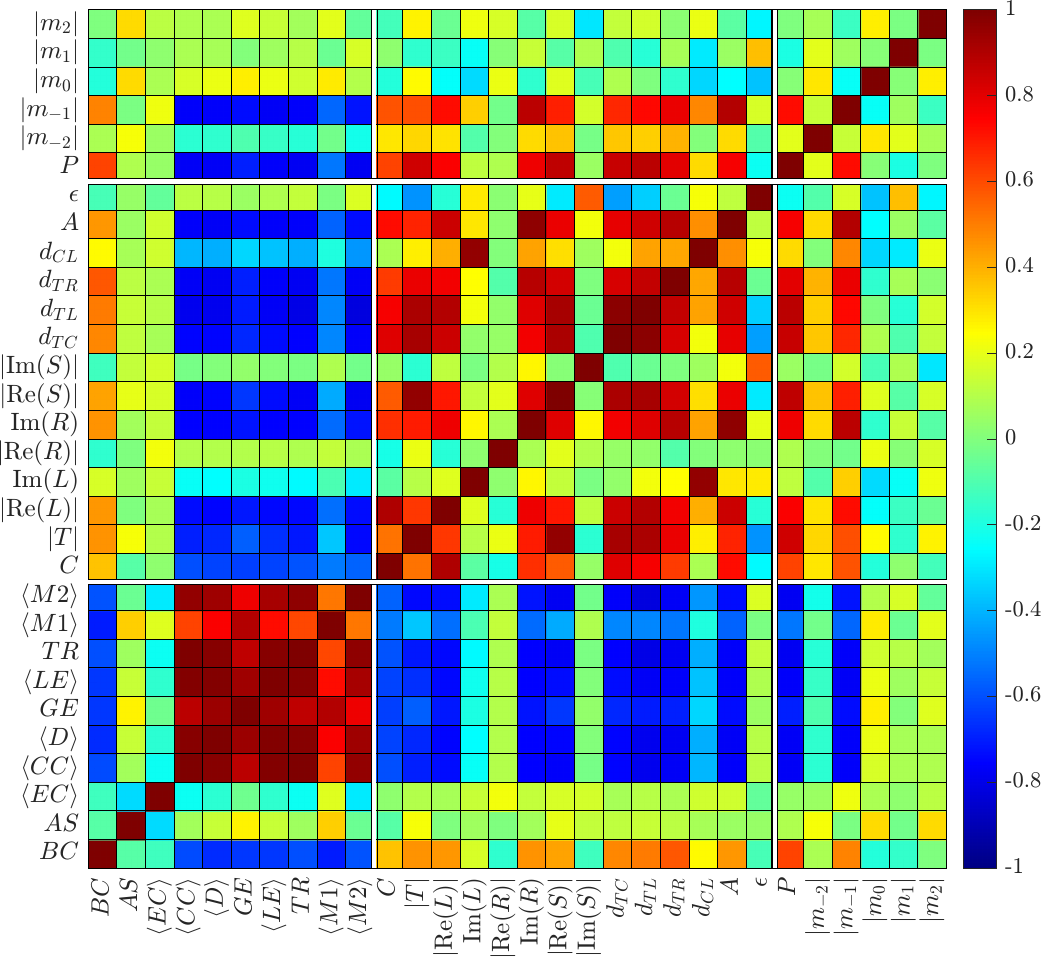}
\caption{}
\end{subfigure}%
\begin{subfigure}{0.43\linewidth}
\centering
\includegraphics[width=0.9\linewidth]{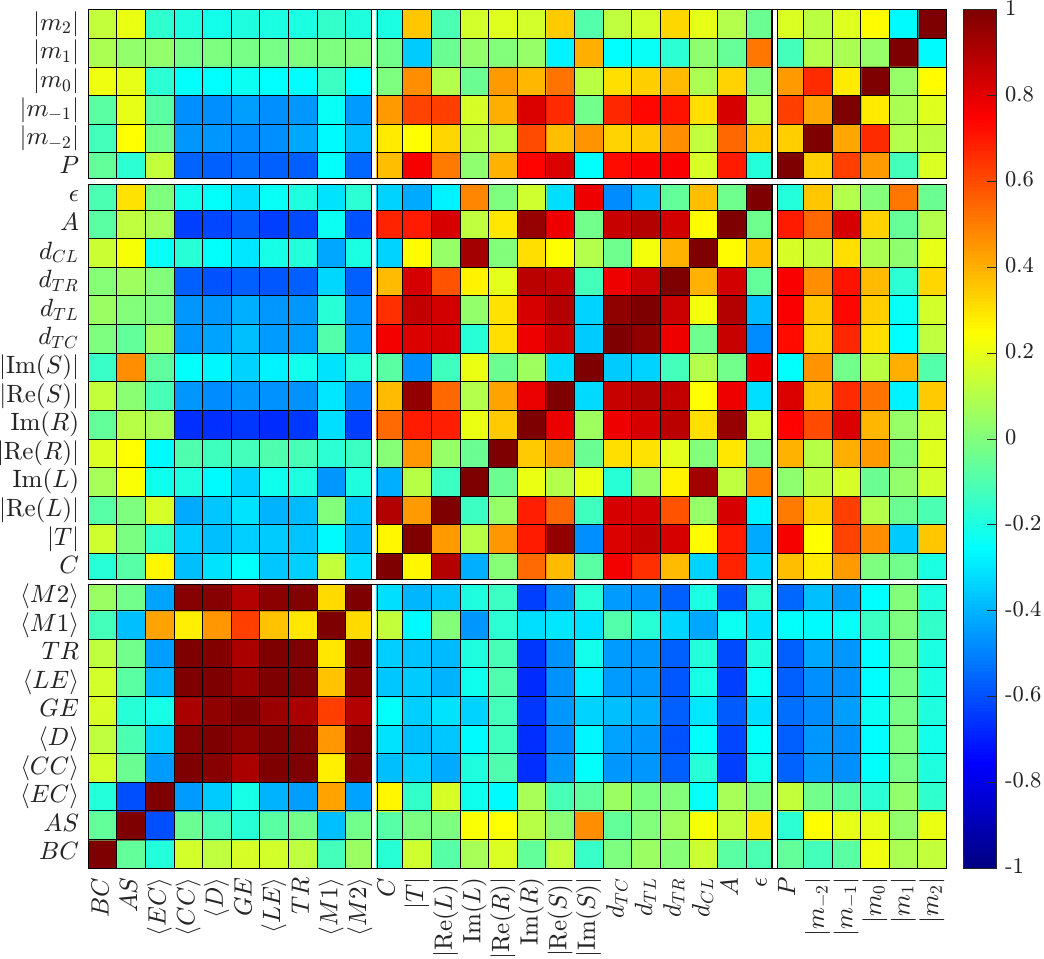}
\caption{}
\end{subfigure}

\begin{subfigure}{0.43\linewidth}
\centering
\includegraphics[width=0.9\linewidth]{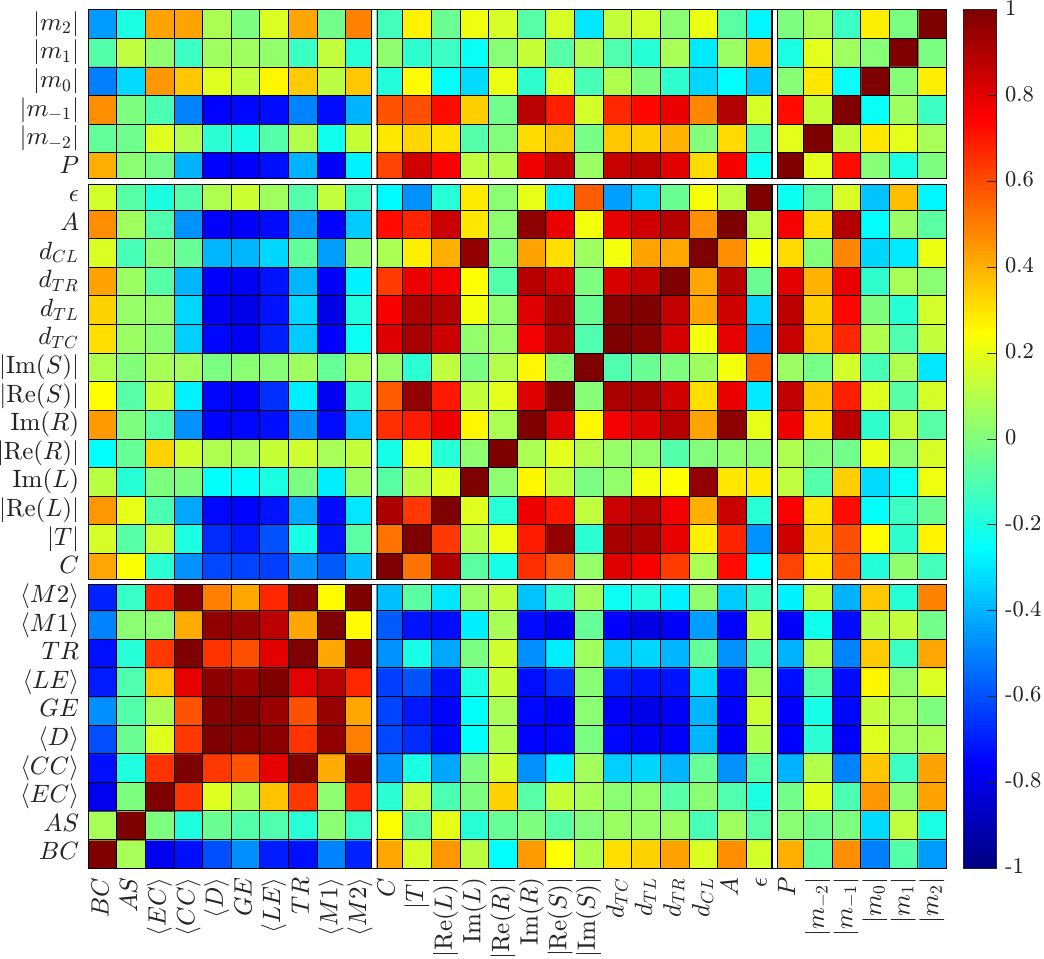}
\caption{}
\end{subfigure}%
\begin{subfigure}{0.43\linewidth}
\centering
\includegraphics[width=0.9\linewidth]{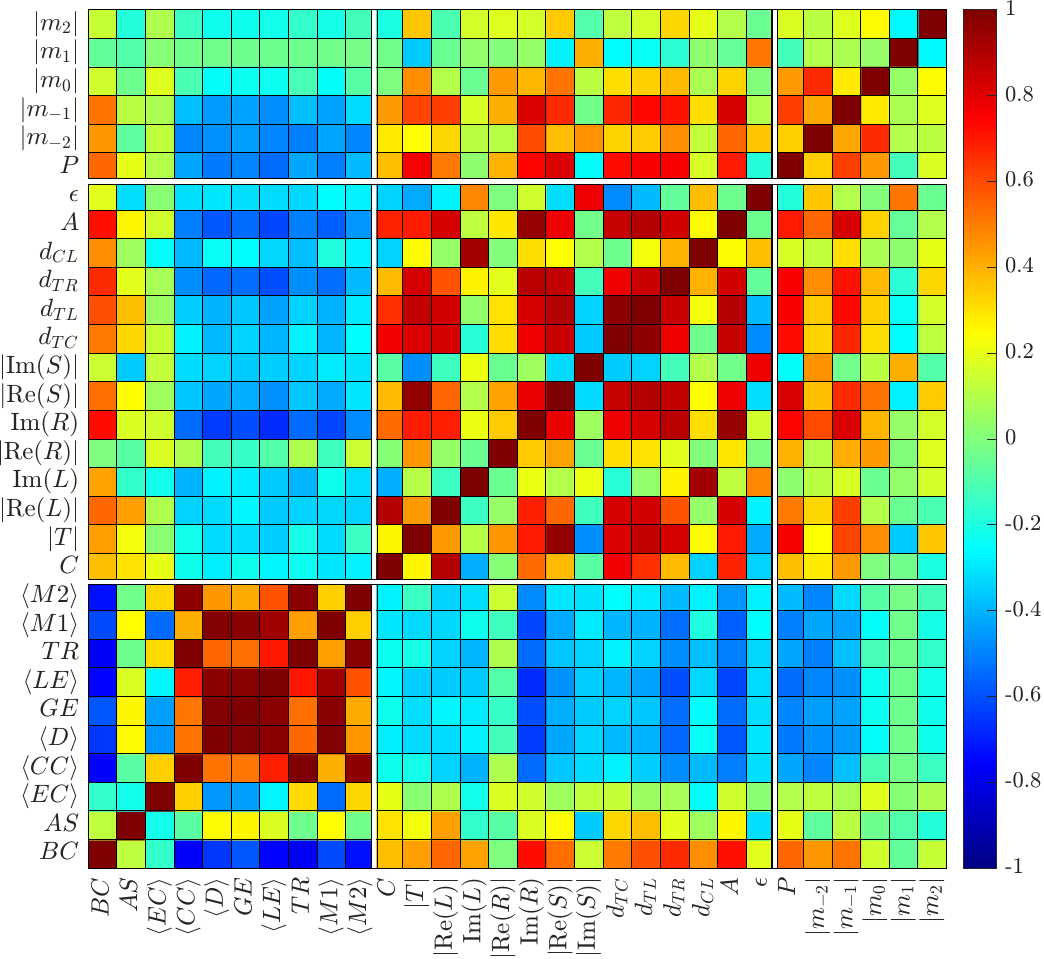}
\caption{}
\end{subfigure}

\begin{subfigure}{0.43\linewidth}
\centering
\includegraphics[width=0.9\linewidth]{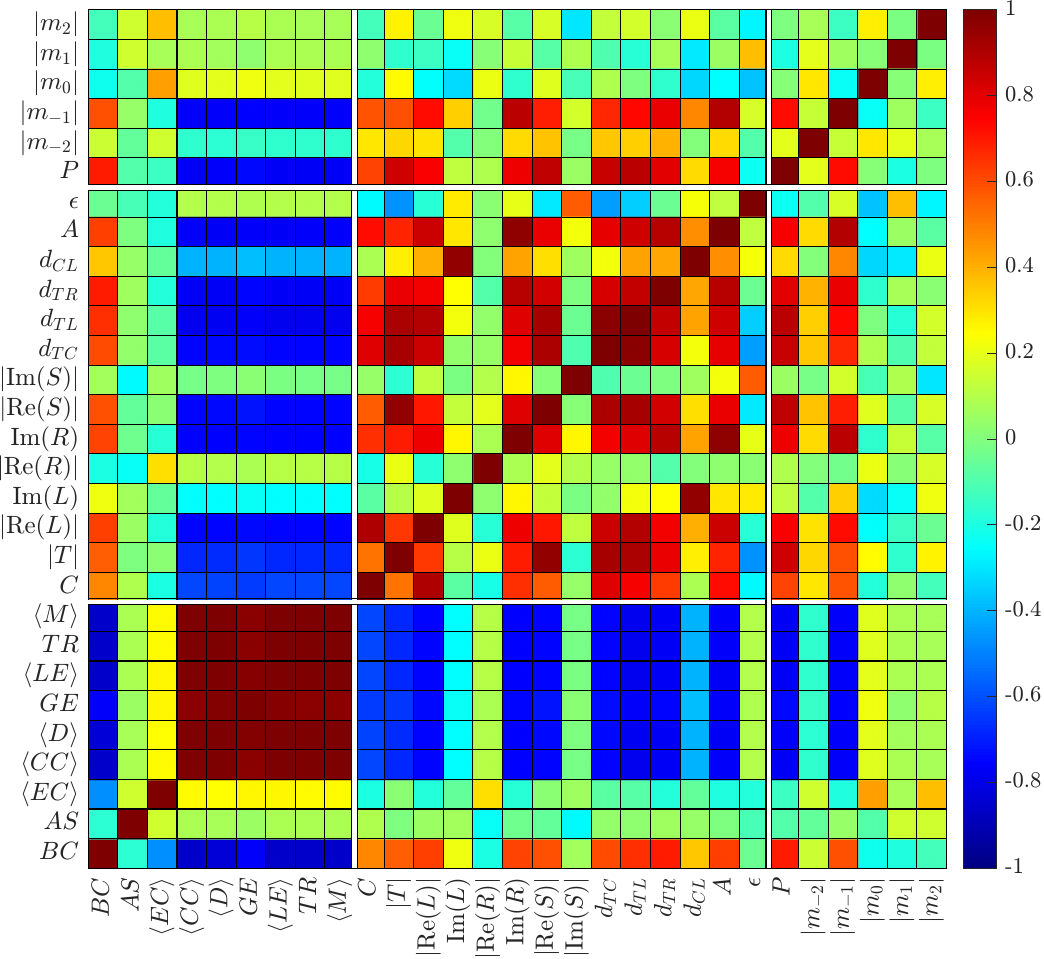}
\caption{}
\end{subfigure}%
\begin{subfigure}{0.43\linewidth}
\centering
\includegraphics[width=0.9\linewidth]{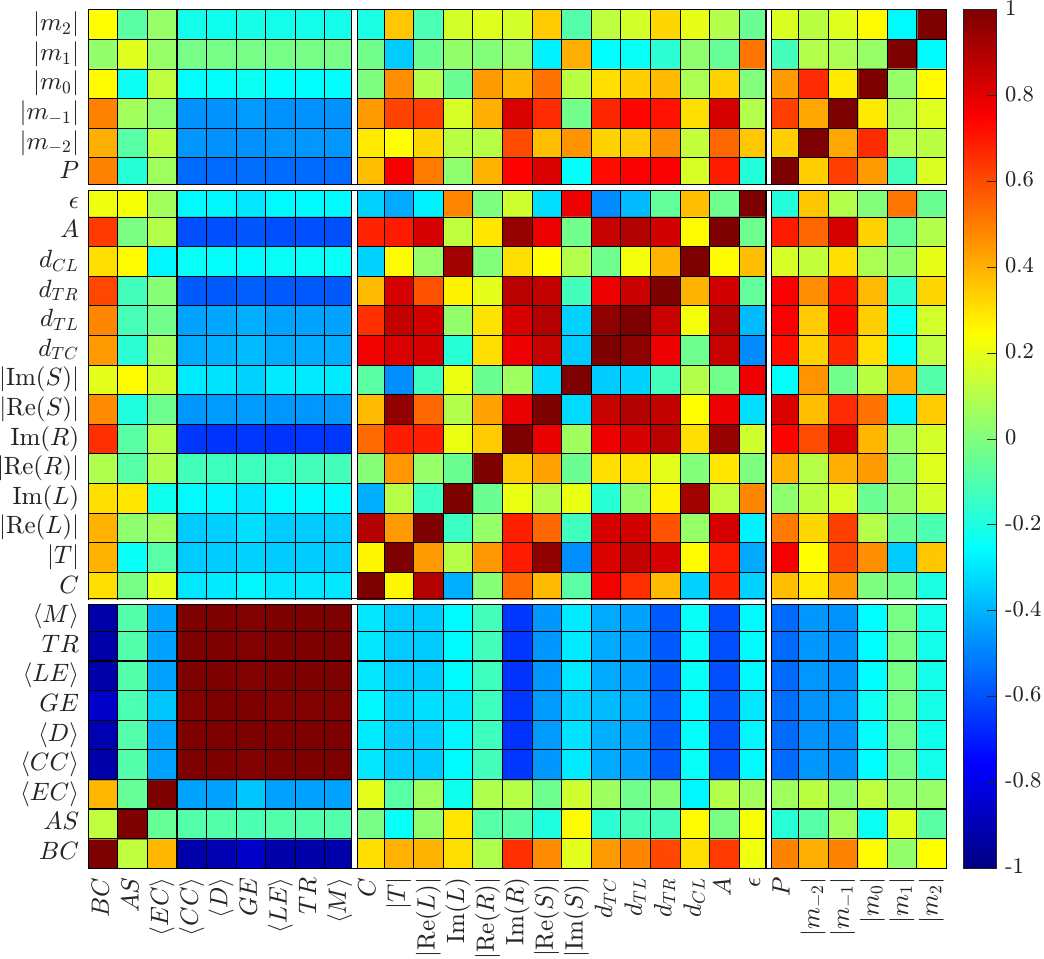}
\caption{}
\end{subfigure}

\caption{\textbf{Correlations} between the graph-theoretical measures of the associated networks, and the geometric measures of the equi-M set of the original network. The left panels \Revision{(a,c,e)} represent Rest connectomes, and the right panels \Revision{(b,d,f)} represent Emotion connectomes. The correlations are presented, top to bottom, for: positive subnetwork, negative subnetwork and absolute value network. In each panel, the measures are listed along both coordinate axes, in the order found in the tables, and in the order introduced in the text. For $C$, Re$(R)$ and $G$, we used their absolute values, to illustrate the distance to the origin. For the complex modes $m_j$, with $-2 \leq j \leq 2$, we use their moduli (to illustrate their strength). In this scheme, the three diagonal squares show correlations between measures within the same category, and the off-diagonal rectangles show correlations between measures in different categories. For simplicity, we only included the $\rho$ values here.}
\label{fig:functional_correlations}
\end{figure*}

A natural next step is to revisit more systematically the relationship between the measures of dynamics (geometric and Fourier measures) and the graph-theoretical measures (for the connectomes' associated networks). By comparing the correlation profiles in~\Cref{fig:structural_correlations}a with those in \Cref{fig:functional_correlations}, one can easily see some similarities, but also crucial differences.

First, let us notice that, within the collection of graph-theoretical measures for the associated networks, some degree of correlation exists (for each of the associate networks), although not as consistent as for the Structural connectome. However, for Rest and Emotion equally, the correlations within the collection of geometric and Fourier-derived measures are lower, making fewer of them significant. This confirms that the shapes of the Functional equi-M sets are more irregular than those for Structural sets (e.g., the position of the cusp is less predictive of the size of the set or of the position of the other landmarks).

Second, the profile of correlations between the geometric measures and the graph-theoretical measures for the associated networks is lower. In the case of Rest connectomes, significant correlations persist primarily between the subsets of measures that are correlated with each other within their own group. The top left panel in \Cref{fig:functional_correlations} shows that, for the positive sub-network, there is a subset of graph measures and a subset of geometric measures that are both ``unified'' by high positive mutual correlations within each set (red blocks). In addition, these are the measures that also show significant negative inter-correlations (blue stripes). The same is true, to a lesser extent, of the negative sub-network and the absolute value network for Rest connectomes. This suggests that some measures of the ``size'' of the equi-M set may still capture graph-theoretical graph patterns in the associate networks of the connectome. However, our experiment of constructing equi-M sets independently for each associate network led to equi-M sets which were identical up to scaling to the traditional Mandelbrot set, across the whole data. This further tells us that the only relevance that the architecture of these networks may have on the global network dynamics is weak (it only impacts the size of the equi-M set, and none of the more subtle shape aspects). This idea is reinforced by the fact that the only significant correlations for Fourier-derived measures appear for the perimeter, which is also a measure of size, and not the modes (which are finer descriptions of the boundary variability of the equi-M set). These effects are even more pronounced in the case of Emotion connectomes (right panels).

\begin{figure}[ht!]
\centering
\begin{subfigure}{0.5\linewidth}
\centering
\includegraphics[width=0.92\linewidth]{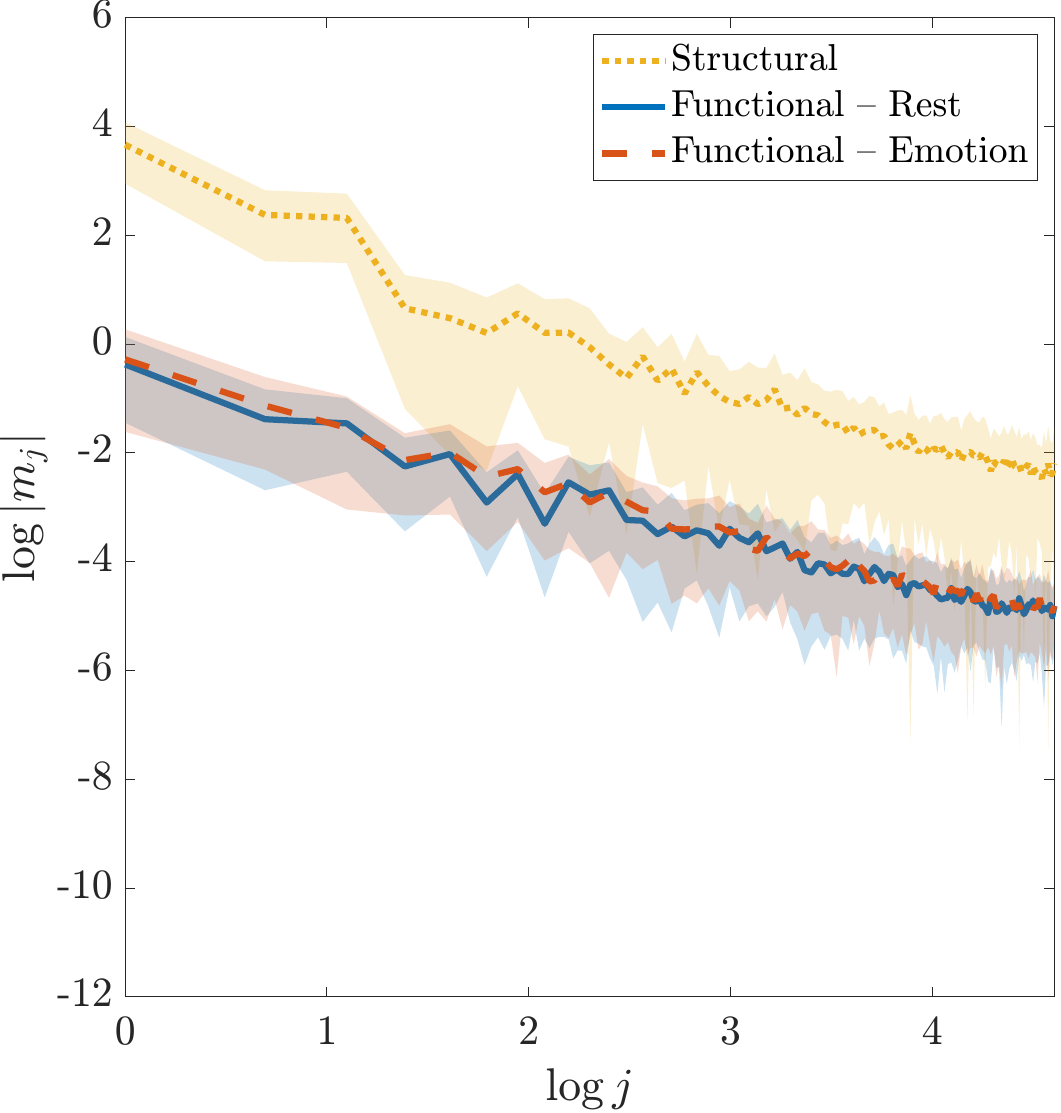}
\caption{}
\end{subfigure}%
\begin{subfigure}{0.5\linewidth}
\centering
\includegraphics[width=0.95\linewidth]{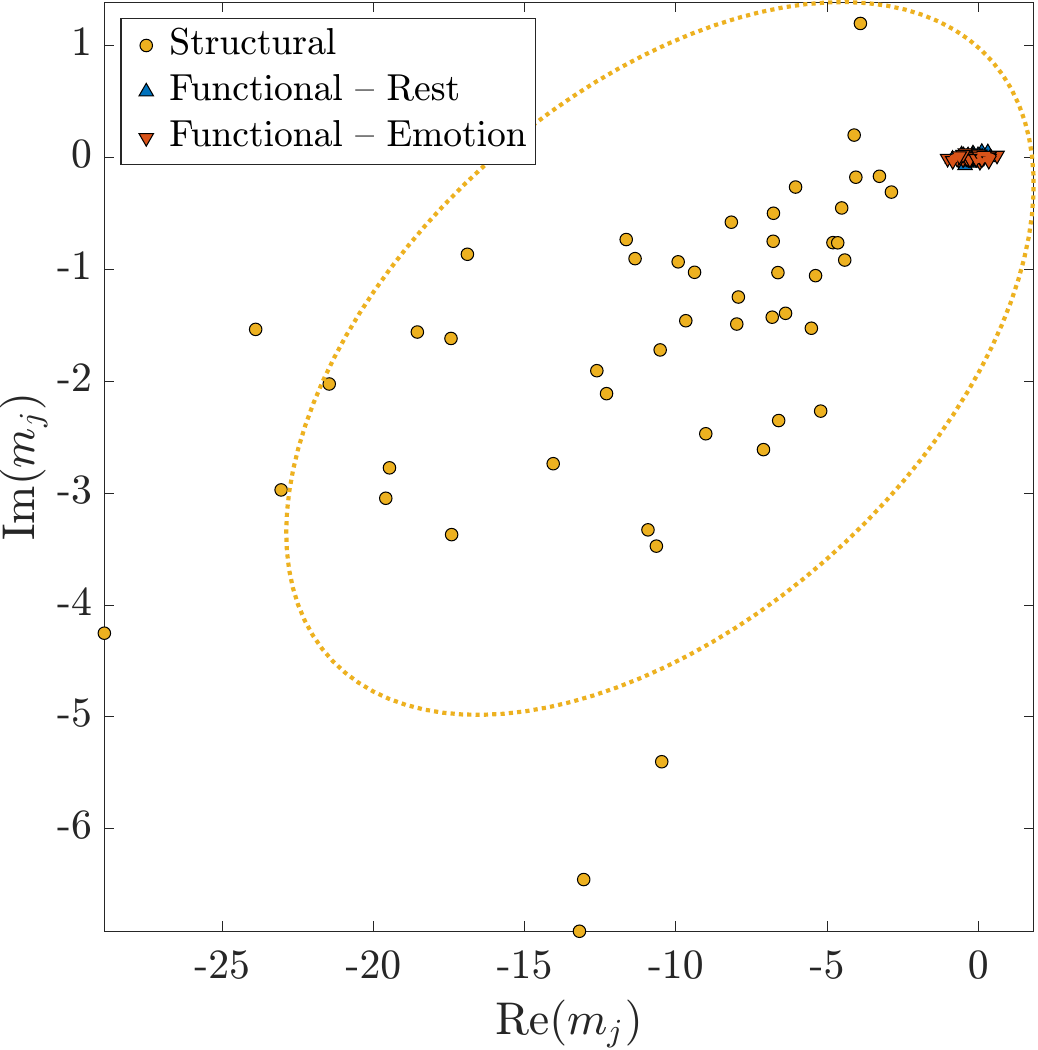}
\caption{}
\end{subfigure}
\begin{subfigure}{0.5\linewidth}
\centering
\includegraphics[width=0.95\linewidth]{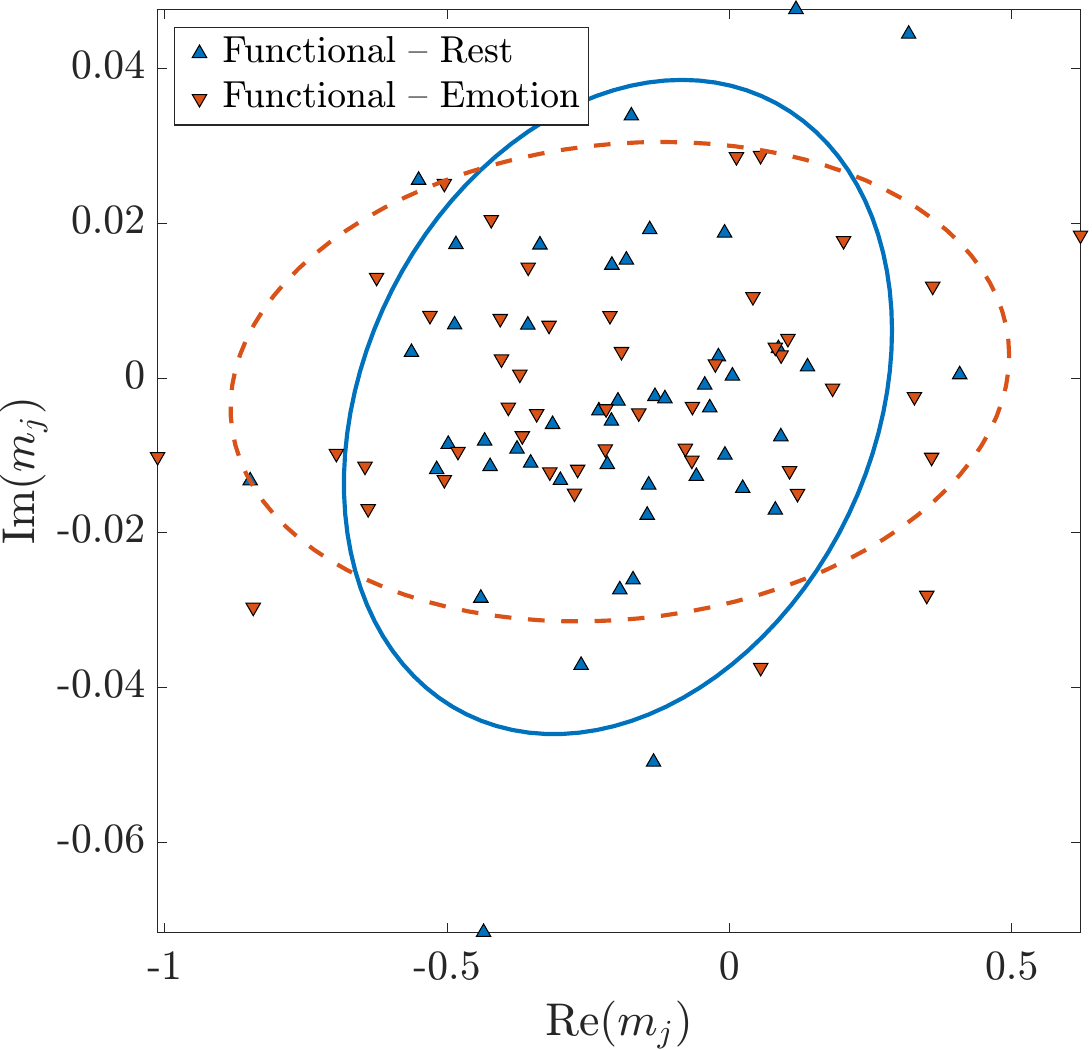}
\caption{}
\end{subfigure}%
\begin{subfigure}{0.5\linewidth}
\centering
\includegraphics[width=0.95\linewidth]{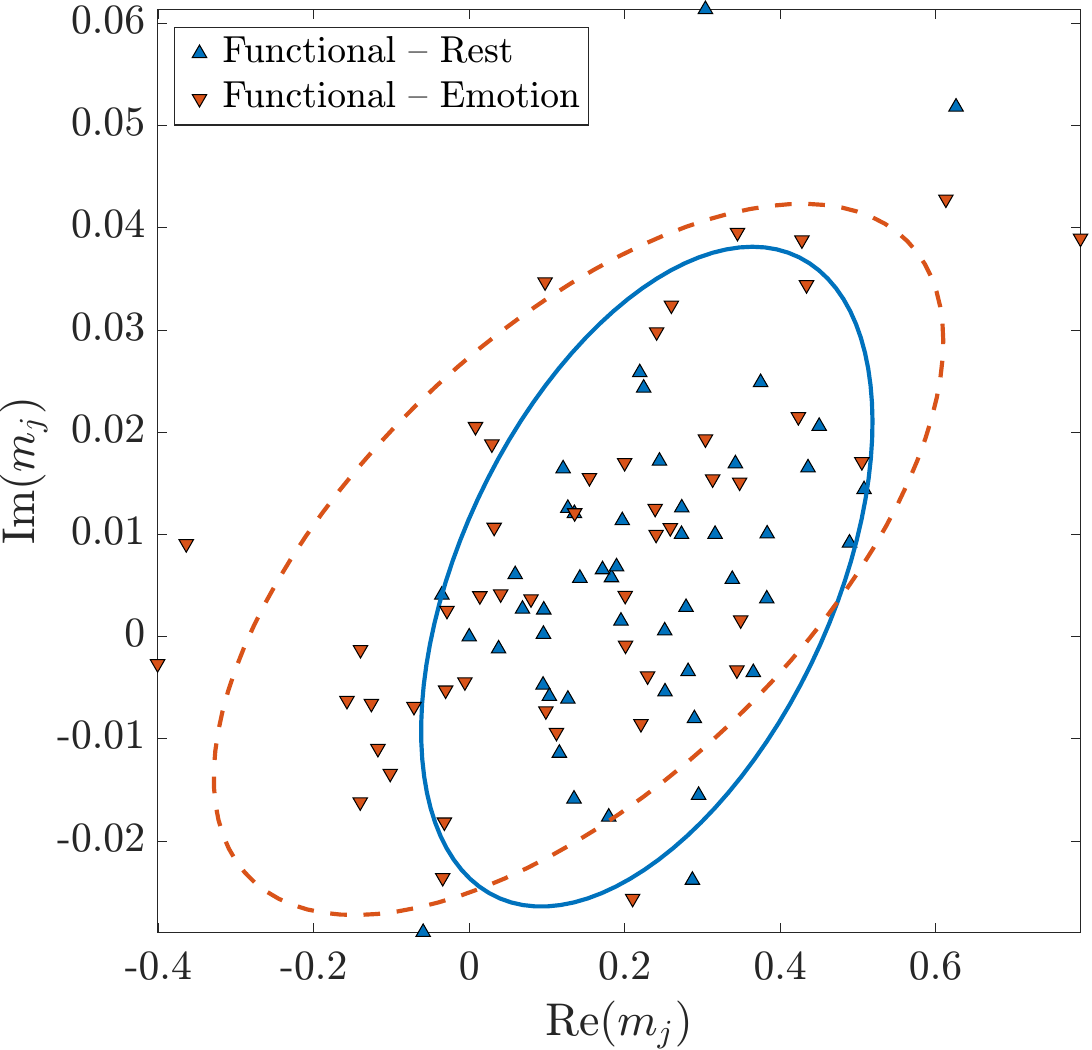}
\caption{}
\end{subfigure}

\caption{\textbf{Differences in modes between connectome types.} \textbf{(a)} log-log representation of the positive power spectra up to $N = 100$. The mean power log-log spectra  ($\log j, \log |m_j|$), for $0 \leq j \leq N$, are shown for each subject, for the equi-M sets of Structural (orange dotted), Rest (blue line) and Emotion (red dashed) connectomes. For each group, one standard deviation is overlaid around the mean. \textbf{(b)} Distributions of the second positive mode $m_1$ in the complex plane for all three connectome types; the scatter plots for Structural (orange circle), Rest (blue triangle up) and Emotion (red triangle down) data are accompanied by the confidence ellipses. \textbf{(c)} Closeup in the complex plane of \textbf{(b)}, showing in more detail the differences in the distributions of $m_1$ between Rest and Emotion data. \textbf{(d)} Differences in the distributions of the third mode $m_2$ between Rest and Emotion connectomes. All scatter plots are shown together with their confidence ellipses.}
\label{fig:fourier_mode1_spread}
\end{figure}

A natural question is whether subtle dynamic invariants such as the modes of the equi-M set boundary can themselves capture relevant information on dynamics, and be used as predictors/classifiers of dynamics. Along this line of inquiry, we investigated if the modes can efficiently differentiate between Structural, Rest and Emotion connectomes.

To illustrate this, the panels of~\Cref{fig:fourier_mode1_spread} offer two perspectives on the differences between the modes for Structural vs Functional connectomes, and also between the modes for Emotion vs Rest connectomes. The first panel shows the log-log power spectra for all connectomes, as well as the group averages. Notice that the magnitudes of the modes for the Structural data are dramatically larger than those for the Functional data (both Rest and Emotion). This is confirmed by the second panel, which illustrates as a specific example the distribution of the second mode $m_2$ in the complex plane, as separate scatter plots accompanied by the corresponding confidence ellipses. The distribution for the Structural data (orange dots) is dramatically more spread out than the distributions for Rest (blue) and Emotion (pink dots), with Levene test significances $p < 10^{-4}$ for both Structural vs. Rest and Structural vs. Emotion.

Notice that the differences between modes for Rest and Emotion connectomes are not as dramatic, but still powerful. Although the Rest vs. Emotion power spectra do not appear significantly different, the distributions of each of these modes in $\mathbb{C}$ present quantifiable differences between Rest and Emotion. These differences are illustrated for $m_2$ and $m_3$ in particular in the bottom \Revision{two} panels. In each case, confidence ellipses are included, emphasizing the differences between their spread and principal components. In addition, a separate Levene test confirms that the distribution of $m_2$ for Rest have significantly lower variability ($\sigma_{Rest}=0.05$) than that for Emotion ($\sigma_{Emotion}=0.11$), with $F=6.70$ and $p=0.01$. Similarly for $m_3$, $\sigma_{Rest}=0.02$ and $\sigma_{Emotion}=0.05$, with $F=8.78$ and $p=0.003$. \Revision{These results are significant even when using a Bonferroni corrected $\alpha = 0.01$ to account for using comparisons over five modes.} This confirms the earlier result that the shape variability of the equi-M sets is higher in the Emotion task than in Rest.

Above all, this shows that the mode distribution can differentiate effectively between types of connectomes. It is not a surprise that the difference is dramatic between Structural and Functional connectomes, since they are fundamentally different types of matrices (positive vs signed). However, we showed that these measures can also capture the finer differences between Rest and Emotion connectomes, which are based on a functional task, and which are harder to pinpoint based on partial graph measures. This comparison is easier to contextualize, since it suggests an idea that we described before: greater variability between subjects during emotional tasks than during rest. We will further comment on this in the Discussion section.

One unifying conclusion is that, while the associated networks may give only restricted information, and should not be used to capture subtleties of the overall dynamics, or functional dynamic variability between subjects. We further suggest that dynamic measures outperform partial graph measures, since (1) they incorporate information on how the positive and negative parts of the network actually interact and (2) they assess how information flows in a working network, rather than simply its functional architecture.

\section{Discussion}


In this paper, we explore how the asymptotic behavior of quadratic iterations can be used to assess the range of dynamics in networks, in a way that is more efficient than conventional graph measures. The idea comes from the need to have a canonical measure of actual dynamics that transcends the hard-wired network architecture. In our work with complex quadratic networks (CQNs), we generate and test theoretical hypotheses about networks in general, but we have also been exploring how this can be applied to support progress in our understanding of natural networks, such as the brain.

An important theoretical direction in this particular study has been to understand the effects of negative signs on the geometry of the equi-M set. The data used was ideal for such an exploration, since it provided Structural connectomes (positive symmetric matrices) and Functional connectomes (signed symmetric matrices) for the same pool of subjects. Our analysis discovered fundamental differences between the equi-M sets for Structural and for Functional networks. One general observation suggests that, for positive connectivity matrices, the equi-M set typically has a cusp point on the right side. We noticed that this cusp moves to the left if sufficient negative connections are included in the network, making the set look ``flipped.'' This phenomenon is difficult to capture and quantify exactly, since the quadratic map is invariant to sign, hence eventually switching all positive connections to negative will return to the original set. In a separate theoretical project, we are working to identify more precisely the critical window in which the set flips, by what mechanism, and to what extent this depends on the exact position and weights of the negative edges.

In addition to the general differences between positive and signed connectomes that are common to all connectomes, we also identified distinctions specifically influenced by the neuroscience characteristics of these networks as brain connectomes, as well as their specialized design for functional brain dynamics. In our previous work, we showed that the geometry of the equi-M set can be very diverse even for positive symmetric networks. The Structural connectomes reflect some of that diversity, but the range of geometries eludes certain properties. For example, all our sets were connected and had similar eccentricities, although their sizes varied widely. In addition, all measures of size and boundary variability were mutually correlated and also significantly correlated with the graph measures we computed for the positive network.

In the past, we used measures of the equi-M set size to distinguish statistically and prototypically between genders, for Structural (positive) connectomes. In this work, we successfully used measures of the size and shape variability of the equi-M set to distinguish between Structural and Functional connectomes, as well as between Rest and Emotion. We explored the advantages of using geometric measures versus using graph measures to assess dynamic range in Functional connectomes. Since graph measures only apply to positive connectomes, for Functional (signed) connectomes, network measures are traditionally computed separately for the positive and negative sub-networks, as well as for the absolute values network (we called these the networks associated with the Functional connectome). Using the equi-M set as a test dynamic model, we showed that considering these associate networks in isolation does not paint a complete picture, and that the interactions between the positive and negative sub-networks play a crucial role in the overall emerging dynamics.

This was done in two ways. First, in addition to computing the equi-M sets for whole Functional connectomes, we also computed the sets corresponding to the associated networks. Although the original equi-M sets exhibited a lot of variability in shape (and not just size), the sets for the associated networks were identical to the Mandelbrot set up scaling (via small detail potentially due to precision). This made it clear that, for the quadratic model, many of the subtleties in dynamics that encode differences between individuals cannot be captured simply by considering the associated networks.

It is not too big a stretch to suggest that this may also be the case for more complex node dynamics (such as neural dynamics). This effect did not come from simply transitioning from a signed Functional connectome to a positive network (since Structural connectomes, although positive, also have shape diversity). The effect is likely due to the sparse distribution of connections comprised in the positive and negative sub-networks, which, taken independently, are insufficient to recover the complexity of the overall connectome, in absence of the interactions between them. This hints towards the fact that the interaction between the positively and negatively correlated regions contribute fundamentally to the emergent properties of the network, rather than simply being an artifact of its structural composition.

To support this idea, we also computed correlations between equi-M geometric and Fourier measures and the graph measures for the associated networks of the Functional connectomes. Although Structural connectomes exhibit strong and significant negative correlations, the correlation profiles were generally lower in Functional connectomes, with significant values only for the measures that reflect the size of the set and not the shape. In particular, no significant correlations were found with the Fourier-derived modes (correlations which were strong in Structural connectomes). To clarify that this is not a shortcoming of the Fourier measures, we showed that these encompass crucial dynamic information, and can be used in and of themselves to differentiate between Rest and Emotion connectomes.

Indeed, our final analysis showed Fourier-based modes to be an efficient differentiator not only between Structural and Functional equi-M sets, but also between our two types of Functional connectomes, in that the modes of the Emotion connectomes were significantly more spread out than those of Rest connectomes. This, together with the equi-M sets themselves showing a higher spread in Emotion than in Rest, suggested increased variability and diversity in dynamics for Emotional task than for Rest. Aside form its relevance as a quantitative assessment, this is an interesting result in the neuroscience context. It is not surprising that subjects appear to have more variable dynamics in Emotional processing than in Resting state (since Rest is known to rely primarily on the default network~\cite{greicius2003functional,raichle2001default,andrews2012brain}, while Emotion tasks may engage a variety of mechanisms, different for each subject). What is surprising and heartening is that the Fourier measures of the equi-M set were able to capture this idea.

Although in this analysis we only looked at the power spectrum overall and at the distribution of the first three nodes (defining the basic shape of the set), higher-level nodes also contribute to more subtle details in the geometry and fractality of the set. In future work, we will use these higher frequency profiles, together with other measures such as fractal dimension and entropy, to further analyze and interpret differences between connectome types.


In this paper, we applied our CQN approach to the Human Connectome Project data. However, equi-M sets are not limited to neuroscience. In fact, we believe their most impactful applications lie across the broader universe of complex systems and artificial intelligence, including artificial neural networks, multi-agent systems, power grid, communication and transportation networks, epidemiological and ecological networks. What unites these domains is the need for interpretable tools that bridge the gap between structure and emergent behavior.

The CQN framework allows one to model infrastructural systems with nonlinear dependencies and load dynamics, and classify dynamic modes or failure risks. Equi-M sets serve the role of canonical, geometry-based representations of dynamic potential. They provide a diagnostic lens for assessing stability, oscillation, or chaotic tendencies in network architectures, and offer an efficient way to evaluate cooperation, divergence, or instability across a system. And most importantly, they reveal how network topology affects long-term outcomes and response to perturbation, and can help detect the boundaries between robust signal propagation and breakdown. Along these lines, our approach provides a \emph{toolkit for interpreting high-dimensional coupled nonlinear dynamics}. Equi-M sets are especially useful in the context of capitalizing on quantitative geometric measures to detect and explain phenomena such as dynamic clustering, sensitivity to topology, phase transitions, and bifurcations. These measures can also be used as features in machine learning models or as regularizers in neural architecture search.


\subsection*{Conclusion: A Canonical Map for Emergent Dynamics}

The original 1970s insights from research on fractals revealed that complexity in nature could arise from simple, deterministic rules. Our development of equi-M sets carries that spirit into the 21st century -- recasting dynamics on networks as fractal geometry, and offering an interpretable, computationally tractable method for understanding how structure drives behavior in complex systems. This is not just a new metric, it is a new mode of analysis. Equi-M sets offer a visual grammar for network dynamics, a diagnostic framework for architecture and function, a bridge between theory and application across disciplines and new toolkit for interpretability and design in intelligent systems. As networked systems become increasingly central to science and technology, we believe equi-M sets will prove essential for those who wish not only to simulate or optimize -- but to truly understand -- how networks behave.

\subsection*{Acknowledgements}

The project received support from an AMS-Simons PUI Faculty Grant (R\u{a}dulescu) and from a joint NSF-UEFISCDI grant: Project No. \#2408407 (R\u{a}dulescu), Project No. ROSUA-2024-0002 (Kaslik, Fikl). Michael Anderson's work on the project was supported by New Paltz RSCA Summer and Academic Year Experience for Undergraduate Programs.

\subsection*{Data Availability}

\Revision{The software developed and used in this study is openly available in the Zenodo repository at \url{https://doi.org/10.5281/zenodo.16921908} under the MIT license. This archive (version 1.1) includes all source code and documentation required to generate the equi-M sets.}

\bibliography{references}

\clearpage

\section*{Appendix A: Equi-M sets for all subjects}
\label{ax:equi_m_sets}

\begin{figure}[h!]
\centering
\includegraphics[width=0.85\linewidth]{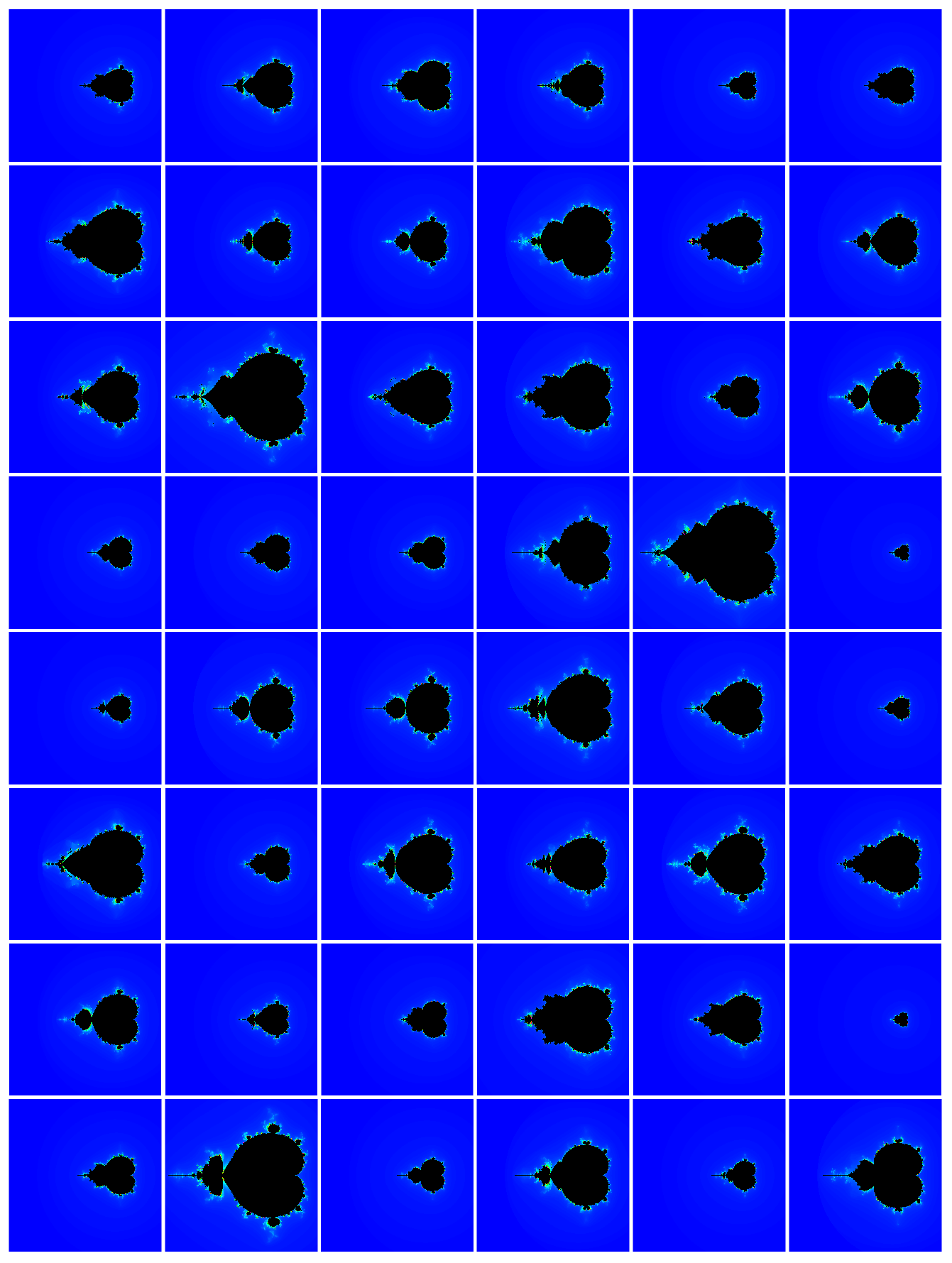}

\caption{\textbf{Structural equi-M sets} for all subjects in our data set. All sets are represented in the complex region $[-1,0.3] \times [-0.65,0.65]$, in resolution $1200 \times 1200$. The equi-M set in each panel appears in black, and the colors represent the number of iterations needed by each point in order to exit the escape disc.}
\label{fig:structural_equi_m_sets}
\end{figure}

\begin{figure}[h!]
\centering
\includegraphics[width=0.85\linewidth]{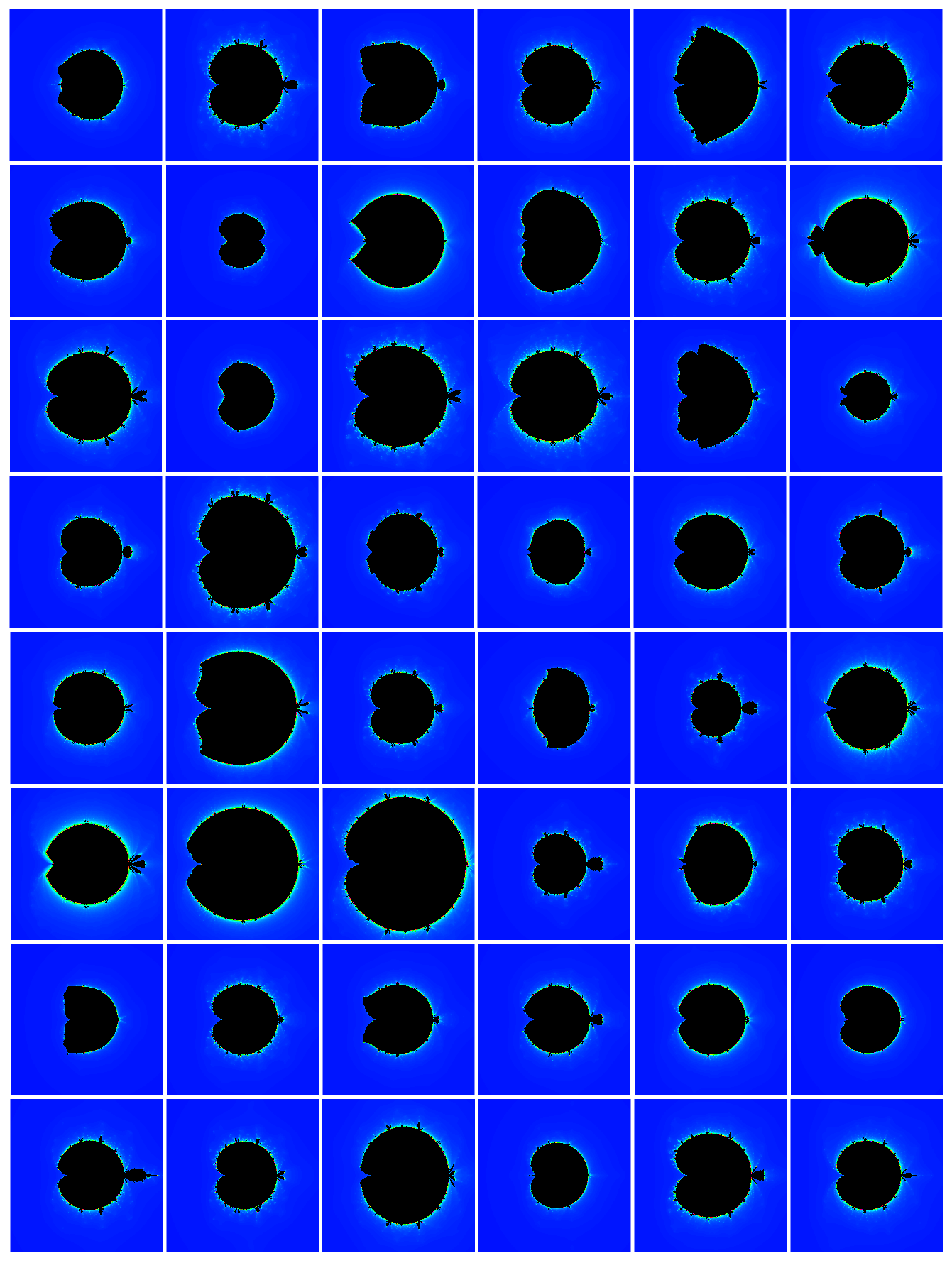}

\caption{\textbf{Equi-M sets for Functional Rest connectomes} for all subjects in our data set. All sets are represented in the complex region $[-0.03,0.03] \times [-0.03,0.03]$, in resolution $1200 \times 1200$. The equi-M set in each panel appears in black, and the colors represent the number of iterations needed by each point in order to exit the escape disc. The sets show a wide variety of shapes across the data set.}
\label{fig:rest_equi_m_sets}
\end{figure}

\end{document}